\documentclass[onecolumn,iop]{emulateapj}

\usepackage{graphics,color}
\graphicspath{{Figures/}{./}}
\usepackage{amsmath}

\newcommand{\lya}{Ly$\alpha$}

\def\2pr{^{\prime \prime}}
\def\sqdeg{{\,{\rm deg}^2}}
\def\deg{{$^\circ$}}
\def\vunits{{\,h^3{\rm Mpc}^{-3}}}

\newcommand{\fnl}{f_{\rm NL}}

\begin{document}

\title{The SDSS-IV extended Baryon Oscillation Spectroscopic Survey:
Overview and Early Data}

\slugcomment{Submitted to AJ}

\author{
Kyle~S.~Dawson\altaffilmark{1},
Jean-Paul~Kneib\altaffilmark{2,3},
Will~J.~Percival\altaffilmark{4},
Shadab~Alam\altaffilmark{5},
Franco~D.~Albareti\altaffilmark{6,7},
Scott~F.~Anderson\altaffilmark{8},
Eric~Armengaud\altaffilmark{9},
\'Eric~Aubourg\altaffilmark{10},
Stephen~Bailey\altaffilmark{11},
Julian~E.~Bautista\altaffilmark{1}
% ...
}

\altaffiltext{1}{
Department of Physics and Astronomy, 
University of Utah, Salt Lake City, UT 84112, USA.
}

\altaffiltext{2}{
Laboratoire d\'astrophysique, Ecole Polytechnique F\'ed\'erale de Lausanne
Observatoire de Sauverny, 1290 Versoix, Switzerland
}

\altaffiltext{3}{
Aix Marseille Universit\'e, CNRS, LAM
(Laboratoire d'Astrophysique de Marseille),
UMR 7326, 13388, Marseille, France
}

\altaffiltext{4}{
Institute of Cosmology \& Gravitation, Dennis Sciama Building, University of Portsmouth, Portsmouth, PO1 3FX, UK.
}

\altaffiltext{5}{
Bruce and Astrid McWilliams Center for Cosmology,
Department of Physics,
Carnegie Mellon University, 5000 Forbes Ave, Pittsburgh, PA 15213, USA.
}

\altaffiltext{6}{
Instituto de F\'{\i}sica Te\'orica, (UAM/CSIC),
Universidad Aut\'onoma de Madrid, Cantoblanco, E-28049 Madrid, Spain.
}

\altaffiltext{7}{
'la Caixa'-Severo Ochoa Scholar
}

\altaffiltext{8}{
Department of Astronomy, University of Washington,
Box 351580, Seattle, WA 98195, USA.
}

\altaffiltext{9}{
CEA, Centre de Saclay, Irfu/SPP,  F-91191 Gif-sur-Yvette, France.
}

\altaffiltext{10}{
APC, University of Paris Diderot, CNRS/IN2P3, CEA/IRFU,
Observatoire de Paris, Sorbonne Paris Cite, France.
}

\altaffiltext{11}{
Lawrence Berkeley National Laboratory, One Cyclotron Road,
Berkeley, CA 94720, USA.
}

% ...

\email{kdawson@astro.utah.edu}

\shorttitle{eBOSS}

\begin{abstract}
In a six-year program started in July 2014, the Extended Baryon Oscillation
Spectroscopic Survey (eBOSS) will conduct novel cosmological observations using
the BOSS spectrograph at Apache Point Observatory.
These observations will be conducted simultaneously with the Time Domain
Spectroscopic Survey (TDSS) designed for variability studies
and the Spectroscopic Identification of eROSITA Sources (SPIDERS)
program designed for studies of X-ray sources.
In particular, eBOSS will measure with percent-level precision the distance-redshift relation with 
baryon acoustic oscillations (BAO) in the clustering of matter.
eBOSS will use four different tracers of the underlying matter density field
to vastly expand the volume covered by BOSS and map the large-scale-structures over the relatively
unconstrained redshift range $0.6<z<2.2$.
Using more than 250,000 new, spectroscopically confirmed luminous red galaxies
at a median redshift $z=0.72$, we project that eBOSS will yield measurements of
the angular diameter distance $d_A(z)$ to an accuracy of 1.2\% 
and measurements of $H(z)$ to 2.1\% when combined with the $z>0.6$ sample of BOSS galaxies.
With $\sim 195,000$ new emission line galaxy redshifts,
we expect BAO measurements of $d_A(z)$ to an accuracy of 3.1\%
and $H(z)$ to 4.7\% at an effective redshift of $z= 0.87$.
A sample of more than 500,000 spectroscopically-confirmed quasars will provide the first BAO
distance measurements over the redshift range $0.9<z<2.2$,
with expected precision of 2.8\% and 4.2\% on $d_A(z)$ and $H(z)$, respectively.
Finally, with 60,000 new quasars and re-observation of 60,000 BOSS quasars, we will
obtain new \lya\ forest measurements at redshifts $z>2.1$;  these new data will enhance the precision of 
$d_A(z)$ and $H(z)$ at $z>2.1$ by a factor of 1.44 relative to BOSS.
Furthermore, eBOSS will provide improved tests of
General Relativity on cosmological scales through redshift-space distortion (RSD) measurements,
improved tests for non-Gaussianity in the primordial density field,
and new constraints on the summed mass of all neutrino species.  
Here, we provide an overview of the cosmological goals,
spectroscopic target sample, demonstration of spectral quality from early data, and
projected cosmological constraints from eBOSS.
\end{abstract}

\section{Introduction}
\label{sec:introduction}
\setcounter{footnote}{0}

The origin of the accelerating expansion of the Universe is arguably the most important unknown
in physics today and has inspired significant efforts to probe beyond the standard
model of high-energy physics through observational cosmology.  The recent
measurements of the Cosmic Microwave Background (CMB) from the Planck satellite
support a picture where the acceleration is driven by ``dark energy''
with density $\Omega_{de} = 0.692 \pm 0.012$ in a spatially flat
universe \citep{ade15a}.  Combining these results with current measurements
from baryon acoustic oscillations (BAO), Type~Ia SNe, and $H_0$, the data imply a constant equation of
state $w=-1.006^{+0.085}_{-0.091}$ at 95\% confidence,
where $w$ is the ratio of pressure to energy density for dark energy.  
Thus, current observations are generally consistent
with the simplest picture where dark energy is described completely by Einstein's
cosmological constant ($\Lambda$).

New precise observations can unravel the origin of the accelerating
universe;  specifically, to determine if cosmic acceleration is caused
by deviations in General Relativity (GR) on large scales or by a new form of (dark) energy.
It is possible to decouple scenarios of acceleration that require dark energy
from those that require modifications to GR by independently probing both cosmic
expansion history and the structure growth rate.
Four primary observational techniques are generally accepted as the most
powerful toward obtaining that goal \citep[e.g.][]{albrecht06a}: SNe~Ia,
weak lensing, galaxy clusters, and BAO.  Wide-field, optical spectroscopy figures
prominently in three of these probes: spectroscopically
observed galaxies improve calibration of photometric redshifts for weak lensing;
direct spectroscopy of cluster galaxies provides precise redshifts and
velocity dispersions as a proxy for cluster mass;
and spectroscopy of galaxies and
quasars provides the atlas of large-scale structure in which the BAO feature is embedded.

The Sloan Digital Sky Survey \citep[SDSS;][]{york00a} at Apache
Point Observatory (APO) has consistently provided the
largest spectroscopic samples for cosmological analysis.  
In the first two generations of SDSS, generally known as SDSS-I and -II,
redshifts of nearly one million galaxies were measured spectroscopically \citep{abazajian09a}.
The Baryon Oscillation
Spectroscopic Survey \citep[BOSS;][]{dawson13a} performed spectroscopic observations
of large-scale structure in
SDSS-III \citep{eisenstein11a};  BOSS recently completed spectroscopy on more than 1.5 million
galaxies as faint as $i=19.9$ and more than 150,000 quasars as faint as $g=22$.
Measurements of BAO with BOSS have led to 1--2\% precision measurements of the
cosmological distance scale for redshifts $z < 0.6$ and $z=2.5$.

With observations that commenced in July 2014, SDSS-IV will continue this legacy in three distinct surveys.
The second generation of the APO Galactic Evolution Experiment \citep[APOGEE/APOGEE-2;][]{majewski15a} will investigate the
formation and chemical history of the Milky Way using high-resolution, infrared spectroscopy  
of 300,000 stars.
Mapping Nearby Galaxies at APO \citep[MaNGA;][]{bundy15a} will measure the internal
structure of approximately 10,000 galaxies using the BOSS spectrograph with
cartridges outfitted for spatially resolved spectroscopy.
The Extended Baryon Oscillation Spectroscopic Survey (eBOSS) is the new cosmological
survey within SDSS-IV.

The eBOSS program will use the same 1000-fiber optical spectrographs as those in BOSS \citep{smee13a}.
We will expand the selection of luminous red galaxies (LRG) beyond that
probed by BOSS and obtain better than a 1.0\% precision distance estimate
when combined with the $z>0.6$ tail of the BOSS galaxy population.
With observations of a new sample of emission line galaxies (ELG) over the period 2016--2018, eBOSS will produce
a 2.0\% precision distance estimate at higher redshifts.
We will obtain a 1.8\% precision distance estimate in
the redshift range $0.9 < z < 2.2$ using quasars that
have luminosities and areal densities well-suited to sensitivity of the BOSS spectrographs.
Finally, we will sharpen the BOSS \lya\ forest measurements by a factor of 1.44 with a new
selection of $z>2.1$ quasars, providing stronger leverage on the history of dark energy.
Concurrent with the eBOSS survey, we will perform two complementary programs
that use $\sim 100$ of the BOSS spectrograph fibers within each
field-of-view. The Time Domain Spectroscopic Survey \citep[TDSS;][]{morganson15a} will target
variable stars and quasars.  The Spectroscopic Identification of
eROSITA Sources (SPIDERS) will target active galactic nuclei, quasars, X-ray emitting stars,
and galaxy clusters identified in X-ray images of wide-area surveys carried out by ROSAT, XMM-Newton, and,
eventually, eROSITA \citep{merloni12a}.

With four classes of spectroscopic targets (LRG, ELG, quasar, \lya\ forest quasar),
eBOSS will enable the first high precision distance measurements in the epochs when dark
energy emerged as the dominant dynamical component of the Universe.
In addition to BAO distance measurements, eBOSS will provide
new tests of GR on cosmological scales through redshift-space distortions (RSD),
new tests for non-Gaussianity in the primordial density field, and new constraints
on the summed mass of all neutrino species.

This paper is one of a series of technical papers describing the eBOSS survey.
Details of the LRG target selection algorithm is described in \citet{prakash15a}
while \citet{myers15a} presents the quasar target selection algorithms.
The ELG selection is still under investigation; \citet{raichoor15a}, \citet{comparat15a} and \citet{delubac15b}
present various approaches and results.
Here, we provide the scientific motivation behind eBOSS, 
summarize the observation strategy for the cosmological tracers,
present the expected quality of spectra for each target class, and 
compute the projected cosmological constraints.
In Section~\ref{sec:cosmology}, we provide a review of the cosmological
signature of BAO and RSD in the matter power spectrum.  We use BOSS measurements to demonstrate the sensitivity
of spectroscopic surveys to both signatures and to demonstrate sources of
systematic errors that must be addressed in eBOSS and future spectroscopic programs.
We present the eBOSS program in Section~\ref{sec:eboss},
including time allocation, expected areal coverage, and requirements for target selection.
In Section~\ref{sec:strategy} we discuss the survey strategy, selection of the
galaxies and quasars that will be used to measure the matter power spectrum, and statistics after fiber assignment.
Favorable weather during SDSS-III led to an early completion of the BOSS program.
A fraction of the remaining time was allocated to an eBOSS pilot program
known as the Sloan Extended Quasar, ELG, and LRG Survey (SEQUELS).
Using the results of those 66 dedicated plates, a demonstration of the data quality for each eBOSS
target class is presented in Section~\ref{sec:sequels}.
We summarize the expected quality of the data with respect to the low-level specifications required
to achieve percent-level precision distance estimates in Section~\ref{sec:performance}.
We also present plans to improve the data reduction software in Section~\ref{sec:performance}.
Finally, the cosmological projections for eBOSS are provided in Section~\ref{sec:projections}
and a summary is provided in Section~\ref{sec:conclusion}.

\section{Signature of BAO and RSD in Spectroscopic Surveys}
\label{sec:cosmology}

Experience from BOSS led to the survey design for eBOSS described in
Section~\ref{sec:eboss} and Section~\ref{sec:strategy}.
BOSS proceeded with the primary goal of obtaining new measurements
of the cosmic distance scale through BAO at redshifts $z<0.6$ and at $z=2.5$.
While there have been other successful BAO programs
\citep[e.g. the 2dF Galaxy Redshift Survey, 6dF Galaxy Survey and WiggleZ Dark Energy Survey;][]{cole05a,beutler11a,blake11b},
we only review BOSS here because it provides a direct demonstration of spectroscopic constraints on
cosmology that we expect from eBOSS, and sets the scene for this survey.
In what follows, we outline the BOSS observational program in
Section~\ref{subsec:boss}, the resulting BAO measurements in Section~\ref{subsec:bao},
and RSD measurements in Section~\ref{subsec:rsd}.

\subsection{Sample of Galaxies and Quasars from BOSS}\label{subsec:boss}

BOSS is described in detail in \citet{dawson13a}, so we provide only a quick summary 
to highlight the essential features that are either inherent or complementary to the eBOSS program.
BOSS and eBOSS use the same telescope as that used in
SDSS-I and -II:  the 2.5-meter Sloan Foundation Telescope
at APO in New Mexico \citep{gunn06a}.
The BOSS spectrographs \citep{smee13a} were built with smaller fibers,
new improved detectors, higher throughput, and a wider wavelength range than
the SDSS spectrographs previously used at APO.
There are two spectrographs, each covering the wavelength range 361 nm -- 1014 nm.
The instrument is fed by 1000
optical fibers (500 per spectrograph), each subtending $2\2pr$ diameter on the sky.

The detection of BAO \citep{eisenstein05a} in the SDSS LRG sample \citep{eisenstein01a}
motivated a similar selection of galaxies as the primary sample of spectroscopic targets for BOSS.
At the beginning of BOSS observations, the techniques for analysis were already advanced to the stage that
the collaboration could reasonably expect to perform BAO distance measurements at roughly 1\% precision with the galaxy sample.
In a more experimental program, more than 150,000 quasars at redshifts
$z>2.1$ were selected to measure fluctuations in the matter density field as observed through the \lya\ forest.
Contrary to the galaxy program, the \lya\ forest program was created to make the first BAO measurements with a new
tracer with an expectation that any detection would be made with somewhat lower significance than the galaxy measurement.

The BOSS survey obtained roughly 10,000 $\sqdeg$ of spectroscopic coverage over a five year period.
At the completion of the main BOSS program, fibers were plugged into 2438
unique spectroscopic plates that each cover a circular field of view with 3\deg\ diameter.
All of the BOSS spectra became public in January 2015 with
Data Release 12 \citep[DR12;][]{alam15a}.
These data include spectra and classification of
343,160 unique low redshift galaxies (denoted ``LOWZ''; $z_{\rm median}=0.32$),
862,735 unique medium redshift galaxies (denoted ``CMASS''; $z_{\rm median}=0.57$),
and 181,605 quasars between $2.1<z<3.5$, of which 160,786 are not tagged as broad absorption line (BAL) quasars.

\subsection{Baryon Acoustic Oscillations}\label{subsec:bao}

While the overall shape of the power spectrum informs our cosmological model,
the most robust cosmological measurements from spectroscopic surveys
derive from the BAO feature in the clustering of matter.
Sound waves that propagate in the opaque early universe imprint a
characteristic scale in the clustering of matter, providing a
``standard ruler'' whose length can be computed using straightforward
physics and cosmological parameters that are constrained by CMB observations.
The characteristic BAO feature has a comoving scale (roughly 150 Mpc) set by the integrated sound speed
between the end of inflation and the decoupling of photons around $z=1000$.
The detection of the acoustic oscillation scale \citep{eisenstein05a} is one of the signature
accomplishments of SDSS.

Measuring the angle subtended by the characteristic BAO feature at a given redshift
provides a means to estimate the angular diameter distance, $d_A(z)$.
Similarly, a measurement of the redshift interval over which the BAO feature extends
provides a means to directly measure the Hubble parameter $H(z)$ at that redshift.
In general, there is a combination of $d_A(z)$ and $H(z)$ denoted by a generalized distance
parameter ``R'' that is better constrained than either of these two quantities alone.
In the specific geometry of the measured clustering of BOSS galaxies,
the best constrained ``distance'' is approximately
$R \equiv D_V = [d_A^2(z) \, cz \, H^{-1}(z)]^{1/3}$ because there are two transverse
dimensions and one line-of-sight dimension.  In the \lya\ forest,
the enhancement of the radial signal gives greater weight to $H(z)$.

At low redshifts, BAO are a powerful
complement to studies with SNe~Ia:  they have low
systematic uncertainties; they measure distances in
absolute units whereas supernovae only constrain relative distances;
and they can directly measure both the cosmic
expansion rate $H(z)$ and the distance-redshift relation $d_A(z)$.
At high redshifts, the large comoving volume
allows the BAO distance method to obtain remarkably
precise measurements of the distance and expansion rate with better precision than SNe~Ia.
The BAO method is reviewed in detail by
\citet[][see their \S 4]{weinberg12a},
including discussions of the underlying theory, the effects
of non-linear evolution and galaxy bias, survey design and
statistical errors, control of systematics, recent observational
results, and complementarity with other probes of cosmic acceleration.

\subsubsection{BOSS Constraints on BAO}

Using data from DR9, the first BAO measurement with the BOSS CMASS sample was
published in 2012 \citep{anderson12a}. The DR9 sample
covered 1/3 of the final BOSS volume,
yet a clear BAO detection allowed a measurement of the distance to $z=0.57$
with a precision of $\sigma_R =1.7\%$. 
The most recent galaxy clustering measurements with BOSS \citep{anderson14a}
produce a clear BAO detection using both the LOWZ and CMASS samples of DR11 galaxies.
The BAO detection in the CMASS sample
provides a 1.0\% measurement of the generalized distance to $z=0.57$,
the most precise distance constraint ever obtained from a galaxy survey.
The LOWZ sample produces a measurement to $z = 0.32$ with 2.1\% precision.

A thorough examination of potential systematics in the clustering of the BOSS CMASS
galaxies revealed a 10\% decrease in the detected number density of
galaxies when comparing areas with high stellar density
to those with low stellar density \citep{ross12b}.
The correlation of galaxy density with stellar density is
the most significant known bias on measured clustering, likely
caused by incomplete deblending of detected objects in crowded fields of the SDSS imaging data.
On the other hand, no significant correlation is seen between number density and
potential imaging systematic errors in the LOWZ sample of galaxies.
The lack of bias is likely because the lower redshift sample
appears brighter and is less vulnerable to imaging artifacts such as the deblending problems
seen in the fainter, high redshift sample.
A weighting scheme based on galaxy surface brightness and stellar density was devised to
reduce the systematic effect of stellar density on the measured clustering signal.
When focusing on scales less than 150 Mpc, tests on mock catalogs reveal 
that the weights produce no significant bias on the mean measured correlation function.
The results imply that systematic errors in galaxy clustering due to artifacts
in imaging data (causing density fluctuations as high as 10\%) can be removed if those artifacts are identified.

The first measurement of clustering in the \lya\ forest of BOSS quasars
was reported in \citet{slosar11a}.
In 2012, the first measurements of BAO using the \lya\ forest
detected in the spectra of $z > 2.15$ quasars were released.
Using 48,640 quasars in the redshift range
$2.15 < z < 3.5$, \citet{busca13a} detected a peak in the correlation function at a separation
equal to $1.01\pm0.03$ times the distance expected for the BAO peak within a concordance
$\Lambda$CDM cosmology.
Using an alternative analysis of the same quasar sample that included some peculiar quasars
and larger wavelength coverage, \citet{slosar13a}
reported a 2\% measurement error on the distance.
In the subsequent DR11 analysis \citep{delubac15a},
137,562 DR11 quasars in the redshift range
$2.1 < z < 3.5$ were used to detect a peak in the correlation function and
constrain the distance to a precision of 2.1\%.
Decomposing the optimized distance, the DR11 measurement yields a measurement $H(z)=222 \pm 7$ km s$^{-1}$Mpc$^{-1}$
and $d_A(z)=1662\pm96$ Mpc at an effective redshift $z = 2.34$.

Modeling the cosmic distance scale at $z>2$ can be enhanced beyond the \lya\ forest auto-correlation measurement
by adding a measurement of the cross-correlation of quasars with the  \lya\ forest absorption field.
In particular, the relatively high bias of quasars provides leverage to
improve precision in the measurement of the angular diameter distance relative to the \lya\ forest auto-correlation.
\citet{font-ribera14a} use the DR11 sample of quasars to measure BAO in the \lya\ forest--quasar cross correlation
both along the line of sight and across the line of sight.
The best fit correlation function at an effective redshift $z_{e}=2.36$ translates to a measurement 
$H(z_{e}) = 226 \pm 8$ km s$^{-1}$Mpc$^{-1}$ and $d_A(z_{e})=1590 \pm 60$ Mpc.
Because the \lya\ forest sightlines are effectively sampled at random, there is no evidence for
systematic errors from imaging artifacts in the derived clustering of either
\lya\ forest auto-correlation or \lya\ forest--quasar cross correlation.

In summary, BOSS provides the most precise BAO measurements to date using
luminous galaxies and correlations in the \lya\ forest from high redshift quasars.
Highlighting the complementarity between BAO and SNe~Ia, \citet{auberg15a}
explore models with the combination of SNe~Ia and BAO calibrated by the sound horizon scale as measured by Planck.
They show that the combined Hubble Diagram \citep{betoule14a} of SNe~Ia from 
the Supernova Legacy Survey \citep{conley11a} and SDSS-II Supernova Survey \citep{frieman08a,sako14a}
can be extrapolated to $z=0$ when anchored at intermediate redshift by BAO measurements.
The BAO distance measurements effectively provide calibration of the absolute magnitude of SNe~Ia
and lead to a 1.7\% precision measurement, $H_0= 67.3 \pm 1.1$ km s$^{-1}$Mpc$^{-1}$.
This value agrees with the value of $H_0$ derived from CMB data alone under an assumption of
flat $\Lambda$CDM but is in mild tension with Cepheid-derived
distances;  for example \citet{riess11a} report $H_0= 73.8 \pm 2.4$ km s$^{-1}$Mpc$^{-1}$.

\begin{figure}[htb!]
\centering
%\vspace{-1.5cm}
\includegraphics[width=0.75\textwidth, angle=0]{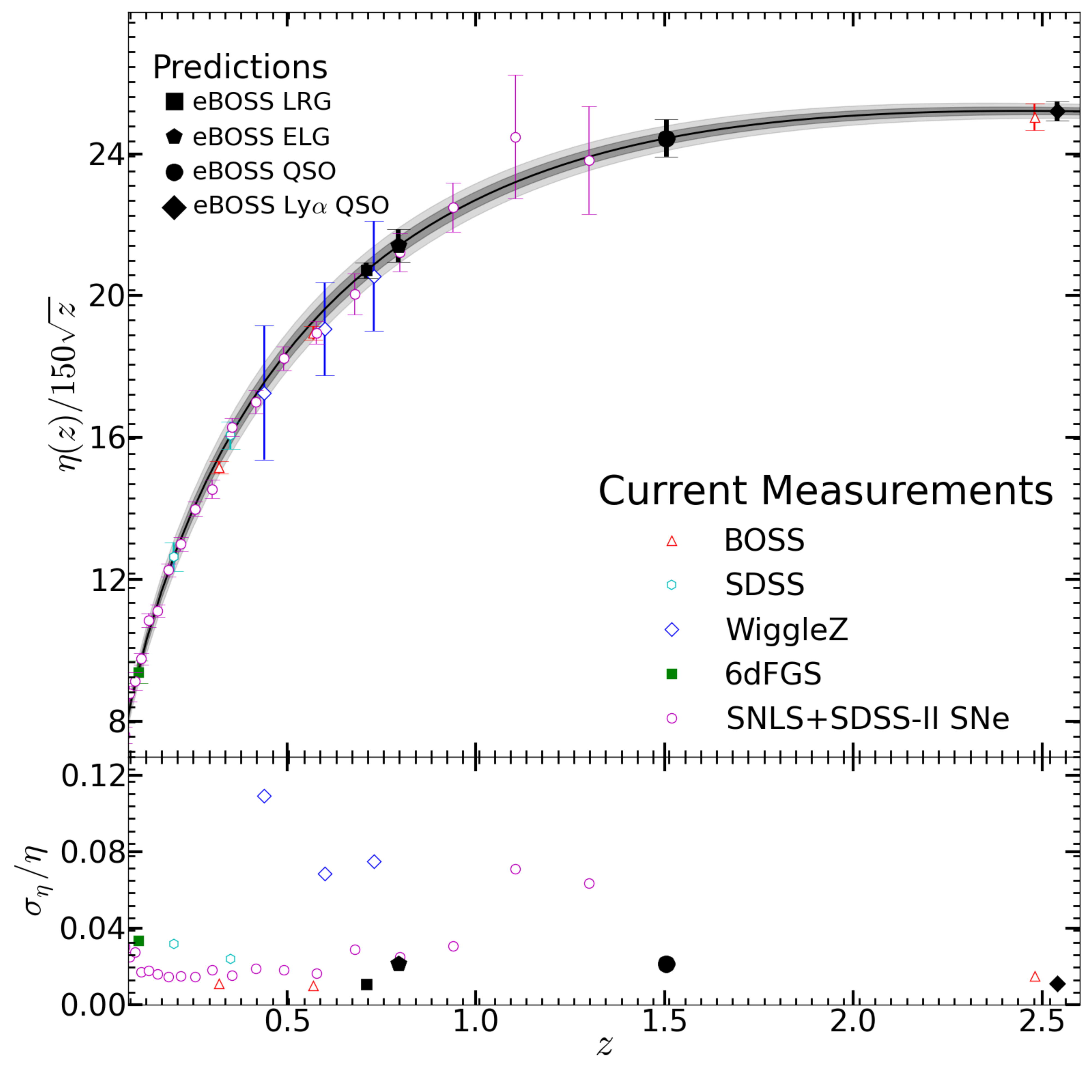}
%\vspace{-0.5cm}
\caption{Projections for eBOSS LRG, ELG, and quasar distance measurements on a Hubble Diagram presented in comoving
distance ($\eta$) versus redshift. 
Current BAO measurements from BOSS, SDSS \citep{xu13a,ross15a}, 6dF Galaxy Survey (6dFGS), and WiggleZ \citep{parkinson12a}
are compared to SNe~Ia measurements \citep{betoule14a} and Planck predictions (solid curve) obtained by marginalizing
over the full likelihood function.
}
\label{fig:baodistance}
\end{figure}

Extending the joint CMB, BAO, and SNe analysis to models of dark energy,
\citet{auberg15a} report a measurement of the constant equation of state
for dark energy $w=-0.98 \pm 0.06$ when curvature is treated as a free parameter.
Allowing an additional free parameter to account for a possible
time-evolving equation of state, the joint BAO-SNe measurements
produce only weak constraints on evolution:  $w_a = -0.6 \pm 0.6$,
where $w(a) =w_0+w_a(1-a)$ \citep[e.g.][]{chevallier01a,linder03b}.
As shown in Figure~\ref{fig:baodistance}, SNe Ia provide only weak constraints on the Hubble Diagram
over the redshift range $1 < z < 2$, and BOSS has no sensitivity to the cosmic interval that lies
between the galaxy and the \lya\ forest samples.
It is in this $0.6 < z < 2$ regime where the Universe is expected to transition from
matter-dominated to dark energy-dominated.   New measurements of the cosmic distance scale
over this redshift range have the potential to improve constraints on models of a time-evolving equation of state for dark energy.

\newpage

\subsection{Redshift Space Distortions and Modified Gravity}\label{subsec:rsd}

Dark energy is often invoked to explain current CMB, SNe, and
BAO observations that imply an accelerating Universe.
It is also possible to explain the accelerated expansion of the Universe by
modifying gravity at large scales.
The galaxy redshifts used in spectroscopic BAO measurements of the
expansion history help differentiate these two possible effects
through measurements of the growth of structure via RSD \citep{kaiser87a}.

RSD arise because the gravitational pull of matter overdensities causes
velocity deviations from the smooth Hubble flow expansion of the Universe.
These peculiar velocities are imprinted in galaxy redshift
surveys in which recessional velocity is used as the
line-of-sight coordinate for galaxy positions.
Although the correlation function of galaxies is isotropic in real
space, the peculiar velocities lead to an increase in the amplitude of radial clustering
relative to transverse clustering when the correlation
function is measured in redshift space.
The resulting anisotropy in the clustering of galaxies is
correlated with the speed at which structure grows; deviations from GR causing slower
or faster growth give smaller or larger
anisotropic distortions in the observed redshift-space clustering.

In general, the amplitude of clustering at a given redshift
is parameterized by $\sigma_8(z)$, the rms fluctuations
in spheres of radius 8\,$h^{-1}$Mpc.
The degree of anisotropy due to RSD depends on the rate of change
of the amplitude of clustering.
This change is typically parameterized as a function of the logarithm of the
expansion scale parameter $f\sigma_8 = \partial \sigma_8/ \partial\ln a$,
where $a=(1+z)^{-1}$ is the dimensionless cosmic expansion factor.
Because RSD measurements are sensitive to the product of the growth rate and the
amplitude of matter fluctuations, a wide range in redshift coverage is essential 
to constrain the evolution in clustering amplitude and directly probe gravity.

\subsubsection{BOSS Constraints on RSD and Modified Gravity}
\label{subsec:boss_rsd}

\begin{figure}[htb!]
\centering
%\vspace{-1.5cm}
\includegraphics[width=0.75\textwidth]{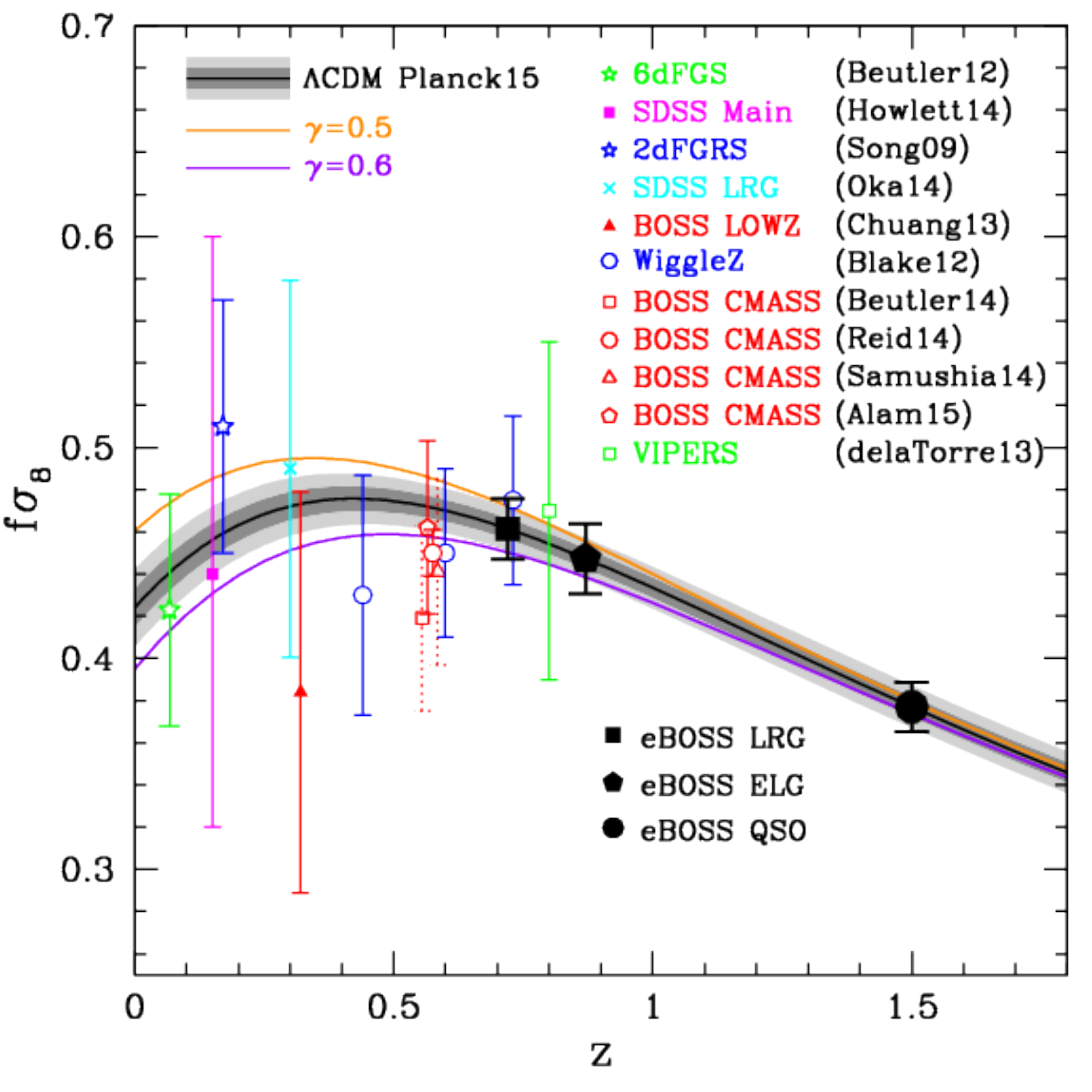}
%\vspace{-3.0cm}
\caption{Current RSD constraints on the growth as a function of redshift compared to the
projected measurements from eBOSS. 
The current measurements include those discussed in Section~\ref{subsec:boss_rsd}
and those for 6dFGS \citep{beutler12a}, the main SDSS sample \citep{howlett15a},
2dFGRS \citep{song09a}, the SDSS LRG sample \citep{oka14a}, a recent result
from the BOSS CMASS sample \citep{alam15b}, WiggleZ \citep{blake12a},
and VIPERS \citep{delatorre13a}.
Various models
of modified gravity are shown, each with the same background expansion, the same comoving BAO position,
and amplitude of the power spectrum normalized to that of the CMB at high redshifts.
The black curve shows the growth in a $\Lambda$CDM universe, assuming the Planck best fit model parameters.
The yellow curve shows $\gamma = 0.5$ where $f=\Omega_M^\gamma$ \citep{linder05a}.
The purple curve shows $\gamma = 0.6$.
}
\label{fig:rsd}
\end{figure}

\citet{reid12a} and \citet{samushia13a} presented the first measurements
and cosmological interpretation of RSD in the BOSS DR9 galaxy sample.
With these results, they constrain the
parameter combination $f\sigma_8=0.43\pm 0.07$.
Using the larger DR11 sample, \citet{samushia14a} constrain the parameter combination
$f\sigma_8=0.447\pm 0.028$
under an assumption of a spatially flat Universe with cosmological constant.
Using the DR10 sample to constrain RSD to smaller scales of 0.8--32 $h^{-1}$ Mpc,
\citet{reid14a} find a model-dependent constraint $f\sigma_8=0.450 \pm 0.011$
obtained in fitting a standard halo occupation distribution model to the anisotropic clustering.
Other measurements from the same data have found similar results.
A summary of current RSD measurements is presented in Figure~\ref{fig:rsd}.

Constraints on gravity from RSD become increasingly powerful as the measurements push
to smaller scales, as evidenced by the higher precision measurements of \citet{reid14a}
relative to \citet{samushia14a} and others.
Cosmological measurements from small-scale clustering are dependent on the accuracy of the
modelling on quasi-linear and non-linear scales.
The development and evaluation of analytic, phenomenological, and halo occupation models for anisotropic
clustering remains a focus with the BOSS galaxy samples \citep[e.g.][]{chuang13a,beutler14a,guo15a}.
A study of several models in configuration-space
tested against mock galaxy catalogs indicates that the clustering signal can
be well characterized on scales in the range $40<s<80h^{-1}$ Mpc \citep{white15a}.
Certain models, such as those based on Lagrangian perturbation theory,
are able to fit the mock clustering samples without significant bias on scales above 25--30 $h^{-1}$ Mpc.
Continued development of theoretical models that allow use of smaller scale data may tighten the current
BOSS constraints still further.

\newpage

\section{The Extended Baryon Oscillation Spectroscopic Survey}
\label{sec:eboss}

As described in Section~\ref{sec:cosmology}, BOSS completed a survey of $10,000 \sqdeg$ 
and enabled the best BAO distance measurements to date at $z<0.6$ and
the most precise distance measurements of any kind at $z \sim 2.5$.
This success motivated the SDSS-IV collaboration to repurpose the BOSS spectrographs for eBOSS.
The primary goal of eBOSS is to extend the BOSS galaxy measurements
to $0.6<z<1$ and to make the first measurements of clustering with quasars as tracers and
the first measurements of BAO at redshifts $1<z<2$.
The strategy to make these measurements is not purely based on an optimized
figure of merit.
Instead, we follow a similar philosophy to BOSS and designed eBOSS to expand a well-understood sample
while exploring new tracers over a fundamentally new redshift regime.
We planned the expanded galaxy sample to achieve 1\% precision on BAO measurements as
in the BOSS galaxy sample and the new tracers to achieve 2\% precision as in the BOSS \lya\ sample.

Experience from BOSS led to the design of an LRG sample to measure clustering over redshifts $0.6<z<1.0$.
As before, the BAO analysis tools for this type of galaxy are well-established.  We
designed the eBOSS LRG program to match the 1\% precision on generalized distance
achieved with the BOSS galaxy sample.
We use the Fisher matrix formalism of \citet{seo07a} to determine BAO-based errors on the angular diameter
distance and the Hubble parameter given an observing volume, number density and bias of galaxies.
We assume a monotonically decreasing number density and a bias of $1.7\sigma_8(0)/\sigma_8(z)$, approximately
correct assuming the sample continues the evolution of the reddest galaxies observed in BOSS \citep{guo13a,ross14a}.
We find that number density can be traded for areal coverage
for number densities around 50 deg$^{-2}$ without degrading the precision on the BAO distance measurement.
This relationship holds true for a fairly wide range of number densities, lending us flexibility in the design of the
LRG program.
Assuming a survey area on the order of thousands of square degrees, the balance of density and survey area leads to the general
requirement that 300,000 LRG's in the redshift range $0.6<z<1.0$ are required
to achieve 1\% precision on the generalized distance.

Just as BOSS introduced the \lya\ forest sample as a new probe of clustering, eBOSS will introduce two
new selections that will expand the legacy of SDSS.
The first of these selections will be quasars in the redshift range $0.9<z<2.2$.
Because quasars are bright and trace large-scale structure with a high bias,
they are the most effective tool to constrain BAO at these redshifts.
The lower bound on redshift is set to have slight overlap with the
eBOSS galaxy samples while the upper bound is set to have slight overlap with the
BOSS and eBOSS \lya\ forest sample.
The potential of this sample is limited by the intrinsic number density of the quasar population.
In a series of spectroscopic observations with BOSS,
1877 quasars were classified over 14.5 deg$^2$ near the celestial equator.
After correcting for targeting efficiency and selecting confirmed quasars at
$0.9<z<2.2$ and $g<22$, the expected surface density of the quasar population is 82.6 deg$^{-2}$ \citep{palanque-delabrouille13a}.
We therefore are forced to design the quasar program with a lower volume density to sample large-scale structure
than the galaxy programs.
We make BAO projections assuming a flat redshift distribution and the bias relationship $b(z) = 0.53 + 0.29(1+z)^2$ determined by
\citet{croom05a} and consistent with SDSS measurements \citep{myers06a, myers07a}.
Unlike the LRG targets, the limited surface density will lead to BAO measurements that are fundamentally limited 
by shot noise.  In order to obtain a 2\% measurement comparable to the BOSS \lya\ sample,
the quasars must sample the intrinsic population with high completeness over as much area as possible.
We establish the maximal areal coverage in Section~\ref{subsec:program} and report the required quasar density there.

Potential risk in any one target class can be mitigated by observing a
number of different galaxy types overlapping in redshift and using
cross-correlation between different populations.
The predominantly passive galaxies in the LRG sample are typically classified
by absorption lines against faint continuum emission, leading to a practical limit that each target's $z$-band magnitude
be brighter than roughly $z_{AB}\sim 20$.
The emission lines in blue galaxies with significant star formation facilitate spectroscopic confirmation 
at high redshifts with an optical spectrograph.
Because ELG's are numerous and have emission line fluxes correlated with the observed $u$ or $g$-band magnitudes \citep{comparat15b},
they can be selected at a much higher density than the LRG sample.
For these reasons, we identified an ELG sample over an area that ensures a BAO
detection with minimal diversion from the LRG and quasar programs.
We conservatively assume that the bias of the sample is $1.0\sigma_8(0)/\sigma_8(z)$,
slightly lower than expected \citep{comparat13a,comparat13b}.
As with the LRG sample, Fisher projections indicate that the 
shot noise and areal coverage contribute
in such a manner that the BAO distance error remains relatively constant for a fixed number of objects
for a sample with density of hundreds per square degree.
We find that a 2\% distance BAO measurement can be achieved with a total sample size
of approximately 190,000 spectroscopically confirmed ELG's
with a uniform redshift distribution in the redshift range $0.6<z<1.0$ or $0.7<z<1.1$.

In the DR11 BOSS sample, the BAO feature at an effective redshift $z=0.57$ was measured at a precision
of 0.9\% using CMASS galaxy targets at a density 85 deg$^{-2}$ \citep{anderson14b}. 
The BAO feature was measured at an effective redshift $z=2.34$ at a precision of 2\% using
\lya\ quasars at an average density 16 deg$^{-2}$ \citep{delubac15a}.
The marginal value of each \lya\ quasar therefore exceeds that of a typical galaxy in
constraining the matter power spectrum at BAO scales,
as long as the \lya\ targets can be reliably identified.
We reserved as many fibers as possible for efficient detection and observations of \lya\ quasars
for the purpose of improving the \lya\ forest measurements obtained in BOSS.

Of equal importance to BAO are measurements of RSD with each of these samples.
However, as discussed in Section~\ref{subsec:rsd}, the current RSD measurement errors are
primarily driven by theoretical uncertainties in the modeling rather than by the statistical
power of the target samples.
{\it We therefore rely entirely on the high-level BAO goals for determining the observing strategy
and properties of the target samples}.
The requirements for target densities are driven by the goal of obtaining 1\%, 2\%, and 2\%
BAO distance measurements on the LRG, quasar, and ELG samples, respectively.
We also present additional requirements on the uniformity of these samples intended to
mitigate potential systematic errors in clustering measurements.
In this section, we present the program that will meet these high-level objectives.

\subsection{The eBOSS Cosmology Program}\label{subsec:program}

Over six years, eBOSS will provide the first percent-level distance measurements with BAO and RSD in
the redshift range $0.6<z<2$, when cosmic expansion transitioned from deceleration to acceleration.
Using LRG, quasars, ELG, and \lya\ absorption as tracers of the underlying density field,
eBOSS will probe the largest volume to date of any cosmological redshift survey.
We designed a program of LRG, ELG, and quasar targets distributed over
the available fibers with a balance that satisfies the high level goals
of obtaining 1--2\% distance measurements while also meeting a series of practical constraints.

In BOSS, we typically finished
observations of a spectroscopic plate in five 15 minute exposures
with 20 minutes of overhead due to field acquisition, calibration exposures, and CCD readout.
These exposures produced spectra of
sufficient quality to classify the targets in the CMASS sample as faint as $i=19.9$
(AB) with over 98\% completeness \citep{bolton12a}.  A series of pilot studies using dedicated plates in 2013 and 2014
proved that the same spectral depth is adequate to classify quasars from $0.9<z<2.2$,
LRGs to $z<0.8$, and ELGs at $0.6<z<1.1$.
Since a typical field can be completed in 1.5 hours of exposure and overhead,
we anticipate that 1800 plates can be completed over the roughly 5400 hours
allocated to eBOSS
if the 50\% weather efficiency experienced during SDSS-III continues for SDSS-IV.
Considering that each plate will contain 100 fibers dedicated to calibration targets
and roughly 100 fibers dedicated to SPIDERS or TDSS targets, a total of 1.44M fibers
will be available for targets designed to trace large-scale structure.
Given this expected time allocation and potential sample size, we designed the program as follows:

\begin{enumerate}

\item
First, we are constrained by the finite area of a spectroscopic plate (7 $\sqdeg$) and the 
requirement of a large-scale structure survey to obtain areal coverage with limited gaps between plates.
As in previous SDSS programs, each eBOSS field center is assigned in the
tiling process \citep{blanton03a}.
We chose a density of field centers to an average of one plate
per 5.0 $\sqdeg$ to avoid gaps in coverage and mitigate inefficiency in fiber assignments caused by density
fluctuations in the target sample.
At this tiling density, 1800 plates can provide a total survey area up to 9000 $\sqdeg$.
The TDSS and SPIDERS projects will each use an average of 50 fibers per plate
and calibration targets require a minimum of 100 fibers per plate.
With 800 remaining fibers per plate, the available density of cosmological tracers is 160 deg$^{-2}$. 
We use this density of fibers to shape the LRG and quasar samples which will be observed
together over the majority of the program.
We isolated the ELG targets to a series of dedicated plates.  Those plates will
contain almost entirely ELG targets except for a few additional objects
at a maximum density of 10 fibers deg$^{-2}$.
To conduct the ELG program, we will reduce the maximal area of the LRG, quasar, and \lya\ quasar samples from
9000 $\sqdeg$ to 7500 $\sqdeg$, thus leaving 300 plates for the ELG program.
As discussed below and in Section~\ref{sec:strategy}, the density and areal coverage of those ELG targets
will be decided based on final target selection algorithms and available imaging data.
By doing so, we only reduce the volume of the primary LRG and
quasar samples by 16\% but obtain a sample of ELG targets 
that will provide a BAO constraint comparable to the DR9 BOSS galaxy results \citep{anderson12a}
over a new redshift range.

\item
At an area of 7500 $\sqdeg$, the Fisher projections predict that $0.9<z<2.2$ quasars at a density of
58 deg$^{-2}$ will produce a 2\% distance measurement.
Because the sample is shot-noise limited, any fractional increase in the number density translates
to the same fractional improvement in the projected BAO uncertainty.
Roughly 13 deg$^{-2}$ of $0.9<z<2.2$ quasars are already known from SDSS-I, -II, or -III,
leaving only 45 deg$^{-2}$ quasars to be identified and spectroscopically confirmed in eBOSS.
Because this sample probes a fundamentally new redshift range, we design the program primarily around the observations
of quasars, allocating fibers at a density of 90 deg$^{-2}$ to ensure that we obtain the minimal target density.
All objects assigned to the target class denoted {\tt QSO\_CORE} form the basis of this sample;
the purity of this core sample needs to be at least 50\% to meet the high-level goal of a 2\% BAO distance measurement
when combined with the sample of known quasars.
Throughout, purity is defined as the fraction of objects that are
reliably classified in the redshift range of interest.

\item
The LRG sample must contain at least 300,000 spectroscopically confirmed galaxies in the
redshift range $0.6<z<1.0$ to obtain the high-level goal of a 1\% BAO distance measurement.
These objects will augment the tail of the BOSS CMASS sample in the range $0.6 < z < 0.7$
and increase the reach of the luminous galaxy sample to $z=1.0$.
Over $7500 \sqdeg$, the corresponding density of galaxies with
a successful redshift in the desired redshift range is 40 deg$^{-2}$.
In terms of projected BAO errors, a fractional change in the number density is equivalent to the same fractional
change in area.
In BOSS, luminous galaxies were reliably identified in imaging data and spectroscopically
confirmed in a sample with purity that typically exceeded 95\%.
The eBOSS sample requires higher absolute luminosity to maintain a reasonable rate
of spectroscopic classification at the higher redshifts.  However, the apparent
magnitudes are still fainter and features are shifted further into the red, likely reducing
redshift efficiency.  Pilot programs and visual inspection during BOSS originally indicated that we could
exceed 80\% purity at $0.6<z<1.0$.
We therefore require that the 40 deg$^{-2}$ LRG sample be reliably classified and at redshifts
$0.6<z<1.0$ from observations of LRG candidates at a density 50 deg$^{-2}$.
As will be shown in Section~\ref{subsec:targets} and Section~\ref{subsec:tiling}, 
the purity of the final sample falls short of requirements, leading to the
only failure we expect in meeting survey design.

\item
Unlike galaxies, the information extracted from each \lya\ quasar scales with the
S/N of the spectra in the \lya\ forest.
Before the beginning of BOSS, \citet{mcdonald07a} presented possible BAO constraints from
the \lya\ forest accounting for area, density of quasars, resolution, and S/N.
As explained in detail in Section~\ref{subsec:lyaspectra}, \citet{mcquinn11a} find that the relative value of
a quasar roughly doubles as the S/N per \AA\ increases from one to two,
but that deeper spectra approach diminishing returns once the depth exceeds
a S/N of three per \AA.
To capitalize on the benefits of deeper spectra,
we will re-observe $z>2.1$ quasars that obtained
$0.75<S/N<3$ per \AA\ in the \lya\ forest in BOSS.
Given that the exposure times in eBOSS will be comparable to those in BOSS,
we expect the typical S/N to increase by 40\% when these quasars
are observed a second time.
The average density of this sample over the BOSS footprint is about 8 deg$^{-2}$.

\item
After fiber allocation to the core quasar, LRG, and repeat \lya\ forest targets,
an average of 12 fibers deg$^{-2}$ remain for additional \lya\ forest quasar targets.
In regions with multiple epochs of SDSS imaging, photometric variability
was used to enhance the selection of \lya\ quasars in BOSS \citep{palanque-delabrouille11a}.
In eBOSS, we will use the same technique applied to multi-epoch imaging
data from the Palomar Transient Factory \citep[PTF;][]{rau09a,law09a}.
PTF $R$-band imaging data with at least five epochs covers 90\% of the
available eBOSS footprint.
We expect an efficiency of about $20\%$ in the selection of \lya\ quasars from variability.
Nominally, we would allocate 12 fibers deg$^{-2}$ to this sample, but instead
allow a target density of $20$ deg$^{-2}$ in the PTF regions with an expectation of
3--4 deg$^{-2}$ new, confirmed \lya\ forest quasars.
We increase the density to account for reduced LRG and quasar sample sizes 
caused by ``fiber collisions'' between objects that lie too close to neighboring targets.

\item
As described above, the ELG program is allocated 300 plates with a goal of
obtaining spectroscopic classification of 190,000 galaxies, where the predicted BAO precision
depends primarily on the total number of ELGs rather than the density
or volume covered.  There remains uncertainty in
the final ELG program, as the exact selection method is still under
investigation.  The algorithm that preferentially selects galaxies
with significant [OII] $\lambda$3727 emission will likely be the one that is most efficient at obtaining
this sample.  Given this uncertainty, we require only that the sample cover
a redshift interval $\Delta z =0.4$ to ensure a cosmologically interesting volume,
lie at a median redshift that is higher than the LRG sample,
and have an upper bound $z<1.1$ to avoid confusion in [OII] line identification in the
wavelength region that is dominated by sky lines.
These constraints effectively limit the sample of 190,000 galaxies to lie at either $0.6<z<1.0$ or
$0.7<z<1.1$.  If targets were to fill the fibers at the lowest density possible, a sample
selected at a density 170 deg$^{-2}$ over $1500 \sqdeg$ would meet the requirements
if attained at a purity exceeding 74\%.  Likewise, a sample at 340 deg$^{-2}$ over 
$750 \sqdeg$ would meet the requirements at the same purity.
As with the LRG sample, a fractional change in the number density has the same
consequence for BAO precision as would the same fractional change in areal coverage.

\end{enumerate}

In total, the final eBOSS spectroscopy will consist of:
LRG targets ($0.6<z<1.0$) at a density of 50 deg$^{-2}$ and desired purity exceeding 80\%,
ELG targets at a somewhat higher redshift over 300 dedicated plates with a desired purity exceeding 74\%,
``clustering'' quasar targets (denoted {\tt QSO\_CORE}) to directly trace large-scale structure ($0.9<z<2.2$)
at a density of 90 deg$^{-2}$ and desired purity exceeding 50\%,
re-observations of faint BOSS \lya\ quasars ($z>2.1$) at a density of 8 deg$^{-2}$,
and new \lya\ quasar candidates identified by variability at an average density of 18 deg$^{-2}$.

\subsection{Target Selection Requirements}\label{subsec:uniformity}

We will only be able to meet the projected cosmological constraints if our final large-scale clustering measurements 
are limited by statistical errors.  This requires stringent control
of systematic errors that can modulate the data on large scales,
such as the impact of stellar contamination and dust extinction on
target selection efficiency, variations in seeing that alter target
selection and redshift success, and so on.  These systematics have
already been extensively studied within BOSS
\citep{ross11a, ross12b, ho12a,pullen12a},
and the greater volume and greater statistical power at large
scales from eBOSS will place new demands on homogeneity of the target samples.
These effects are important
primarily for measurements (such as neutrino masses, the effective
number of neutrino species, and inflation) that use the full shape
of the galaxy power spectrum.  BAO measurements are largely protected because
they rely on a relatively sharp feature, but with the high precision of
eBOSS, one must be careful to extract the BAO
signature in a way that is immune to large-scale modulations.

Based on experience from BOSS, we apply a strict list of requirements to
the selection of the LRG, ELG, and clustering quasar target classes to limit
these systematic effects. 
Because \lya\ quasars provide random sightlines to sample the foreground
density field, we do not require uniformity in the \lya\ quasar sample.

\begin{itemize}

\item
Statistical and systematic uncertainties in redshift
estimates dampen the BAO feature in the radial direction.
We require that the data reduction pipeline provide
a combined precision and accuracy of $\sigma_v < 300\,$kms$^{-1}$ RMS at all redshifts,
where $\sigma_v$ is defined as $c \sigma_z/(1+z)$.
Increasing the tolerance in redshift estimates to $< 540\,$kms$^{-1}$ RMS would
increase the projected BAO and RSD uncertainty by 10\%.
The identifying features in galaxy spectra are typically well-characterized
emission or absorption features with widths less than $300\,$km s$^{-1}$.
We expect to have redshift precision significantly better than this requirement
for all classified galaxies;  we expect occasional line confusion to
introduce larger redshift errors that would be classified as ``catastrophic'',
as addressed in the next item.

The prominent emission lines in quasars are broad and subject to velocity
shifts with respect to the systemic redshift \citep[e.g.][]{hewett10a,richards11a}.  Measurements of clustering
in the BOSS quasar sample imply redshift errors as large as $\Delta z/(1+z) = 0.003$ \citep{white12a}.
In addition to inflated redshift errors, \citet{font-ribera13a} provides compelling evidence for apparent velocity
biases in the $z>2.1$ BOSS quasar sample of roughly $160\,$kms$^{-1}$.
Comparing automated redshifts to redshifts based on MgII $\lambda$2799, we believe we can
improve the redshifting of quasars from those in BOSS and meet
the strict redshift accuracy and precision requirements for quasars at $z<1.5$.
To acknowledge the difficulty in redshift accuracy for the high redshift region of the clustering
quasar sample, we relax the requirement to $[300+400(z-1.5)]$\,kms$^{-1}$ for objects at $z>1.5$.
Any errors from the MgII redshifts would contribute an additional degradation in the redshift accuracy.
We estimate that degrading the redshift accuracy by 70\% (by the same factor across all redshifts)
increases the projected BAO and RSD uncertainty in the quasar sample by factor of 1.1.

\item
Redshift errors exceeding $1000\,$kms$^{-1}$ can potentially bias the derived
clustering signal by falsely projecting classified spectra onto the wrong redshift, possibly
through line confusion or contamination by sky lines.
As long as the fraction of objects with catastrophic redshift failures is
small compared to the number of true tracers at any redshift, the misidentified fraction should
lead to small enough changes in clustering to not bias the estimate of the BAO position.
To minimize the impact of catastrophic errors, we require that the spectroscopic pipeline classify
spectra with fewer than 1\% catastrophic errors, where the redshifts are not known to be in error.
As with the redshift accuracy requirement, we relax the definition of a catastrophic failure
in the quasar sample to unknown errors in excess of $3000\,$km s$^{-1}$.

\item
The maximum absolute variation in expected galaxy density as a function of imaging survey
sensitivity, stellar density, and Galactic extinction must be less than 15\% (peak to peak).
As discussed in Section~\ref{subsec:bao}, variations smaller than 10\% have been mitigated in BOSS by straightforward
weighting schemes.  Based on this experience, we expect that we can extend galaxy weights to
regions with 15\% variation in target density with little effect on measurements in the power
spectrum in the range $0.02 h $Mpc$^{-1} < k < 0.3 h $Mpc$^{-1}$.
Any areas with fluctuations greater than 15\% could be discarded in the cosmology analysis
and thus degrade the final precision of the clustering measurement.

\item
Finally, for all samples used to directly constrain clustering, we require that our target selection
procedures be robust against variations in the imaging zeropoints.
We require that sample densities vary by less than 15\% for the estimated uncertainties in zeropoint
for the input imaging data in each bandpass.  Unlike the previous requirement for maximum variation in density,
this requirement can be computed directly by evaluating the derivative
of target density with respect to an offset in photometry in a single band.

\end{itemize}

\section{Survey Design}
\label{sec:strategy}

Several developments since the beginning of BOSS allow us to identify
the new samples of tracers for eBOSS cosmological measurements.
First, the recent release of infrared sky maps from the
Wide-field Infrared Survey Explorer \citep[WISE;][]{wright10a}
makes it possible to expand the selection of LRG targets beyond that
probed by BOSS.
Second, recent spectroscopic observations prove that we can select targets from
imaging data and spectroscopically confirm star-forming galaxies with high efficiency
from the 2.5-meter Sloan Foundation Telescope \citep{comparat13a}.
The galaxies with strong emission lines make it possible to further extend
the galaxy redshift survey.
Selection for these targets is not yet finalized and we plan to conduct the ELG
observations in the third and fourth years of eBOSS.
Third, spectroscopic observations reveal that quasars in the redshift range
$0.9 < z < 2.2$ can be efficiently identified from spectroscopy \citep{palanque-delabrouille13a}.
Finally, improved photometric target selection and time-domain imaging data
provide a dense sample of $z>2.1$ quasars to significantly
sharpen the cosmology measurements from the \lya\ forest relative to BOSS.

The procedures to select targets and complete a program of spectroscopy are well established from BOSS.
We will filter the photometric catalogs, design the spectroscopic
plates, and acquire the spectroscopic data in a nearly identical fashion to the
procedures outlined in \citet{dawson13a}.
However, there are changes to each step in the process, listed below.

\subsection{Imaging and Catalog Creation}\label{subsec:pre-selection}

The baseline data set for eBOSS target selection is the
well-understood photometry obtained from the SDSS imaging camera \citep{gunn98a}
in $ugriz$ filters \citep{fukugita96a}.
The median 5-$\sigma$ depth for photometric observations of point sources is $u = 22.15$, $g=23.13$,
$r=22.70$, $i=22.20$, $z=20.71$.
Photometry for each filter $X$ is presented in PSF magnitudes (denoted $X_{\rm PSF}$), 
fiber2 magnitudes ($X_{\rm fib2}$) to represent the fiber aperture losses, model magnitudes ($X_{\rm model}$), or cmodel magnitudes
($X_{\rm cmodel}$), as discussed in online SDSS documentation\footnote{http://www.sdss.org/dr12/algorithms/magnitudes}.
The SDSS photon counts and object detections used in eBOSS target selection
algorithms stem from the DR9 public release \citep{ahn13a}.
The BOSS DR9 photometry is internally calibrated using the ``ubercalibration''
procedure described in \citet{padmanabhan08a}, with residual systematic errors
of approximately 1.5\% in $u$ and 1\% in the other four bands.
eBOSS has recalibrated the flat fields in all five filters and zero-points in the $griz$ bands
using a combination of PanSTARRS-1 \citep{kaiser10a} and SDSS stellar photometry.
The residual systematics are reduced to 0.9, 0.7, 0.7 and 0.8\% in 
the $griz$ bands, respectively \citep{finkbeiner15a}.  
In addition, some poorly-constrained zero-points
with errors exceeding 3\% in the DR9 data are now significantly improved.

The photometric selection for the BOSS galaxy sample was based entirely on imaging from SDSS.
Photometry from the UKIRT Infrared Deep Sky Survey
\citep[UKIDSS;][]{lawrence07a} and the Galaxy Evolution Explorer
\citep[GALEX;][]{martin05a} were used to enhance the selection of \lya\ forest quasars \citep{bovy12a}.
The use of additional imaging resources to identify the primary cosmological sample set a precedent
that we follow in eBOSS.  The primary source of external imaging data will be the highly-uniform,
infrared (IR) photometry from the WISE satellite.
The catalog of IR sources is derived using forced photometry informed by positions
of sources in the SDSS imaging data.  The full process is explained in \citet{lang14a}.
The resulting 3.4$\mu$m and 4.6$\mu$m ($W1$ and $W2$, respectively) magnitudes
can be found in the public release\footnote{http://unwise.me/}.

The imaging sample to be used for selection of ELG targets is still under investigation.
ELG targets could be selected using SDSS $griz$ and $U$-band imaging
from the South Galactic Cap U-band Sky Survey (SCUSS) conducted at the 2.3m
Bok Telescope on Kitt Peak.  The reductions of those two imaging samples are complete
and photometric catalogs are available over the entire SGC area.
ELG targets could also be selected from deeper $grz$ imaging obtained
from the Dark Energy Camera \citep[DECam;][]{flaugher12a}.
The deeper imaging data allows more precise identification of strong emission-line
galaxies in the redshift range of interest.
We currently have catalogs based on preliminary data reductions of first year DECam observations.
Through a large program approved for the Dark Energy Spectroscopic Instrument \citep[DESI;][]{levi13a},
we expect to cover the majority of the SGC with DECam in $grz$.  Known as the
DECam Legacy Survey (DECaLS\footnote{http://legacysurvey.org/}),
the raw data from those observations are immediately public and the DESI team plans to publicly
release full reductions of these data.
However, observations are scheduled over
2014 -- 2017 and it is not yet clear how much area will be available for selection of eBOSS targets
on the timescales needed for SDSS-IV observations.

Finally, we use multi-epoch imaging data from the Palomar Transient
Factory \citep[PTF;][]{law09a} and data from the Faint Images of the Radio Sky at Twenty-Centimeters \citep[FIRST;][]{becker95a}
program to increase the number of quasars.
PTF covers roughly 90\% of the potential eBOSS footprint at a depth that is sufficient to produce and
characterize lightcurves.
These data are made publicly available by the PTF team\footnote{http://irsa.ipac.caltech.edu/Missions/ptf.html}.
The average field is observed over four years in 45 60-second exposures.
Lightcurves are constructed with a customized image processing pipeline
that produces coadditions of all images acquired in a single year.
The depth of a resulting epoch is typically comparable to the depth of SDSS photometry.
Earlier SDSS photometry is used to anchor the lightcurves and
provide a longer time baseline to measure variability.
The catalog of sources from FIRST is used to identify potential quasars matched to SDSS optical counterparts.
Sources are selected from the June 5, 2013 version of the FIRST 
point source catalog\footnote{\url{http://sundog.stsci.edu/first/catalogs/readme\_13jun05.html}}.

For all imaging samples, magnitudes are corrected for Galactic extinction using the \citet{schlegel98a}
models of dust absorption.
Galactic extinction coefficients have been updated as recommended in
\citet{schlafly11a}.  The extinction coefficients $R_u$, $R_g$, $R_r$, $R_i$, and $R_z$
are changed from the values used in BOSS (5.155, 3.793, 2.751, 2.086, 1.479) to (4.239, 3.303, 2.285, 1.698, 1.263), respectively.
An extinction coefficient $R_{W1}=0.184$ is used for the WISE 3.4$\mu$m band
and $R_{W2}=0.113$ is used for the 4.6$\mu$m band \citep{fitzpatrick99a}.

In BOSS, we used a series of imaging flags to remove possible contaminants
from the target samples.
The standard CALIB\_STATUS SDSS imaging flags ({\tt SATUR} etc.) become increasingly poor at distinguishing real objects
from artifacts at fainter magnitudes.
We therefore make no explicit cuts on our galaxy or quasar targets based on the standard SDSS imaging flags.
We do remove quasar targets with IMAGE\_STATUS flags identified as ``BAD\_ROTATOR'', ``BAD\_ASTROM'',``BAD\_FOCUS',
``SHUTTERS'', ``FF\_PETALS'', ``DEAD\_CCD'', or ``NOISY\_CCD''.
We rely on our use of multi-wavelength photometry (such as imaging from WISE)
to identify true astrophysical objects as candidate sources for spectroscopy.
We expect that some objects close to bright stars might have contaminated colors,
causing spurious objects to enter into the target selection.
Areas close to bright stars represent a tiny fraction of the sky
and it is not obvious a priori how close to bright stars the photometry remains reliable,
so we do not explicitly remove targets.
Clustering analyses will likely either mask or re-weight the galaxy and quasar 
densities due to the loss of objects near bright stars, as has been done for the BOSS
clustering analyses \citep[e.g.][]{anderson12a}.

\subsection{Target Selection Algorithms}\label{subsec:targets}

The targets for eBOSS spectroscopy will consist of:
LRGs at $0.6<z<1.0$, ELGs at somewhat higher redshift,
``clustering'' quasars to directly trace large-scale structure ($0.9<z<2.2$),
re-observations of faint BOSS \lya\ quasars ($z>2.1$), and new \lya\ quasars ($z>2.1$).
The selection algorithms for identifying the LRG and quasar samples are complete and
observations of these targets have begun.
The selection algorithm for ELG sample is under investigation.
Here, we summarize the selection scheme and statistics of each sample.

\begin{table}[t]
\centering
\caption{
\label{tab:LRGQSOdensities}
Expected redshift distribution for the LRG, quasar, and \lya\ quasar samples.
The surface densities are presented in units of deg$^{-2}$ assuming that 100\% of the objects
in the parent sample are spectroscopically observed.
Entries highlighted in bold font denote the fraction of the sample that satisfies the high-level
requirement for the redshift distribution of the sample.}
\begin{tabular}{l c c c c c c c}
\hline\hline
& CMASS & LRG  & LRG  & {\tt QSO\_CORE} & {\tt QSO\_CORE} & \lya\ Quasar & Quasar  \\
& Known & ($z_{\rm conf}>0$)\tablenotemark{a}  & ($z_{\rm conf}>1$)\tablenotemark{a} & New  & Known\tablenotemark{b}  & Reobservation & PTF\tablenotemark{c}  \\ \hline
Poor Spectra  & -- & 4.0 & 6.7  & 3.2 & 0.9  & 0 & 0.7   \\
Stellar       & -- & 5.3 & 5.3  & 2.8 & 0.3 & 0 &  10.7  \\
Galaxy        & -- & --  & --   & 6.6 & 0.4 & 0 &  1.5  \\
$0.0 < z < 0.5$  & 27.3 & 0.6 & 0.6  & 1.0 & 0.4 & 0 & 0.2   \\
$0.5 < z < 0.6$  & 45.7 & 6.2 & 5.9   & 1.1 & 0.4 & 0 &  0.1  \\
$0.6 < z < 0.7$  & {\bf 19.4} & {\bf 15.2} & {\bf 14.8}  & 1.4 & 0.7 & 0 & 0.2   \\
$0.7 < z < 0.8$  & {\bf 3.5} & {\bf 15.3} & {\bf 14.7}  & 1.4 & 1.3 & 0 &  0.2  \\
$0.8 < z < 0.9$  & {\bf 0.2} & {\bf 9.4}  & {\bf 8.7}    & 2.2 & 1.5 & 0 &  0.2  \\
$0.9 < z < 1.0$  & {\bf 0.03}& {\bf 3.2}  & {\bf 2.7}    & {\bf 3.6} & {\bf 1.0} & 0 &  0.3  \\
$1.0 < z < 1.2$  & 0 & 0.6   & 0.5   & {\bf 8.4} & {\bf 1.8} & 0 &  0.5  \\
$1.2 < z < 1.4$  & 0 & 0  & 0 & {\bf 10.3} & {\bf 1.8} & 0 &  0.4  \\
$1.4 < z < 1.6$  & 0 & 0  & 0 & {\bf 10.3} & {\bf 2.1} & 0 &  0.6  \\
$1.6 < z < 1.8$  & 0 & 0  & 0 & {\bf 9.9} & {\bf 2.0} & 0 &  0.4  \\
$1.8 < z < 2.0$  & 0 & 0  & 0 & {\bf 9.2} & {\bf 1.9} & 0 &  0.3  \\
$2.0 < z < 2.1$  & 0 & 0  & 0 & {\bf 4.0} & {\bf 1.0} & 0 &  0.2  \\
$2.1 < z < 2.2$  & 0 & 0  & 0 & {\bf 2.2}\tablenotemark{d} & {\bf 1.6}\tablenotemark{d} & {\bf 0.5}  & {\bf 0.2}   \\
$2.2 < z < 2.4$  & 0 & 0  & 0 & 1.8\tablenotemark{d} & 4.5\tablenotemark{d} & {\bf 2.9} &  {\bf 0.5}  \\
$2.4 < z < 2.6$  & 0 & 0  & 0 & 1.1\tablenotemark{d} & 3.1\tablenotemark{d} & {\bf 1.9} &  {\bf 0.5}  \\
$2.6 < z < 2.8$  & 0 & 0  & 0 & 0.7\tablenotemark{d} & 1.4\tablenotemark{d} & {\bf 1.0} &  {\bf 0.6}  \\
$2.8 < z < 3.0$  & 0 & 0  & 0 & 0.3\tablenotemark{d} & 0.8\tablenotemark{d} & {\bf 0.7} &  {\bf 0.5}  \\
$3.0 < z < 3.5$  & 0 & 0  & 0 & 0.4\tablenotemark{d} & 1.2\tablenotemark{d} & {\bf 0.9} &  {\bf 0.5}  \\
$z > 3.5$ & 0 & 0 & 0  & 0.1\tablenotemark{d} & 0.1\tablenotemark{d} & {\bf 0.3} &  {\bf 0.2}  \\
\hline
Total Targets & 23  & 60 & 60  & 81.8 & 30.4 & 8.3 & 20   \\
Total Tracers & {\bf 23.1}  & {\bf 43.1} & {\bf 41.0}   & {\bf 57.9} & {\bf 13.1} & {\bf 8.3} & {\bf 3.2}   \\ \hline
\end{tabular}
\tablenotetext{a}{The redshift distribution for the LRG sample is determined by visual inspection.
It is not clear how well the automated pipeline will perform in the final analysis, so we include
the results from visual inspections with $z_{\rm conf}>0$ in the first entry as the more optimistic
estimate and results with $z_{\rm conf}>1$ in the second entry as the less optimistic estimate.}
\tablenotetext{b}{Objects that satisfy the {\tt QSO\_CORE} selection that have reliable spectra from previous
incarnations of SDSS will not be assigned fibers in eBOSS.}
\tablenotetext{c}{The density of PTF-selected quasars is only computed over areas where
targets are selected.  Roughly 10\% of the eBOSS footprint will not have PTF-selected quasar targets.}
\tablenotetext{d}{The population of $z>2.1$ quasars that appear in the target selection for the
clustering sample will be used for \lya\ forest studies.}
\end{table}

\subsubsection{LRG Samples}

A full investigation of the LRG selection from imaging data is presented in the companion
paper \citep{prakash15a}.
The final sample was designed to extend the BOSS galaxy sample to higher redshifts
following initial studies of an optical and infrared selection of LRGs \citep{prakash15b}.
LRG candidates will be observed (assigned a fiber) at an average density of 50 deg$^{-2}$ as
governed by the survey design explained in Section~\ref{subsec:program}.
Targets are selected at a density to oversubscribe the fiber budget and ensure a high overall
efficiency of fibers assigned to cosmological tracers.
Given the fiber assignment statistics described in Section~\ref{subsec:tiling},
we find that a parent sample with density of 60 deg$^{-2}$ is adequate to satisfy the requirement
to spectroscopically observe objects at a density of 50 deg$^{-2}$.

At the redshifts of the LRG sample ($z>0.6$), the 4000 \AA\ break is shifted into the SDSS $i$ filter,
increasing the error of $ugr$ photometric estimates.
In addition, the degradation of photometry at higher redshifts provides motivation to include
ancillary photometric information beyond SDSS to improve the selection efficiency.
The WISE $W1$ filter centered at 3.4 microns provides restframe coverage of
the ``1.6 micron bump'' that results from a reduction in the opacity of H$^{-}$ ions
exhibited by old stellar populations \citep{john88a}.
The final LRG selection algorithm is tuned to identifying objects that are red in $r-i$, $i-z$ and
$r-W1$ colors as demonstrated in \citet{prakash15a}.

As will be explained in Section~\ref{subsec:LRGspectra}, the estimate of the redshift distribution
stems from visual inspections that are difficult to translate to the final pipeline performance.
For that reason, we report maximum and minimum number densities that correspond to more or less optimistic
interpretations of the visual inspections.
Between 68.3\% and 71.8\% of the targets identified by this selection lie between $0.6<z<1.0$,
depending on which level of confidence is assumed in the visual inspection process.
The surface density of the sample is shown as a function of redshift in Table~\ref{tab:LRGQSOdensities}.
Objects with spectra that fail to produce a reliable classification are labeled ``Poor Spectra''.

The density of M-stars and of galaxies at $0.5<z<0.6$ are both higher than what was expected
following the initial pilot programs in BOSS.
Because of these two contaminants, the parent population represented by the SDSS/WISE selection fails to meet
the goal of 80\% purity for galaxies in the redshift range $0.6<z<1.0$.
Roughly 24\% of the CMASS galaxies in BOSS lie at redshifts $z>0.6$ and roughly 4\% lie
at redshifts $z>0.7$.  While not originally intended, those CMASS galaxies can be combined with
the new eBOSS LRG galaxies to increase the overall sample size and obtain a 1\% precision distance measurement.

We assess uniformity of the target sample through a regression analysis of
surface density against tracers of potential systematics.  We focus on systematics associated with
imaging data and astrophysical effects such as dust extinction and stellar density.
The resulting regression fits reveal that 92\% of the potential imaging area in eBOSS
has predicted surface density that varies by less than 15\%, thus satisfying the imaging
uniformity requirements outlined in Section~\ref{subsec:uniformity}.
The 8\% of the area that fails to meet this requirement must be re-examined when spectroscopic
observations are complete to assess variations after identification of contaminating sources such as stars.
Likewise, we assess the variations in number density associated with errors in the zeropoint
calibration.  We find that zeropoint errors of 0.01 magnitude in the $rizW1$ filters
cause fractional changes in the number density of LRG targets of 2.26\%, 2.5\%, 6.24\%, and 0.6\%, respectively.
Likely due to the fact that it is used for both a color boundary and a magnitude boundary, the target selection
is most sensitive to the uncertainty in the $z$-band calibration;  non-uniformity with
15\% peak-to-peak amplitude occurs in regions where the $z$-band zeropoint is in error by
$\pm 0.012$ magnitudes.  Assuming a Gaussian dispersion of 0.008 magnitudes in the $z$-band
calibration as discussed in Section~\ref{subsec:pre-selection} and in \citet{finkbeiner15a},
13.3\% of the eBOSS volume exceeds the 15\% uniformity requirement.
This effect is not expected to be correlated with imaging systematics and
in the worst case scenario, will have to be addressed through an independent census of zeropoints
in each SDSS field.
Analysis through mock catalogs will inform the manner in which zeropoint
uncertainties are modeled in the clustering measurements.
The full methodology of the uniformity tests is presented in \citet{prakash15a}.

\subsubsection{Quasar Samples}

The algorithm to identify quasars in BOSS was intended to build
a sample of $z>2.1$ quasars to map the large scale distribution of neutral hydrogen via absorption
in the \lya\ forest.
As explained in \citet{ross12a}, the most uniform quasar sample for BOSS was identified from
a density of 20 targets deg$^{-2}$ using the ``Extreme Deconvolution''
\citep[XDQSO;][]{bovy11a,bovy11b} selection.   This selection was
performed after estimating the relative density of stars and quasars as a function of
color, magnitude, and photometric uncertainty.
The probability that an object is a quasar is determined by the fraction of objects
with similar photometric properties expected to be a quasar and not a star.

Following that precedent, the XDQSO algorithm
will be again used to identify quasars for the {\tt QSO\_CORE} sample in eBOSS,
but using the XDQSOz version of the algorithm \citep{bovy12a} that can be applied to select quasars in any redshift range.

To improve the completeness of the sample, objects with
much lower XDQSO probabilities will be included in eBOSS relative to BOSS.
The inclusion of lower probability objects will also tap into the population
of $z>2.1$ quasars that were missed in BOSS and enhance the sample for \lya\
forest clustering measurements.

Colors between optical and WISE passbands can be used to
distinguish quasars from stellar objects characterized by blackbody spectra \citep[for example][]{stern12a}.
Morphology selection helps reduce the number of lower luminosity, extended AGN sources in
favor of compact, point-like quasars which map onto higher mass halos.
In these ways, WISE imaging will be used to decrease stellar contamination
and increase the overall efficiency of fiber assignments.

The final selection algorithm for quasars in the {\tt QSO\_CORE} sample is described 
in the companion paper on quasar target selection \citep{myers15a}.
The final selection when applying XDQSOz and the SDSS/WISE colors results in a target
density of 115 deg$^{-2}$.
The average density of targets falls to roughly 90 deg$^{-2}$ after removing objects
that were confidently classified as a star, galaxy, or quasar in SDSS or BOSS spectra.

As shown in Table~\ref{tab:LRGQSOdensities}, pilot studies in BOSS indicate that the selection identifies
71 quasars deg$^{-2}$ over $0.9 < z < 2.2$, of which 13 will already
be known from previous observations.  The selection will also increase the
number of \lya\ quasars:  6.6 new $z>2.1$ quasars deg$^{-2}$ are expected.
In addition to stars, we list the rate of galaxy contamination
in the quasar sample;  roughly 8\% of the new targets turn out to be galaxies with significant line emission.
If targeted at 100\% completeness, this shot noise-limited sample would
exceed the goal of 58 deg$^{-2}$, leading to potential improvements of up to 20\%
over the high-level goal of a 2\% precision distance measurement with this sample.

Performing the same regression analysis as on the LRG sample, we find that
$\sim 90$\% of the eBOSS area satisfies the requirement of $<15\%$ peak-to-peak variation
in the quasar target density.
The quasar target selection is very robust against zeropoint errors.
We find that zeropoint errors of 0.01 magnitudes in any of the $ugrizW1W2$
filters cause fluctuations in target density of less than 1\%.  The largest contributor
to density fluctuations is the $g$-band calibration;
a 0.01 magnitude error leads to a 0.86\% change in target density.
Given the expected RMS error in flux calibration is only 0.009 magnitudes,
we expect an RMS scatter in target density of only 0.77\% due to calibration errors.
Only a negligible fraction of sky will see density fluctuations as large as 15\%.

Quasars selected for \lya\ forest studies are not subject to the same strict requirements of uniformity
as those used in direct clustering work.  For this reason, a series of selections was used to
increase the total number of $z>2.1$ quasars.
The first selection was based on known quasars from BOSS.
A known quasar is included in the eBOSS selection if it obtained a low signal-to-noise in the BOSS
observation.  The mean signal-to-noise ratio is computed over the range $1040 < \lambda < 1200$ \AA.
Objects that did not have broad absorption lines identified in visual inspection and
that have $0.75<S/N<3$ or $S/N=0$ are observed again in eBOSS.
As explained in \citet{myers15a} and in Table~\ref{tab:LRGQSOdensities}, we also identify unique new \lya\
forest quasars in the PTF data at a density of 3.2 deg$^{-2}$ where PTF imaging is available to generate 
sufficient lightcurves.  At redshifts $z>2.5$, the PTF selection identifies quasars at a density
2.7 deg$^{-2}$, of which 2.3 deg$^{-2}$ are unique to variability selection.
Finally, a small number of objects from the FIRST catalogs are included as possible quasars.
Because the target density is only around 1 deg$^{-2}$, we do not include the statistics from that sample in
Table~\ref{tab:LRGQSOdensities}.

\subsubsection{ELG sample}

The ELG survey will begin in Fall 2016, the third year of eBOSS observations.
By starting the ELG program two years later than the LRG and quasar programs, the eBOSS team has given
itself time to perform deeper analysis of the potential selection algorithms.
We conducted a series of observations in Fall 2014 to test possible techniques for the selection of ELG
targets.  At the time of writing, the tests are being used to evaluate the redshift success rates, redshift distributions,
and rates of stellar contamination from four different selection algorithms. 

Each of these four options relies on selecting targets from a subset of imaging
data selected from SCUSS $U$-band, SDSS $griz$, WISE $W1$ and $W2$, or $grz$ imaging with DECam.
The tests will also inform the final choice for imaging data to be used in the ELG selection.
The expected redshift distribution for
each candidate selection is presented in Table~\ref{tab:ELGtargets}.

\begin{table}[htp]
\centering
\caption{
\label{tab:ELGtargets}
Expected redshift distribution for the four potential ELG selection algorithms.
The surface densities are presented in units of deg$^{-2}$ assuming that 100\% of the objects
in the parent sample are spectroscopically observed.
Entries highlighted in bold font denote the fraction of the sample that satisfies the high-level
requirement for the redshift distribution of the sample.}
\begin{tabular}{l c c c c}
\hline\hline
  & $gri\,+\,Uri$ & Fisher Discriminant & DECam\tablenotemark{a} (Low Density) & DECam\tablenotemark{a}(High Density)   \\ \hline
Poor Spectra    & 64.7        & 19.8         & 17.1         &  24.1 \\
Stellar         & 4.0         & 1.5          & 0.0          &  0.0  \\
$0.0 < z < 0.1$ &  2.1        & 2.9          & 0.6          &  0.6 \\
$0.1 < z < 0.2$ &  2.7        & 2.0          & 1.7          &  2.4 \\
$0.2 < z < 0.3$ &  3.6        & 2.5          & 1.6          &  2.0 \\
$0.3 < z < 0.4$ &  4.1        & 1.7          & 1.0          &  1.2 \\
$0.4 < z < 0.5$ &  4.8        & 2.8          & 0.6          &  0.7 \\
$0.5 < z < 0.6$ &  9.4        & 7.2          & 1.2          &  1.7 \\
$0.6 < z < 0.7$ & {\bf 27.6}  & {\bf 25.4}   & 3.3          &  3.7 \\
$0.7 < z < 0.8$ & {\bf 42.8}  & {\bf 46.2}   & {\bf 40.7}   & {\bf 44.1} \\
$0.8 < z < 0.9$ & {\bf 25.9}  & {\bf 40.4}   & {\bf 65.3}   & {\bf 74.1} \\
$0.9 < z < 1.0$ & {\bf 10.8}  & {\bf 17.0}   & {\bf 30.9}   & {\bf 43.7} \\
$1.0 < z < 1.1$ &  2.5        & 6.5          & {\bf 11.1}   & {\bf 16.9} \\
$1.1 < z < 1.2$ &  0.4        & 2.6          & 5.1          & 7.7        \\
$z>1.2$         &  2.2        & 2.9          & 9.5          & 16.9       \\
\hline
Total Targets   &  203.9      & 182   & 190.4 & 241.3  \\
Total Tracers   &  {\bf 107.1}      & {\bf 129.0} & {\bf 148.0} & {\bf 178.8}  \\ \hline
\end{tabular}
\tablenotetext{a}{The desired range for DECam-selected ELGs is $0.7<z<1.1$ instead of $0.6<z<1.0$.}
\end{table}

The first candidate for ELG selection uses SCUSS imaging for deeper $U$-band photometry to complement SDSS
$gri$ photometry.  Objects for the spectroscopic sample are taken from the union of a $g-r$ and $r-i$ selection
in SDSS imaging and a $U-r$ and $r-i$ selection in SCUSS and SDSS imaging.
In addition to several cuts designed to reject point sources and ensure good photometry,
objects satisfying $g-r<0.8$ and $r-i>0.8$ are included in the SDSS $gri$ color selection.
These cuts lead to a mean target density of 132.5 deg$^{-2}$.
Objects satisfying $i - U > -3.5*(r - i) + 0.7$ and $r-i>0.7$ are included in the SCUSS and SDSS $Uri$ selection
at a density of 84.0 deg$^{-2}$.
In all cases, photometry is assumed in extinction-corrected model magnitudes.
The selections are described in detail in \citet{comparat15a}.
The combined algorithm allows the selection of 180 objects per square degree over any region of the South
Galactic Cap.  The density in the test region was somewhat higher and had median redshift $z=0.72$.
Many objects in this selection lie near the SDSS detection limit; photometric errors are
sometimes as large as 0.5 magnitudes.  The large errors reduce the precision of the color estimates and 
effectively broaden the redshift distribution.
Only 52.5\% of the targets produce reliable redshift estimates in the redshift range intended for clustering studies.
In addition, the large photometric errors allow fainter objects into the selection, thus reducing the
bias relative to a sample represented by a more precisely defined magnitude limit.

The second candidate selection relies on the addition of the WISE photometry to the SDSS and SCUSS photometry.
This scheme imposes a cut on Fisher discriminant quantities \citep{fisher36a} 
instead of cuts in color-color diagrams.
In this case, the Fisher discriminant quantity is a linear combination of colors taken from $UgrizW1$. 
The exact definitions of the linear combination of colors, the
training from a spectroscopic sample, and the expected results are described in \citet{raichoor15a}.
The selection algorithm can be used to identify targets up to a density of 200 deg$^{-2}$, somewhat
higher than the density in the test region.
The test observations demonstrate a significant improvement over the SDSS$+$SCUSS ($gri + Uri$) selection.
Roughly 71\% of these objects are expected to produce reliable redshifts in the range
$0.6<z<1.0$.  The initial tests of this selection approach the requirement of 74\% purity presented
in Section~\ref{subsec:program}.

The final two candidates for ELG selection use DECam $grz$ photometry instead of SDSS,
SCUSS, or WISE photometry.  
Both of these selections exceed the 74\% purity requirement as detailed in \citet{comparat15a}.
The advantage of these selection schemes is that deeper DECam
photometry allows more precise selection of star-forming galaxies at somewhat higher redshifts
than the SDSS and SCUSS selections.  For this reason, we increase the required redshift range
for the spectroscopically-confirmed tracer population from $0.6<z<1.0$ to $0.7<z<1.1$.
The disadvantage of this selection scheme is that it relies on a relatively new imaging camera and
a reduction pipeline that is in a state of development.
The two candidate algorithms were tuned from spectroscopic observations of a broad
color selection that produced targets to $g<22.8$ at a density of 597 deg$^{-2}$ in the test region.
The DES imaging is reduced using magnitudes determined in a $4\2pr$ diameter aperture and in
photometry similar to model magnitudes from SDSS.  Extinction-corrected magnitudes are used throughout.
Both selection algorithms apply the following cuts to reduce stellar contamination to nearly zero:
$g_{\rm APER,4} - g_{\rm DETMODEL} < 2$,
$r_{\rm APER,4} - r_{\rm DETMODEL} < 2$, and
$z_{\rm APER,4} - z_{\rm DETMODEL} < 2$,
where $X_{\rm APER,4}$ is derived from an aperture two arcseconds in diameter and $X_{\rm DETMODEL}$ is
a DES measurement determined from SExtractor \citep{bertin96a} similar to modelmag in SDSS.
The point source rejection also limits the number of quasars entering into the sample;
the expected quasar contamination is expected to be less than 1\% at $z<1.1$.

The first of the DECam selections (denoted ``Low Density'') was tuned to achieve a target density of roughly 190 deg$^{-2}$.
An equivalent to model magnitudes is used in the selection.
The color cuts that define that sample are as follows:
\begin{itemize}
\item $21.6< g < 22.8$
\item $0.3< g-r< 0.7$
\item $0.25 < r-z < 1.4$
\item $r-z > 0.45*(g-r) + 0.4$
\item $r-z < 0.7*(g-r) + 0.8$
\end{itemize}
The resulting redshift distribution is shown in the fourth column of Table~\ref{tab:ELGtargets}.
77.8$\pm1.1\%$ of objects lie at $0.7<z<1.1$ with a median redshift $z=0.86$.

The second of the DECam selections was tuned to achieve a target density of roughly 240 deg$^{-2}$.
The color cuts that define that sample are as follows:
\begin{itemize}
\item $21.5< g < 22.8$
\item $0.2< g-r< 0.7$
\item $0.25 < r-z < 1.4$
\item $r-z > 0.45*(g-r) + 0.4$
\item $r-z < 0.8*(g-r) + 1$
\end{itemize}
The redshift distribution from this ``High Density''
DECam selection is shown in the last column of Table~\ref{tab:ELGtargets}.
74.1$\pm0.5\%$ of objects lie at $0.7<z<1.1$ with a median redshift $z=0.87$.

While it is possible to increase the density above 240 deg$^{-2}$ using DECam data, initial tests reveal that
contamination from galaxies outside the desired redshift range reduces the efficiency of the selection. 
In addition, as the selection approaches 300 deg$^{-2}$, the density peaks around $3 \times 10^{-4}$ h$^3$ Mpc$^{-3}$.
At this density, BAO measurements become dominated by sample variance rather than shot noise
and observing time is more efficiently spent expanding the survey volume than by increasing the density.
When comparing the low and high density DECam selections presented above, one must also consider
the strength of the [OII] and [OIII] emission line fluxes.
The weighted mean of the [OIII] line flux is 6.6 and $6.9 \times 10^{-17}$ erg cm$^2$ s$^{-1}$
for the low and high density selections, respectively.
Likewise, the [OII] line flux is 8.15 and $8.5 \times 10^{-17}$ erg cm$^2$ s$^{-1}$.
Although the high density selection produces a somewhat smaller rate of galaxies at
$0.7<z<1.1$, the typical line strengths are 5\% higher than in the low density selection,
thus making it more robust to automated classification.

Tests of uniformity, sensitivity to zeropoint uncertainty, and average target density
are underway \citep{delubac15b}.  A final decision on the selection to be used for eBOSS is expected in early
2016.  The sample that is able to produce a uniform target density, redshift classification
exceeding 74\% efficiency over the appropriate redshift range, and imaging area sufficient for a total
sample of 300 plates will be chosen.  If more than one selection meets these requirements, the selection
algorithms producing the highest median redshift will be the one used for the ELG sample.

\subsection{Tiling and Fiber Assignment}\label{subsec:tiling}

The goal of survey tiling is to create a spatial distribution of tiles that
maximizes the number of targets observed with the minimum number of tiles.
We define the tiling completeness as the fraction
of objects in a given class that were assigned fibers.
We refer to the fiber efficiency as the fraction of available fibers used for
unique science targets. In BOSS, the quantity of highest priority was
the tiling completeness for `decollided' targets. The physical size of
the ferrules that support each fiber in the plug plate limits the
proximity of neighboring targets to $62\2pr$.  Groups of targets that
lie within $62\2pr$ of one another are denoted ``collision
groups''. The decollided set contains all targets that are not within
collision groups, combined with the subset of collided targets that
can be assigned fibers on a single plate.  A collision pair
contributes one galaxy to the decollided set because, in all cases,
one target from the pair will be assigned a fiber. If the pair is located within a
region observed by more than one tile, the second object may be
assigned a fiber as well. Thus the completeness of the collided set
will be non-zero.

The spatial distribution of tiles in BOSS was set such that the
decollided completeness of galaxy and \lya\ forest quasar samples was
100\%. Due to the inhomogeneity of the target list, it is not possible
to reach 100\% decollided completeness and 100\% fiber efficiency. 
Since the tiling completeness was the
higher priority, the BOSS fiber efficiency for LOWZ, CMASS, and quasar
targets was $\sim 90\%$.

In eBOSS, we adopted a tiered-priority system for survey targets
to maximize both tiling completeness and fiber efficiency. All non-LRG targets
receive maximal priority and are assigned fibers first. The tiling
solution must satisfy the requirement of 100\% tiling completeness for the
decollided set of all non-LRG targets. For LRGs, we no longer require
full decollided completeness. Rather, the density of LRG targets
intentionally oversubscribes the remaining fiber budget.
The input sample is tiled at a lower density due to
collisions with higher-priority targets, collisions between multiple
LRGs, and lack of available fibers for LRGs in the decollided set. We
refer to collisions of LRGs with higher-priority targets as {\it knockouts}
in order to differentiate them with collisions between two LRGs.
To achieve the survey goal of candidate LRG targets observed at 50 deg$^{-2}$, the
input target catalog is increased to a target density of 60 deg$^{-2}$. Thus, after
high-priority targets are assigned fibers, plates that have a surplus
of unused fibers can sometimes be used to observe LRGs at a density
higher than 50 deg$^{-2}$. The quantity of
interest for LRG tiling is the completeness of targets that are both
decollided (i.e., with respect to other LRGs) and no-knockout (ie, with
respect to high-priority targets). The former effect can be corrected
in clustering measurements by proper weighting of tiled targets. The
latter is essentially a mask overlayed on the LRG portion of the
survey. In eBOSS, we require that the completeness of decollided,
no-knockout LRGs be at least 95\%. This layered tiling scheme
allows 8\% more area to be covered than would be possible if the eBOSS fiber
efficiency were the same as in BOSS.

In the first round of fiber assignments---the non-LRG targets---we
must specify the priority for fiber assignments within collision
groups. Because the quasar targets are significantly higher density than
TDSS and SPIDERS targets, quasar-TDSS/SPIDERS collisions are fractionally
more common for TDSS/SPIDERS target classes. Thus collisions are
resolved in the following order (highest to lowest priority): SPIDERS,
TDSS, reobservation of known quasars, clustering quasars, and
variability-selected quasars.  Quasars found in the FIRST survey and
white dwarf stars that can be used as possible calibration standards
are given the final priorities for resolving fiber collisions. 
Because the density of clustering
quasar targets is comparable to the galaxy sample in BOSS, and because
the fiber assignments require 100\% completeness on the decollided
sample, the resulting sample of clustering quasars follows tiling
statistics that are very similar to the galaxies in BOSS.

\begin{figure}[h!]
\centering
%\vspace{-1.5cm}
\includegraphics[width=0.75\textwidth]{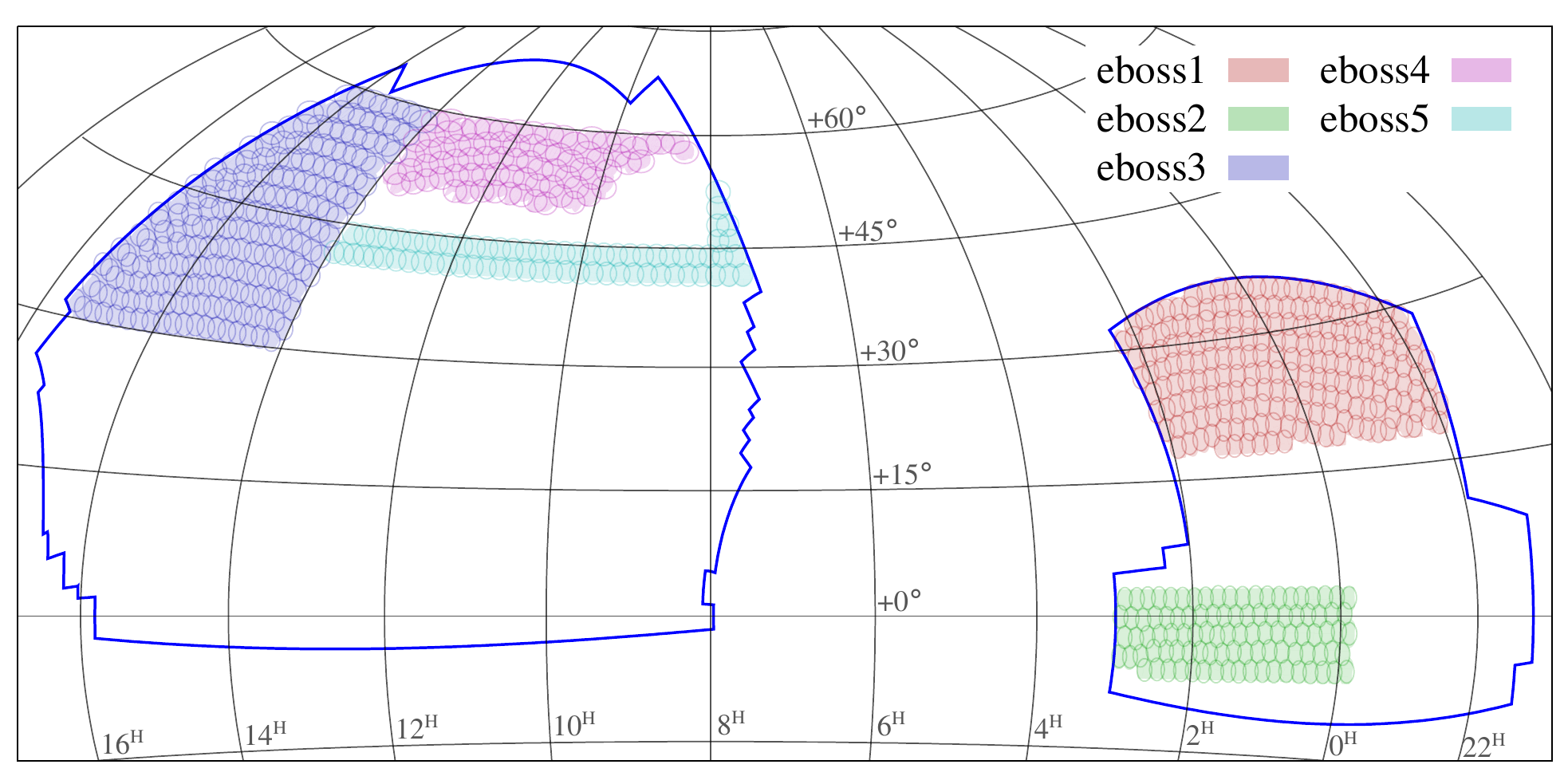}
\caption{ Field centers for eboss[1--5].  The SEQUELS area is clearly
  defined by white space between the boundaries of eboss4 and eboss5.
  The area covered here is the area that was tiled in the beginning
  of SDSS-IV and the approximate survey area expected to be
  completed in the first two years of observation.  }
\label{fig:ebosschunks}
\end{figure}

Five distinct regions of sky (denoted ebossN, where N is a number
ranging from 1 -- 5) were tiled in the first year of eBOSS.  The area
and tile centers for those regions are shown in
Figure~\ref{fig:ebosschunks}.  These regions convey the average tiling
statistics that can be expected for the quasar samples over the
duration of the program.  The statistics of the quasar samples for
each region are presented in Table~\ref{tab:completeness}.  The input
target density for the {\tt QSO\_CORE} sample of quasars ranges between 73
and 88 deg$^{-2}$, a bit lower than the expected density of 90
deg$^{-2}$.  The density of previously-observed objects that satisfies
the primary quasar selection is 17, 32, 31, 41, and 32
deg$^{-2}$ for eboss[1--5], respectively.  Region eboss1 is outside
the final SDSS-II footprint, so fewer SDSS spectra exist and the number
of known targets is reduced by a factor of two. Generally, the variation in
the density of known objects explains the variation in the new quasar sample;
the total {\tt QSO\_CORE} density only ranges from 113.1 to 118.2  deg$^{-2}$
for eboss[2--5].  The total {\tt QSO\_CORE} density in
eboss1 is 100.9 deg$^{-2}$, significantly lower than the average of the other
chunks. Regression tests indicate that this suppression in surface density
is likely due to the higher extinction in eboss1 relative to the rest of the
SDSS imaging area.
See \citet{myers15a} for a full discussion of non-uniformity in the quasar
target selection.

\begin{table}[htp]
\centering
\caption{\label{tab:completeness} Fiber assignment statistics in the first five tiled regions for the quasar programs.
  Surface densities are all presented in units of deg$^{-2}$.  The ``target density'' corresponds to the density of objects from
each class that satisfy the photometric selection criteria.  The ``fiber density'' corresponds to the density of targets
that were assigned fibers in the tiling process.}
\begin{tabular}{l c c c c c c c c}
\hline\hline
{Chunk} & {Number of} & {Area} & {Total CORE\tablenotemark{a}} & {New CORE\tablenotemark{b}} & {New CORE\tablenotemark{b}}
& {Reobserved \lya} & {PTF\tablenotemark{c} Target} & {PTF\tablenotemark{c} Fiber}   \\
{} & {Plates} &  {(deg$^2$)} & {Target Density} & {Target Density} & {Fiber Density}
& {Fiber Density} & {Density} & {Density}  \\ \hline
eboss1     & 199 & 987.9 & 100.9 & 84.10   &   80.47   &   6.19    &   18.59   &   17.23     \\
eboss2     & 98 & 469.4  & 113.1& 80.84   &   77.08   &   9.47    &   10.56   &   9.67      \\
eboss3     & 204 & 984.7 & 118.2 & 87.52   &   83.69   &   8.55     &   21.24   &   19.80     \\
eboss4     & 80 & 397.1  & 114.4 & 73.47   &   69.86   &   9.16   &   28.70   &   26.36     \\
eboss5     & 70 & 353.7  & 116.4 & 84.19   &   79.68   &   8.49    &   14.51   &   13.30     \\ \hline
\end{tabular}
\tablenotetext{a}{The sample denoted ``Total CORE'' corresponds to all targets satisfying the
{\tt QSO\_CORE} selection criteria, regardless of prior observations.}
\tablenotetext{b}{The sample denoted ``New CORE'' corresponds to all targets satisfying the
{\tt QSO\_CORE} selection criteria that were not observed previously and are candidates for fiber assignment
in eBOSS.}
\tablenotetext{c}{The sample denoted ``PTF'' corresponds to all targets satisfying the
variability selection criteria using PTF imaging data that are exclusive of all other samples.}
\end{table}

On average, the input target sample of clustering quasars is assigned
fibers at a completeness of 95\%, thus reducing the expected number
density of $0.9<z<2.2$ quasars by 5\% relative to the parent sample
presented in Table~\ref{tab:LRGQSOdensities}.  A similar trend is
found for the PTF-selected quasars, but at slightly lower completeness
($\sim$92\%) due to their lower priority in resolving fiber
collisions.  Note that the repeated observations of \lya\ quasars are
by definition exclusive of the {\tt QSO\_CORE} selection but the
PTF-selected quasars do have significant overlap with the {\tt QSO\_CORE}
sample.  In Table~\ref{tab:LRGQSOdensities} and Table~\ref{tab:completeness},
we have assigned PTF-selected targets only to the {\tt QSO\_CORE}
sample when overlap occurs.  The PTF-selected densities presented in
the table therefore reflect the density of unique PTF-selected targets.  Because
the FIRST sample only produces quasars at 1 deg$^{-2}$, we do not
include the statistics of that sample in Table~\ref{tab:completeness}.

The area of the first five regions that is covered by only a single
tile is 2054 deg$^2$, 64\% of the total 3193 deg$^2$ currently tiled.
In these areas, there is no way to capture objects lost to fiber
collisions with other targets.  The remaining 1139 deg$^2$ is covered
by two or more tiles.  In these regions, the completeness of the
collided objects increases significantly, leading to a total
completeness on the quasar sample of close to 100\%.  

In the second round of tiling, LRGs are the only target species and
require no prioritization to resolve fiber collisions.  The statistics of the
LRG sample are presented in Table~\ref{tab:LRGtiles}.  Because the
redshift distribution of the LRG sample only overlaps $\sim 8.5$ deg$^{-2}$
new {\tt QSO\_CORE} objects at redshifts $z<1$, the masked region is
mostly uncorrelated with clustering in the LRG sample.
There is negligible overlap between the LRG and TDSS/SPIDERS samples.
To first order, the areas of sky restricted from observing LRG targets can be treated in a
similar manner to regions lost to bright stars or imaging artifacts in
BOSS.  Quantifying the full consequences of the overlapping samples
will be done on mock catalogs as part of future clustering analyses.

\begin{figure}[htb!]
\centering
%\vspace{-1.5cm}
\includegraphics[width=0.48\textwidth,trim=0 -2in 0 0, clip]{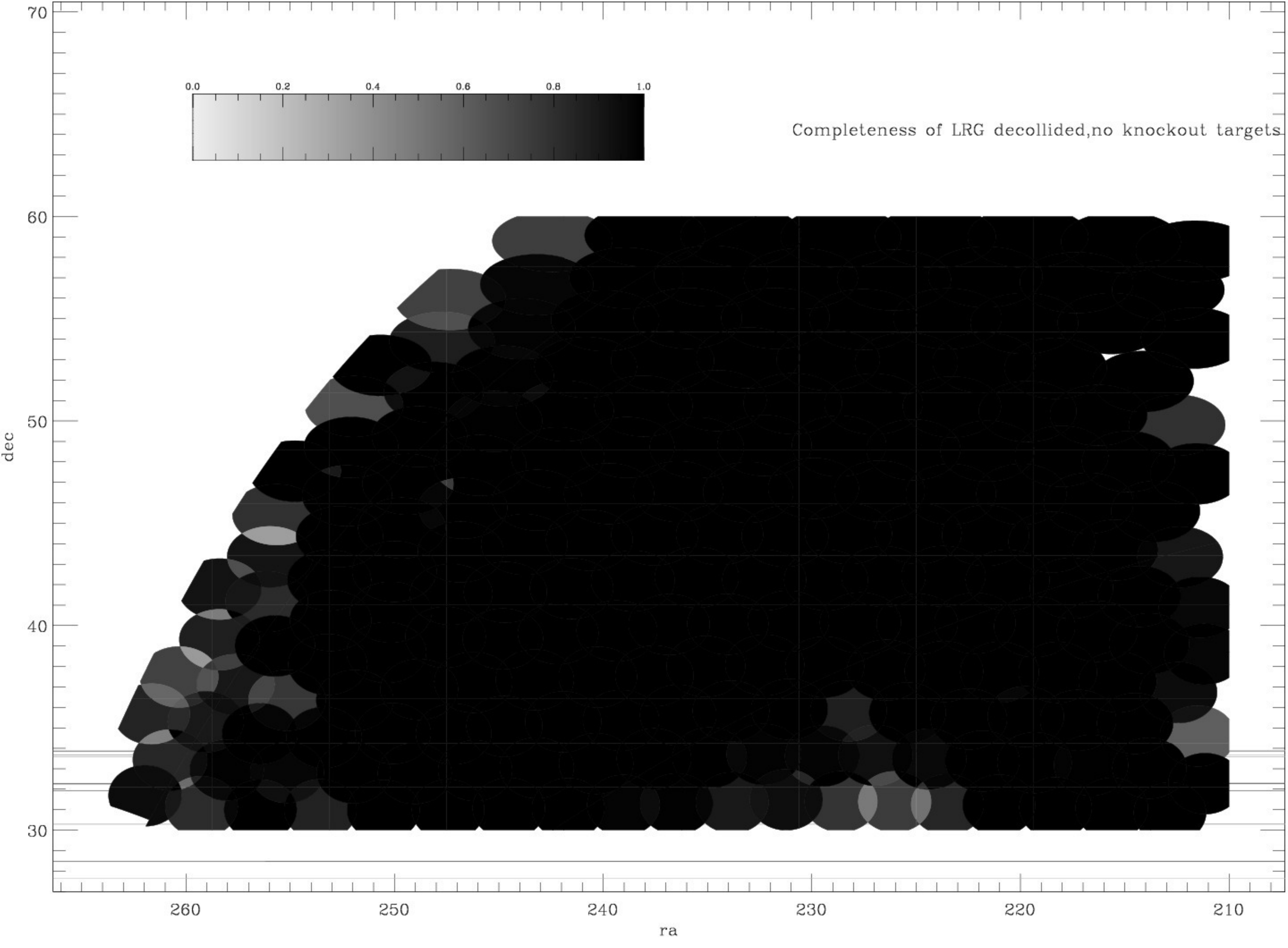}
\includegraphics[width=0.48\textwidth,trim=0 2.25in 0 1.4in, clip]{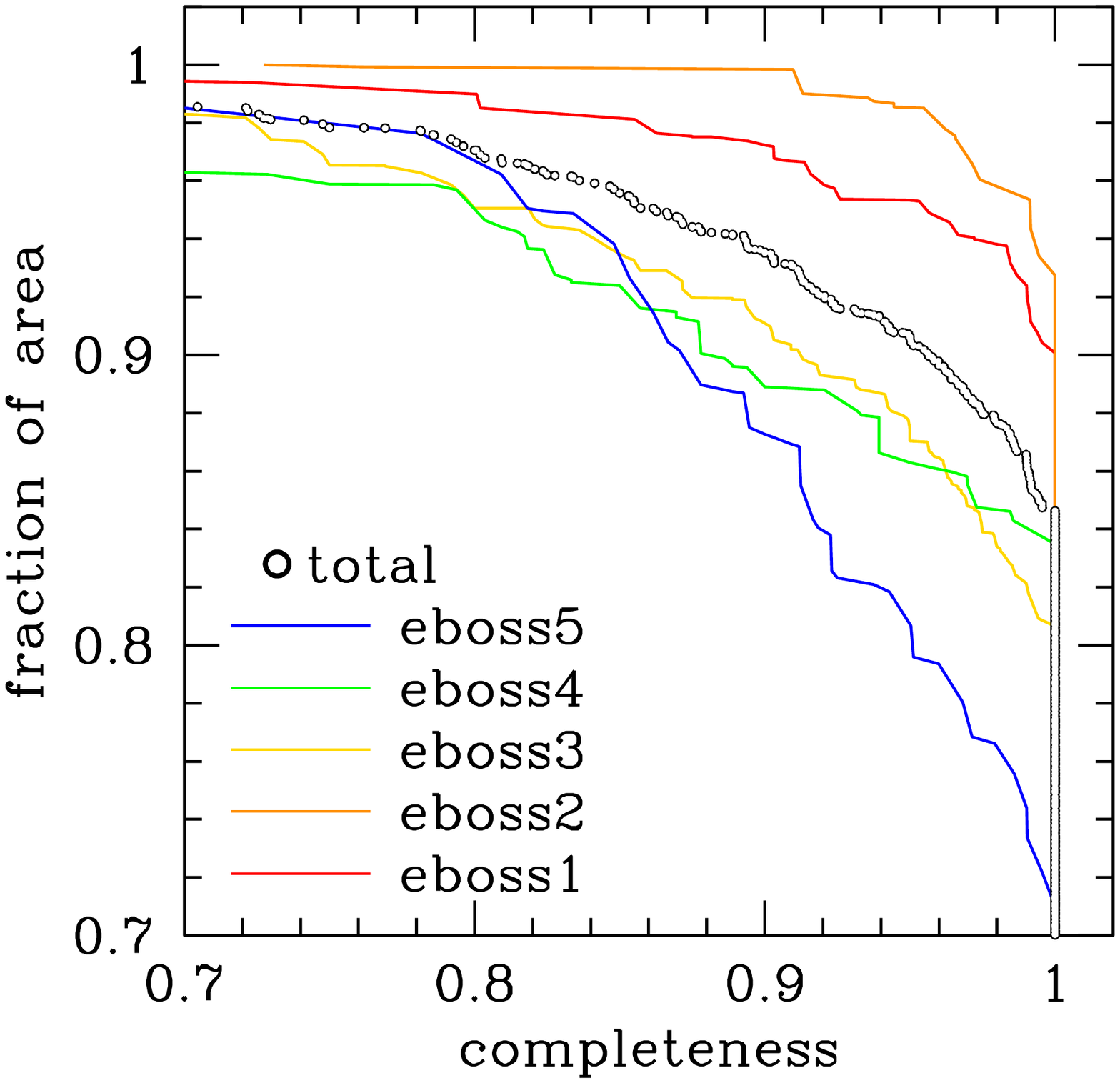}
\caption{ {\bf Left: }The completeness of the noknock, decollided LRG
  sample over the eboss3 region.  {\bf Right: } The cumulative
  distribution of completeness in the noknock, decollided LRG sample.
  The distribution is weighted by the area of each independent sector
  defined by areas covered by overlapping and unique tiles. }
\label{fig:lrgtiles}
\end{figure}

On average, the total completeness of the LRG sample is 87\% while the
completeness of decollided, noknock (no-knockout)
targets is 98\%.  The average
density of LRG targets given a fiber is 52 deg$^{-2}$, slightly
exceeding the goal of 50 deg$^{-2}$.  The resulting fiber assignments
occasionally produce local fluctuations that may have insufficient
completeness for clustering analysis.  An example of the fluctuations
of the decollided, noknock LRG targets tiled in eboss3 is shown in the
left hand panel of Figure~\ref{fig:lrgtiles}.  The area-weighted
cumulative histogram of completeness over the entire region covered by
the first five regions is shown in the right hand panel of
Figure~\ref{fig:lrgtiles}.  Assuming these five regions are
representative of the completeness we expect in the full eBOSS
footprint, 5\% of the area will fall below 85\% completeness in the
decollided, noknock LRG targets.  These areas exceed the 15\%
uniformity requirement that we have generally assumed and may be
excised from the clustering analysis depending on the results
of future tests on mock catalogs.

\begin{table}[htp]
\centering
\caption{\label{tab:LRGtiles} Fiber assignment statistics in the first five regions tiled for eBOSS.
  Surface densities are all presented in units of deg$^{-2}$.}
\begin{tabular}{l c c c c c}
\hline\hline
{Chunk}  & {LRG Target} & {LRG Knockout} & {LRG noknock, collided} & {LRG Fiber}  & {Noknock, decollided}  \\
{} & {Density} & {Density} & {Density} & {Density}  & {Completeness} \\ \hline
eboss1     &  64.66   &   4.41  &    3.56  &    56.64   &  98.9\%  \\
eboss2     &  58.73   &   3.87  &    2.53  &    52.90   &  99.7\%  \\
eboss3     &  60.53   &   4.11  &    2.75  &    52.74   &  97.3\%  \\
eboss4     &  57.12   &   4.07  &    2.48  &    49.24   &  96.6\%  \\
eboss5     &  59.37   &   4.81  &    2.66  &    50.37   &  96.4\%  \\  \hline
\end{tabular}
\end{table}

In the 300-plate ELG program, all ELG targets will be assigned equal
priority and ranked ahead of other targets.
Because they have not been tiled over a cosmologically useful volume, we do not present
the statistics of the ELG sample.  Instead, those statistics
will be included in a future paper dedicated to the final ELG target selection algorithm.

\subsection{Plate Design}

In the tiling stage described above, we assign science targets to fibers based on the
input from target selection algorithms and the available fiber budget.
In plate design, we convert those positions from celestial coordinates to the
coordinate system of the telescope focal plane.  We also assign additional 
targets to the reserved fibers to provide reference for sky subtraction and 
flux calibration.

Tests in BOSS reveal a measurable increase in sky-subtraction residuals when the number of sky fibers
drops below 80 per plate.  We therefore
maintain the density of at least 80 sky fibers per plate in eBOSS.
In addition, we now require that at least 30 sky fibers are assigned to each spectrograph.

Standard stars for flux calibration in BOSS were selected at a density of 20 per plate over the magnitude range
$15 < r_{\rm fib2} < 19$.
We maintain a density of 20 F-stars selected as standard stars, but now require
that at least six stars be assigned to both spectrographs.
The F-stars at the faint limit in BOSS
produced spectra that were of marginal quality for flux calibration, while the brightest stars had
nearly three orders of magnitude higher flux than the faintest science targets and increase the
risk of contamination from cross-talk between neighboring fibers.  For this reason, we choose
eBOSS F-stars to have a narrower magnitude distribution, with $16 < r_{\rm fib2} < 18$.
To compensate for the reduced density in candidate F-stars, we increase the scaled
distance in extinction-corrected color space from the color of a fiducial F star by a factor
of two relative to BOSS.  We now require that $m_{\mathrm{dist}} < 0.16$, where

%\begin{align*}
\begin{eqnarray}
m_{\mathrm{dist}} &=& \left[ \left((u-g)-0.82\right)^2 + \left((g-r)-0.30\right)^2  \right. \nonumber \\
                && \left. + \left((r-i)-0.09\right)^2 + \left((i-z)-0.02\right)^2  \right]^{1/2} .
\end{eqnarray}
%\end{align*}

In BOSS, after accounting for predicted atmospheric differential refraction (ADR) for
each plate, galaxy targets were centered in the focal plane to maximize throughput
for 5400 \AA\ light.  
The hole positions for quasar targets were centered to maximize throughput at 
4000 \AA\ to increase the S/N in the \lya\ forest.
In eBOSS, the {\tt QSO\_CORE} sample is selected to lie at $z>0.9$ to provide direct clustering
measurements while the other quasar samples are selected to lie at $z>2.1$ to increase the
\lya\ forest signal with respect to BOSS.
All objects selected in the {\tt QSO\_CORE} sample are centered in the focal plane
at a position corresponding to the focus of 5400 \AA\ light.
These targets will have appropriate flux calibration derived from the F-stars.
All $z>2.1$ quasars known from BOSS and all quasar candidates selected by variability
are centered in the focal plane
at a position corresponding to the focus of 4000 \AA\ light.
While not appropriate for flux calibration in the current data reduction pipeline,
the routines developed in \citet{margala15a} can be applied to the \lya\ target spectra
to improve the broadband distortion introduced by the ADR offset.
As before, the wavelength that determines the center of the hole position is recorded
in the quantity {\tt LAMBDA\_EFF}.
In BOSS, washers manufactured with an adhesive were applied to the back of the plates at the location of quasar
targets.  Washer thickness varied according to distance from the plate
center to account for the 0--300 micron difference in focus between 4000 \AA\ light and
5400 \AA\ light.   These washers matched the quasar fibers to the focal plane for 4000 \AA\ light for
optimal focus by offsetting the fiber tip in the direction perpendicular to the surface of the plate.
However, due to weather, guiding, and other stochastic processes, we were unable to confirm
the ability of the sticky washers to significantly improve the S/N in the \lya\ forest.
Those washers are therefore not applied in eBOSS.

\subsection{Observing Sequence}\label{subsec:strategy}

Simple data reductions are performed in real time
to provide quick feedback to the observers and to track the depth of each exposure.
The observers acquire signal on each field until the accumulated depth
exceeds an empirically derived threshold for each camera.
In BOSS, the depth was tuned to reach desired redshift success rates on the
highest redshift galaxy targets in the shortest exposure time possible. 
Exposures were tuned so that the typical square of the signal-to-noise per pixel
was at least 22 for an object with native SDSS magnitude $i_{\rm fib2}=21$ over the wavelengths covered
by the $i$-band filter.
We also required the square of the signal-to-noise per pixel
to be at least 10 for an object with $g_{\rm fib2}=22$ over the wavelengths covered
by the $g$-band filter.
In eBOSS, we will maintain the same thresholds for the beginning of the survey.
To ensure that we define plate completion criteria so that we complete the goal of 1800 plates,
we will use the first year of observation to empirically determine new exposure thresholds.
As was done in BOSS, the depths will be tuned to ensure that we can reliably classify
targets in the shortest exposure possible.
It is likely that different thresholds will be used for
the quasar and LRG plates than those used for the ELG plates.

Because quasars and LRGs are the primary focus of the eBOSS program,
the majority of fibers and survey area are dedicated to these two target classes.
The total area covered by these samples will be $7500\,\sqdeg$ divided over one
contiguous region in the SGC and one contiguous region in the NGC.
The eboss1 region presented in Figure~\ref{fig:ebosschunks} was tiled to cover
an SGC area at high declination that is easily observed.
The eboss2 region was chosen to obtain 
spectroscopy in a 500 deg$^2$ region that overlaps with the Dark Energy Survey (DES) footprint.
The NGC is tiled in one contiguous region at declination below 60 degrees.

As was explained in Section~\ref{subsec:targets}, the sample selection for
the quasar and LRG samples is complete.
The first two years of observation will be dedicated almost exclusively
to these primary fields.
After two years, approximately 600 plates should be completed for the LRG and quasar targets,
producing a sample comparable in area to the DR9 CMASS cosmology sample.
Under the assumption of average weather conditions, we expect to cover almost the
entire area covered by eboss[1--5] and shown in Figure~\ref{fig:ebosschunks}.

The third and fourth years will be split evenly between observations of ELG
plates and observations of the LRG and quasar plates.
The scheduled time in which the SGC is visible will be dedicated to ELG targets
while the NGC time will be dedicated to the LRG and quasar targets.
At a rate of 300 plates per year, we expect the ELG program to be completed by the summer
of 2018.
Because the selection of ELG targets remains uncertain, we do not define the exact
footprint of that survey at this time.
For 2016--2018 LRG and quasar observations, we expect 300 plates to be completed in a 1500 $\sqdeg$ area of the NGC that lies
just below the eboss3 and eboss5 regions.  The exact area will depend on how much progress is made
in the 2014--2015 and 2015--2016 observing seasons.

Observations of ELG targets will be completed before the summer of 2018, regardless of whether
we meet the goal of 300 plates.  By completing the ELG program in the fourth year,
we establish a meaningful cosmological sample in the shortest possible time.
The final two years will be dedicated entirely to
the LRG and quasar targets, with the goal of covering as much SGC and NGC area as possible.

\section{Spectroscopic Data Quality}
\label{sec:sequels}

\begin{figure}[h!]
\centering
\vspace{-3.75cm}
\includegraphics[trim=1in 2.8in 0 0]{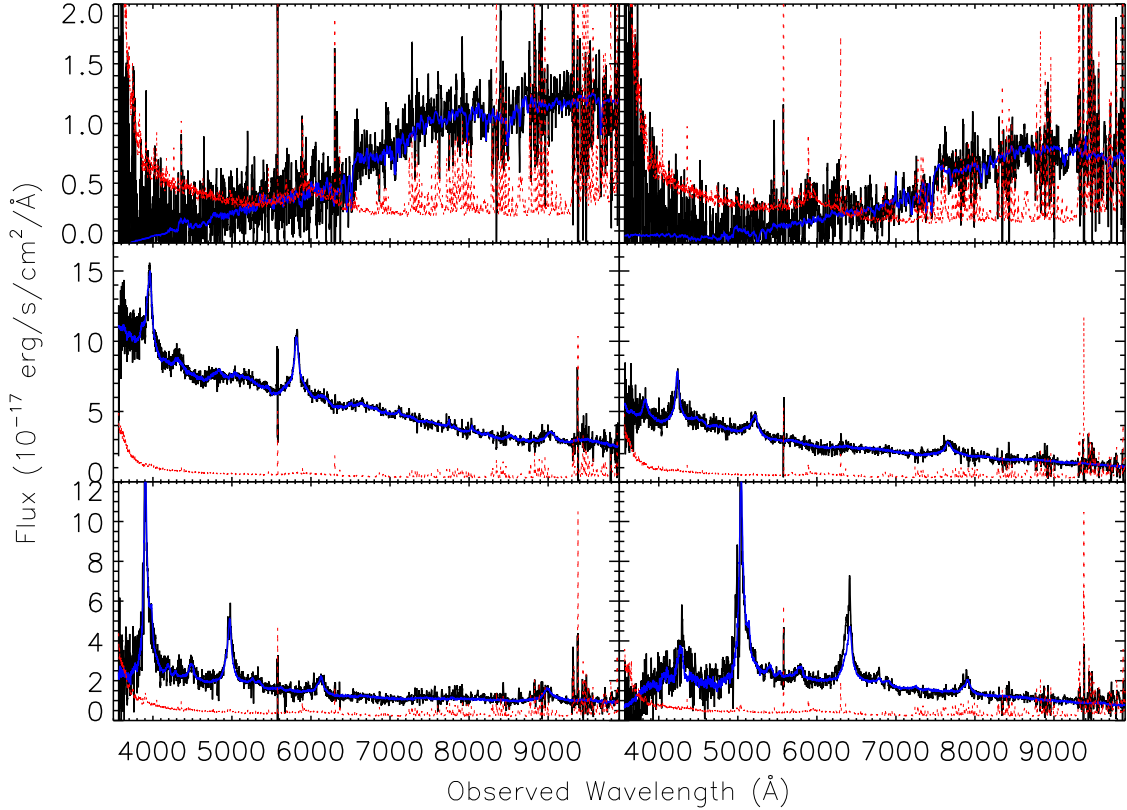}
\vspace{-6.9cm}
\caption{
Examples of SEQUELS spectra that span the range of redshifts expected in the LRG, clustering quasar,
and \lya\ forest quasar samples.  In each, the data are represented in black, the flux errors on each pixel in red, and the template
in blue.  A boxcar smoothing kernel of width 5 pixels has been applied to the data
for illustrative effect.  Each spectrum is classified with high confidence by the automated data reduction pipeline.
{\bf Top Left:  }An LRG at $z=0.64$.
{\bf Top Right:  }An LRG at $z=0.88$.
{\bf Middle Left:  }A quasar at $z=1.08$ identified by the {\tt QSO\_CORE} selection algorithm.
{\bf Middle Right:  }A quasar at $z=1.74$ identified by the {\tt QSO\_CORE} selection algorithm.
{\bf Bottom Left:  }A quasar at $z=2.21$ identified by the {\tt QSO\_CORE} selection algorithm.
{\bf Bottom Right:  }A quasar at $z=3.15$ identified by variability in the PTF imaging data.
}
\label{fig:goodspectra}
\end{figure}

As explained in the appendix of \citet{alam15a}, good fortune during 2010--2014 allowed the main BOSS program
to finish early, leaving time for new dedicated spectroscopic observations.
128 plates were drilled for the SEQUELS program,
covering the $466\,\sqdeg$ region bounded by eboss4 and eboss5 in Figure~\ref{fig:ebosschunks}.
In this section, we use the public SEQUELS data to present the quality of the LRG, clustering quasar,
and \lya\ forest quasar spectra expected in eBOSS.  A mosaic of randomly selected spectra spanning the
relevant range of redshift for each of these three samples is presented in Figure~\ref{fig:goodspectra}.

SEQUELS was designed as a pilot survey for eBOSS, using a slightly broader selection for LRG,
clustering quasars, and \lya\ forest quasars that was guaranteed to include the final
eBOSS selection for these classes. 
66 SEQUELS plates were completed before the end of SDSS-III and are included in DR12.
Details for accessing these data are online\footnote{http://www.sdss.org/dr12/} and in \citet{alam15a}.
The remaining 62 plates will be observed at highest priority during the first year of eBOSS.
The 466 $\sqdeg$ area will not be re-tiled with eBOSS-selected targets;  instead, the SEQUELS
targets will be integrated into future clustering analysis according to the same
selection algorithms as those in eBOSS.

All SEQUELS targets are tracked by the dedicated {\tt EBOSS\_TARGET0} bitmask.
In the 66 plates that were observed, fibers were placed on 11,687 unique targets
from a modified version of the final eBOSS LRG selection algorithm.  
Although the LRG selection in SEQUELS is broader than in eBOSS, the eBOSS targets
can be identified by objects with {\tt LRG\_RIW} (corresponding to bit number 2),
extinction-corrected magnitude $i_{model}<21.8$, and extinction-corrected magnitude $W1 < 17.6$ (approximately 20.3 in AB).
10,873 of the SEQUELS LRG targets satisfy the final eBOSS LRG selection algorithm.
Likewise, 19,461 unique clustering quasar targets, 6,479 PTF-selected quasar targets,
and 1,368 reobserved \lya\ quasars are found in SEQUELS.
The clustering sample and the reobserved \lya\ quasar sample are selected in an identical fashion to those
in eBOSS, identified by {\tt QSO\_EBOSS\_CORE} (bit 10) and {\tt QSO\_REOBS} (bit 12), respectively.
The PTF sample contains a higher density of objects than in eBOSS and is difficult to
reproduce without the variability parameters for each object.  The variability parameters will
be found in the final BOSS quasar catalog \citep{paris15a} when it is complete.

The SEQUELS spectra are the source of the redshift distributions
for quasars presented in Table~\ref{tab:LRGQSOdensities}.
Only SEQUELS data are presented in this section for the quasar samples.
Both SEQUELS and proprietary eBOSS spectra
were used to determine the redshift distributions
for LRGs presented in Table~\ref{tab:LRGQSOdensities},
although the discussion of pipeline performance and the example spectra are based
only on SEQUELS.
The spectra of a large number of these objects were visually inspected to
help characterize pipeline performance and to settle on the final redshifts of the samples.
Pilot observations and visual inspection during the first few months of SDSS-IV led to the
estimates of ELG redshift distributions presented in Table~\ref{tab:ELGtargets}.
Since the final ELG program is not decided, we postpone further discussion of those samples until future publications.
Below, we describe the typical quality of LRG and quasar spectra 
and the interpretation of the automated redshifts and visual inspections that led to the $N(z)$ estimates assumed for eBOSS.

\subsection{LRG Spectra from SEQUELS}\label{subsec:LRGspectra}

The SEQUELS data are reduced by the same ``idlspec2d''
routines as those used in BOSS.  The spectral templates described in \citet{bolton12a} are fit
to each reduced spectrum to derive a redshift and object classification.
These templates were tuned for optimal performance in the BOSS galaxy sample
and exceeded the predicted redshift success rates for even the most distant
galaxies.  The automated classification of the final BOSS CMASS galaxy sample shows
98.4\% of objects with {\tt ZWARNING\_NOQSO} equal to zero, indicating that the automated
redshift estimate is reliable when using a template suite that excludes quasar templates.

The choice of spectral templates was sufficient for BOSS, but is not optimized for the fainter,
higher redshift LRG galaxies that comprise the eBOSS LRG sample.
Of the 10,873 eBOSS LRG targets in SEQUELS, 7,796 produced a {\tt ZWARNING\_NOQSO} value equal to zero
in the automated reductions.
A statistical error on the redshift estimate is provided by the automated classification routine
for each of these galaxies.
Tests in BOSS using repeat observations showed that these errors were underestimated
by up to a factor of 1.34 for galaxies \citep{bolton12a}.
Assuming that the SEQUELS/eBOSS errors are underestimated by a factor of 1.34,
we find that the median redshift error is only 58 km s$^{-1}$ and that only ten objects have
a redshift error larger than 200 km s$^{-1}$.  The automated classification characterizes
redshifts with a precision much better than required.

\begin{figure}[h!]
\centering
\includegraphics[width=0.45\textwidth]{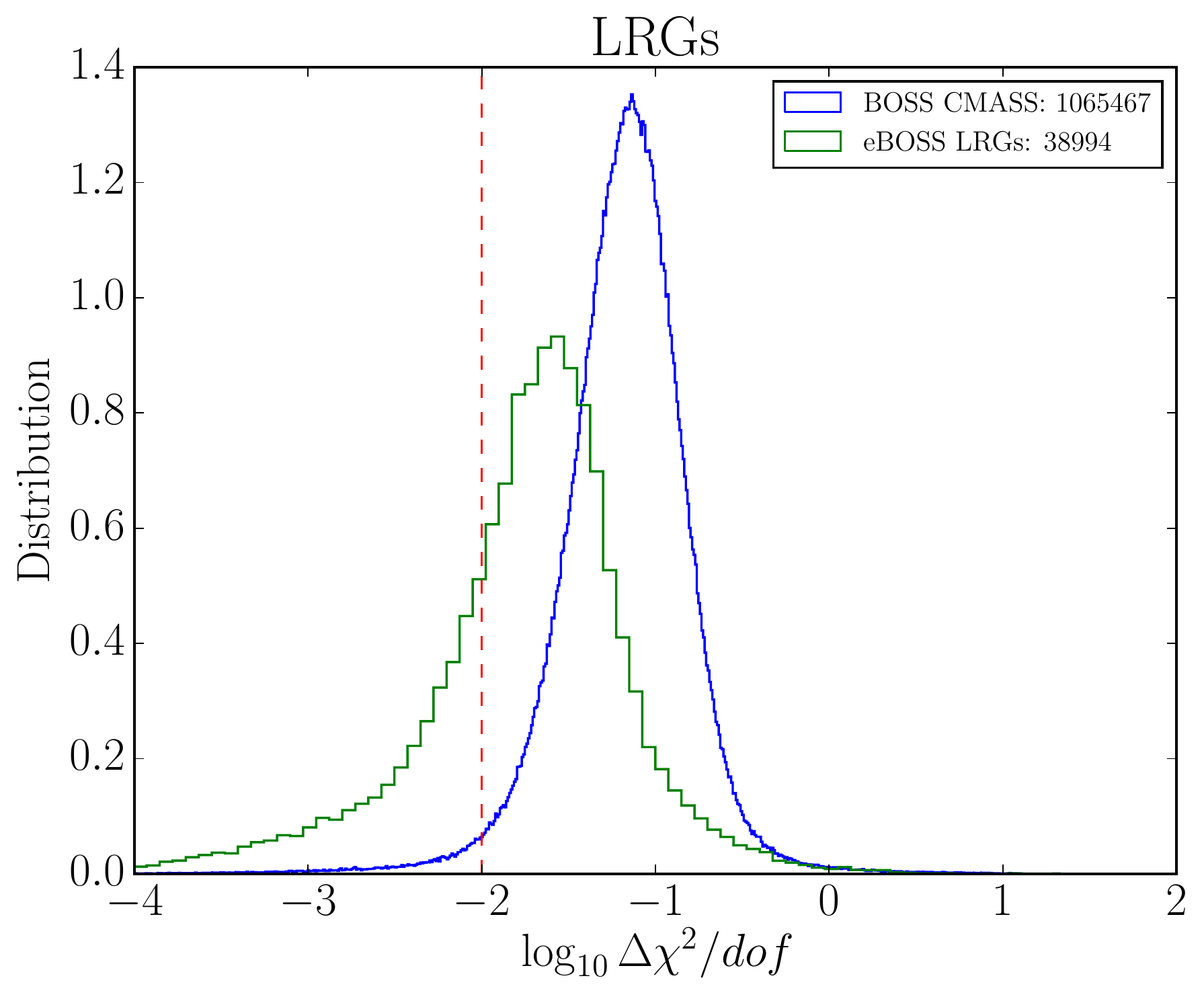}
\includegraphics[width=0.45\textwidth]{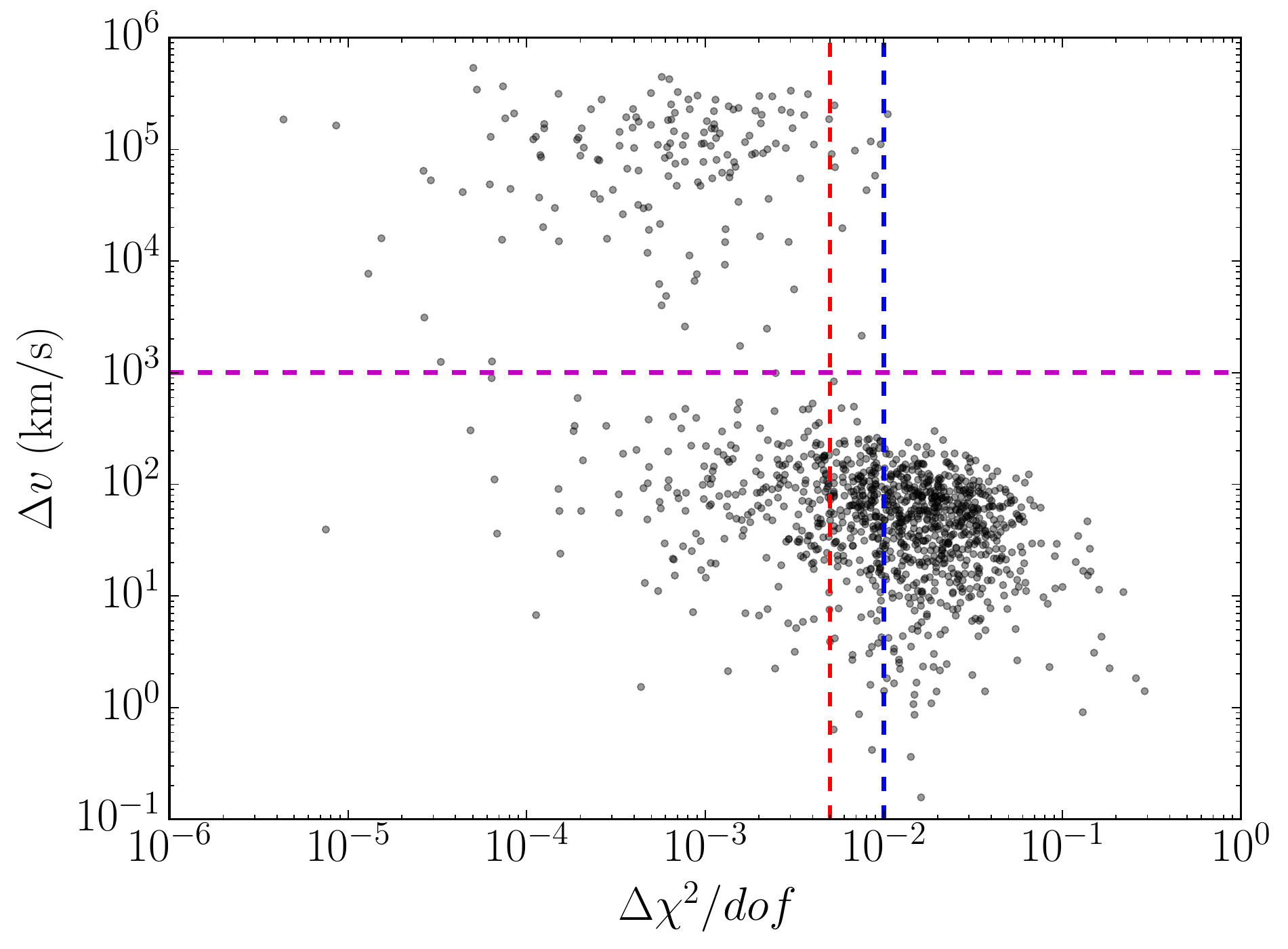}
\caption{
{\bf Left:  }The distribution of $\Delta \chi^2$ per degree of freedom for the BOSS sample of CMASS galaxies
and the eBOSS LRG sample.  The $\chi^2$ per degree of freedom value represents the difference between the best fit spectral template and
the second best spectral template when quasar templates are excluded from the fits.
{\bf Right:  }A scatter plot showing the difference in redshift ($km\,s^{-1}$) between pairs of observations taken of the
same LRG target.  The two vertical lines represent threshold for the current $\Delta \chi^2$ {\tt ZWARNING\_NOQSO}
(blue) and the more lenient threshold that increases the rate of redshifts that are considered reliable (red).
The horizontal line (magenta) represents the limit at which redshift discrepancy is considered a catastrophic failure.
}
\label{fig:zwarning}
\end{figure}

The single largest failure mode for the automated classifications is due to the inability
to discriminate between best fit templates at different redshifts.  Objects that fail
according to this criteria are assigned a warning flag equal to two, which means
the $\chi^2$ difference between the best spectral fit and the second best spectral
fit is less than 0.01 per degree of freedom.
This failure mode is not unexpected; the eBOSS LRG targets are intrinsically fainter than the
BOSS galaxies and thus have a lower S/N measurement of flux at each pixel in the spectrum.
The faint spectra result in less pronounced absorption features that would otherwise provide a good redshift estimate
and a reliance on broad band flux that is harder to fit as a function of redshift.
The overall reduction of the $\chi^2$ difference with respect to the BOSS galaxy sample
is shown in the left panel of Figure~\ref{fig:zwarning}.  The full distribution clearly
shifts toward lower $\Delta \chi^2$, leading to a higher fraction of objects failing to
meet the required difference of 0.01 per degree of freedom.

Of the 7,796 objects that were given a {\tt ZWARNING\_NOQSO} value equal to zero, only 6,096 fall in the redshift range $0.6<z<1.0$.
The corresponding 56.1\% combined targeting and spectroscopic efficiency would fall below our requirement (80\%)
to obtain 300,000 spectroscopically-confirmed LRG tracers.
In particular, the fact that 1700 of the 10,873 spectra are given {\tt ZWARNING\_NOQSO} equal to zero and do not lie in the
redshift range of interest makes it nearly impossible to meet the requirement for 80\% completeness.

We further explored the quality of the automated classification in two ways to see how close the selection
is to the required tracer density.
First, we identified 591 LRG targets that were tiled on more than one plate and produced
multiple spectra.  We supplemented this sample by re-running the data reduction pipeline
on four plates that were observed to more than twice the normal depth, dividing those exposures
into unique coadds of the data.  The data split produced 630 targets with multiple spectral
classifications.  We assessed the rate of catastrophic redshift failures by counting the fraction
of objects for which a pair of observations produced discrepant redshifts.  For the sample of 2442
total observations, we found 166 discrepant redshifts, corresponding to a catastrophic failure
rate of 6.8\% if we assume that consistent redshifts are correct.
We further explored the rate of discrepant redshifts as a function of $\chi^2$ per degree of freedom.
As shown in the right panel of Figure~\ref{fig:zwarning}, the vast majority of discrepant redshifts
occur for {\tt rchisq\_NOQSO} $<0.005$, implying that the threshold of 0.01 may be too strict.
Considering that a spectrum has 4400 pixels over the range $3600<\lambda<10000$ \AA\ before accounting for masked artifacts,
reducing the threshold to 0.005 corresponds roughly to $\Delta \chi^2 = 20$.
Filtering on targets with this new threshold, we find that the rate of {\tt ZWARNING\_NOQSO} failures drops
from 28\% to 16\%.
Of the 1650 observations of objects where both repeated spectra were fit with {\tt rchisq\_NOQSO} $>0.005$, only ten
produced discrepant redshifts, implying a catastrophic failure rate of 0.6\%, below the maximum allowed (1\%).

When reducing the threshold for {\tt rchisq\_NOQSO}, we reduce the rate of {\tt ZWARNING\_NOQSO} failures
by nearly a factor of two and appear to meet the requirements for catastrophic redshift failures.
However, even with this change we are still well below the required rate of
spectroscopic confirmation.  As a second test, we proceeded to visually inspect the LRG spectra from 18 different
plates that obtained the deepest observations.  The visual inspections were intended to provide insight into the
spectral quality, the properties of the templates producing the best and second best fits, and to
gauge the true redshifts of the LRG sample.  

While it is impossible to run visual inspections in a reproducible
and consistent fashion, we attempted to provide a scheme by which we could assess the range of likelihoods that a redshift
is correct.  We established a simple four-tiered redshift confidence metric $z_{\rm conf}$.  

\begin{itemize}

\item
$z_{\rm conf} = 0$ denotes a spectrum for which it is impossible to classify the type or determine a redshift.  Of the
LRG target spectra that were visually inspected, 6.6\% were assigned $z_{\rm conf} = 0$.   
88\% of these spectra were given {\tt ZWARNING\_NOQSO} $>0$ in the automated classification.

\item
$z_{\rm conf} = 1$ denotes a spectrum that is ambiguously classified, where more than one of the best-fit models
is a possible fit. The redshift provided by the inspector is intended to be a best estimate of the correct redshift.
4.5\% of the spectra were determined to be galaxies and assigned $z_{\rm conf} = 1$.
48\% of these spectra were given {\tt ZWARNING\_NOQSO} $>0$ in the automated classification.

\item
$z_{\rm conf} = 2$ denotes a spectrum where the redshift estimate is likely to be correct as there is no
other template that provides an equally compelling fit.  However, some degree of uncertainty remains
because the best-fit spectral template does not reproduce all of the expected features.
For example, the template may fit the continuum but certain absorption features may be ambiguous in the data.
12\% of the spectra were determined to be galaxies and assigned $z_{\rm conf} = 2$.
23\% of these spectra were given {\tt ZWARNING\_NOQSO} $>0$ in the automated classification.

\item
Finally, $z_{\rm conf} = 3$ denotes a case where the redshift is estimated at very high confidence.
68\% of the spectra were determined to be galaxies and assigned $z_{\rm conf} = 3$.
4.8\% of these spectra were given {\tt ZWARNING\_NOQSO} $>0$ in the automated classification.

\item
8.9\% of the objects were classified as stars.  M-stars were the dominant stellar
contaminant and were very easy to identify in visual inspection.  92\% of the stars were given
$z_{\rm conf} = 3$, but 42\% were given {\tt ZWARNING\_NOQSO} $>0$.  The high rate of 
{\tt ZWARNING\_NOQSO} was later found to be caused by a bug in the version of the pipeline
used for DR12.  This will be fixed for the next public data release.

\end{itemize}

Two redshift distributions are presented in Table~\ref{tab:LRGQSOdensities}.
The more conservative estimate (higher rate of ``Poor Spectra'') assumes that any target given
$z_{\rm conf} >1$ is given the correct redshift.  The less conservative estimate
corresponds to all objects with $z_{\rm conf} >0$.  It is likely that the true
distribution lies between these two bounds.

\begin{figure}[h!]
\centering
\vspace{-2cm}
\includegraphics[width=0.45\textwidth, trim=4cm 0 2cm 0]{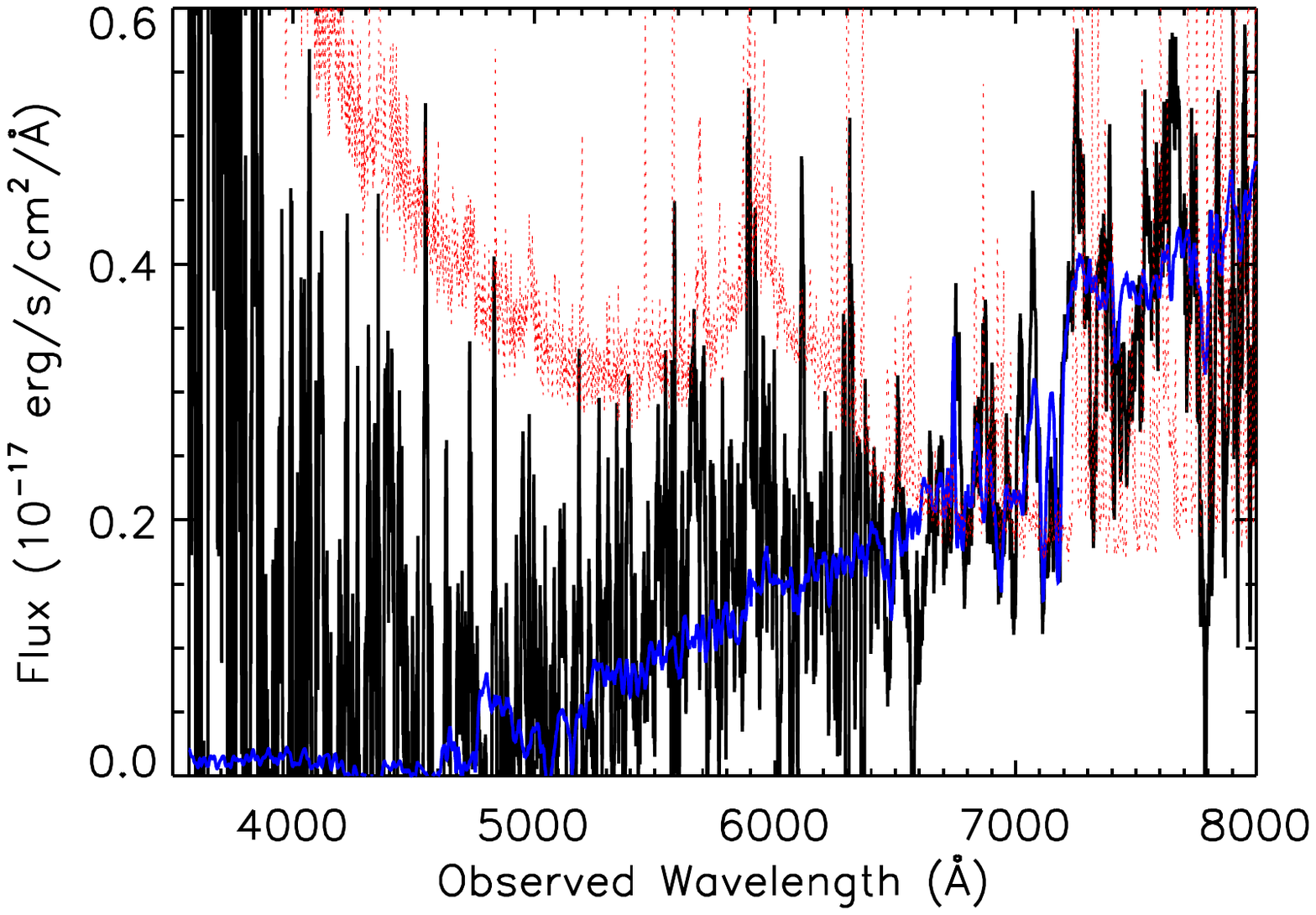}
\includegraphics[width=0.45\textwidth, trim=2cm 0 4cm 0]{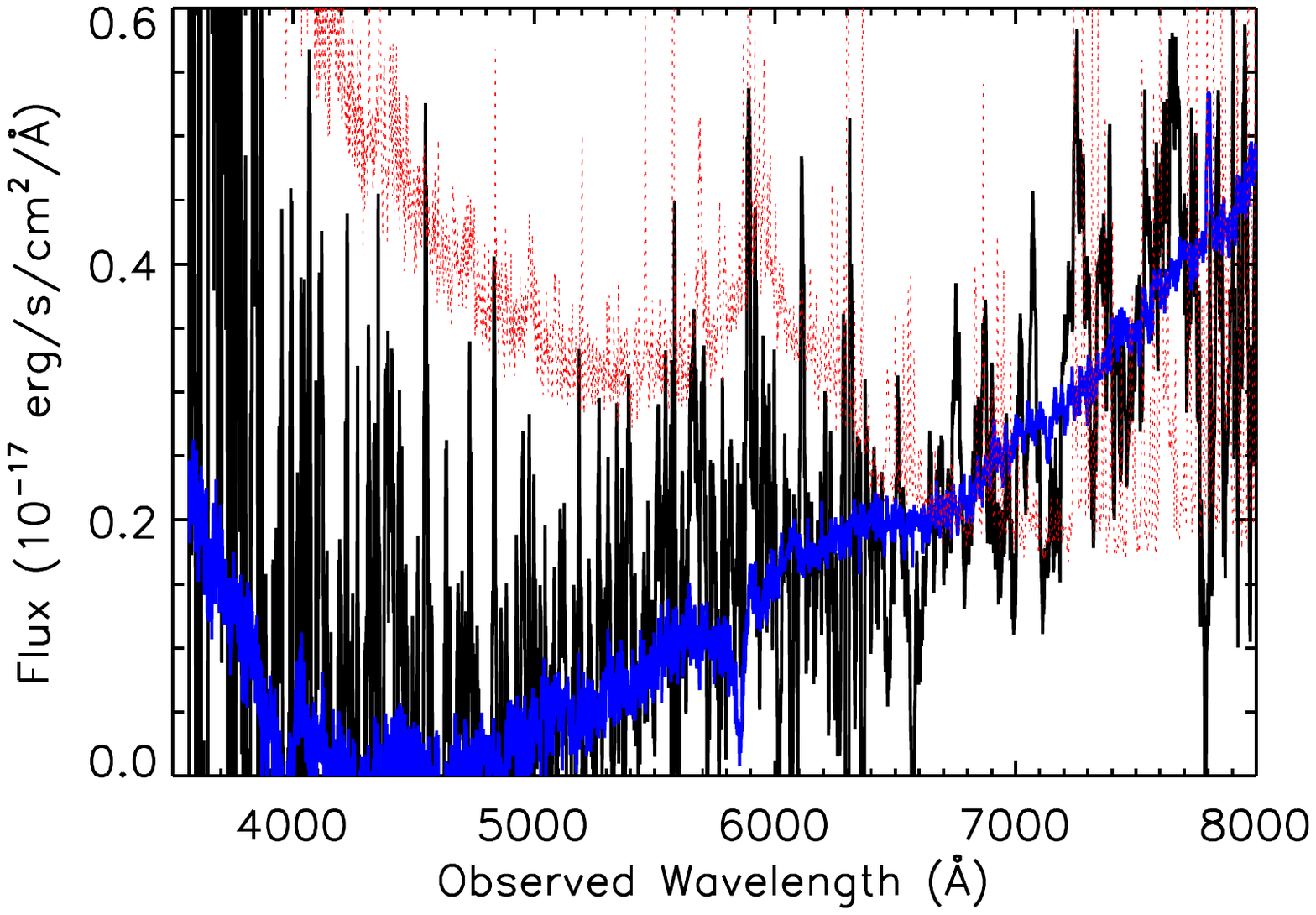}
\vspace{-6.5cm}
\caption{
Example of the influence on classification by non-physical flux at extrema of spectrum.
The data are represented in black, the flux errors on each pixel in red, and the template
in blue.  A boxcar smoothing kernel of width 11 pixels has been applied to the data
for illustrative effect. {\bf Left:  }The template corresponding to the visual inspection redshift
is fit to the data with a $\chi^2$ per degree of freedom of 1.017 at a redshift $z=0.80831$.
{\bf Right:  }An incorrect template is fit to the data with $\chi^2$ per degree of freedom of
1.022 at $z=1.09220$, a difference from the first template small enough to trigger the {\tt ZWARNING\_NOQSO} flag.
}
\label{fig:fluxdistort}
\end{figure}

Visual inspection reveals at least three failure modes among the targets with small
$\Delta \chi^2$.
One failure mode occurs when the spectrum possesses a non-physical gradient in the flux
at the bluest wavelengths or the reddest wavelengths.
An example of one such failure is shown in Figure~\ref{fig:fluxdistort}.
One notes a clear identification of the Ca H\&K features around observed wavelength 7200 \AA\
in the left hand panel.  However, the data deviate significantly from the template over the range
3600--4000 \AA, leading to an inflated $\chi^2$ estimate.
The template in the right hand panel does not capture the narrow absorption features because the
redshift is incorrect.  Instead, the template is better fit
to the excessive UV flux that is likely due to imperfect spectral extraction and not
of astrophysical origins.
In cases such as this, the polynomial term that accounts for uncertainties in flux calibration
appears to be better coupled to incorrect galaxy templates than to a galaxy template
at the correct redshift.

\begin{figure}[h!]
\centering
\vspace{-2cm}
\includegraphics[width=0.45\textwidth, trim=4cm 0 2cm 0]{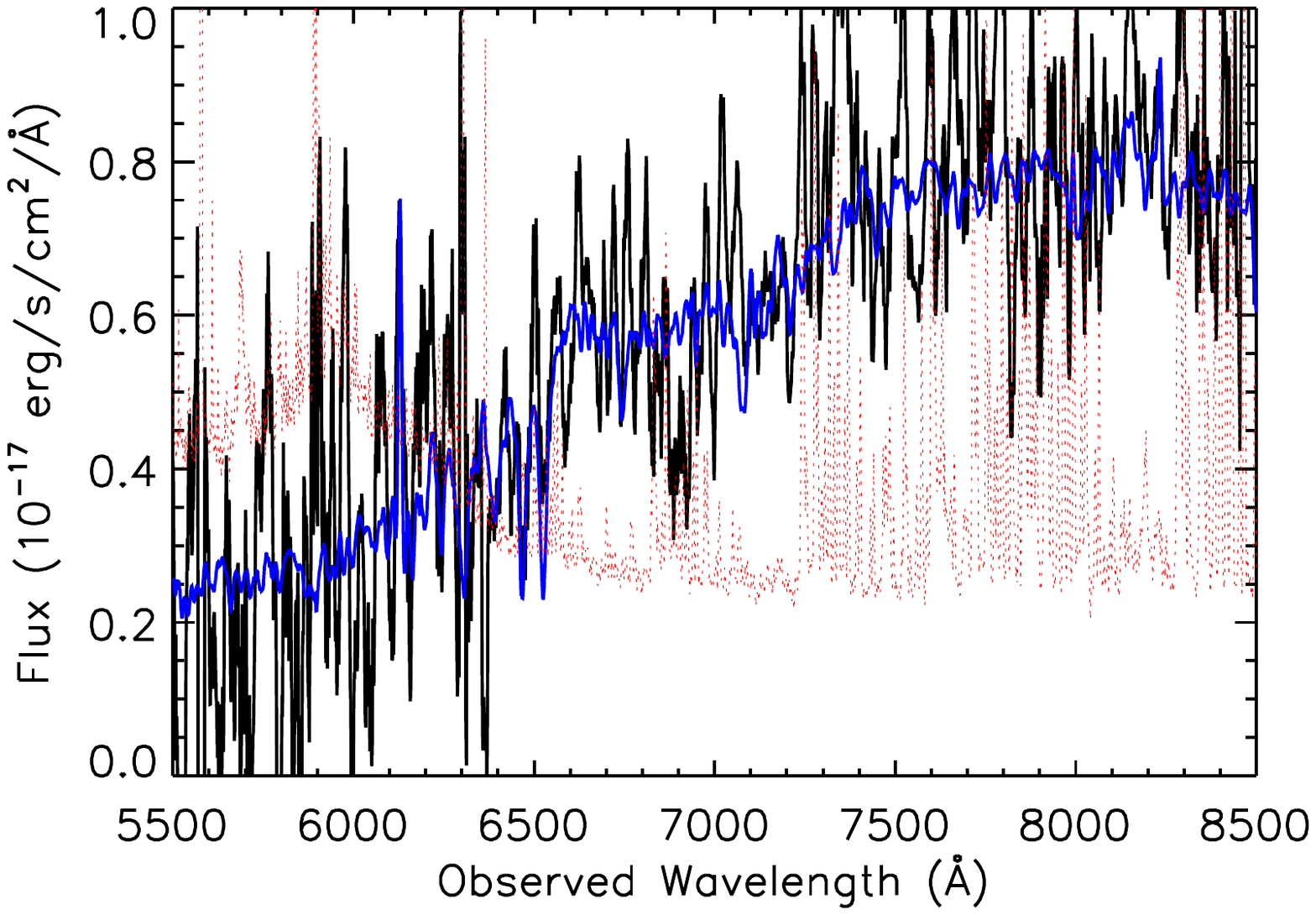}
\includegraphics[width=0.45\textwidth, trim=2cm 0 4cm 0]{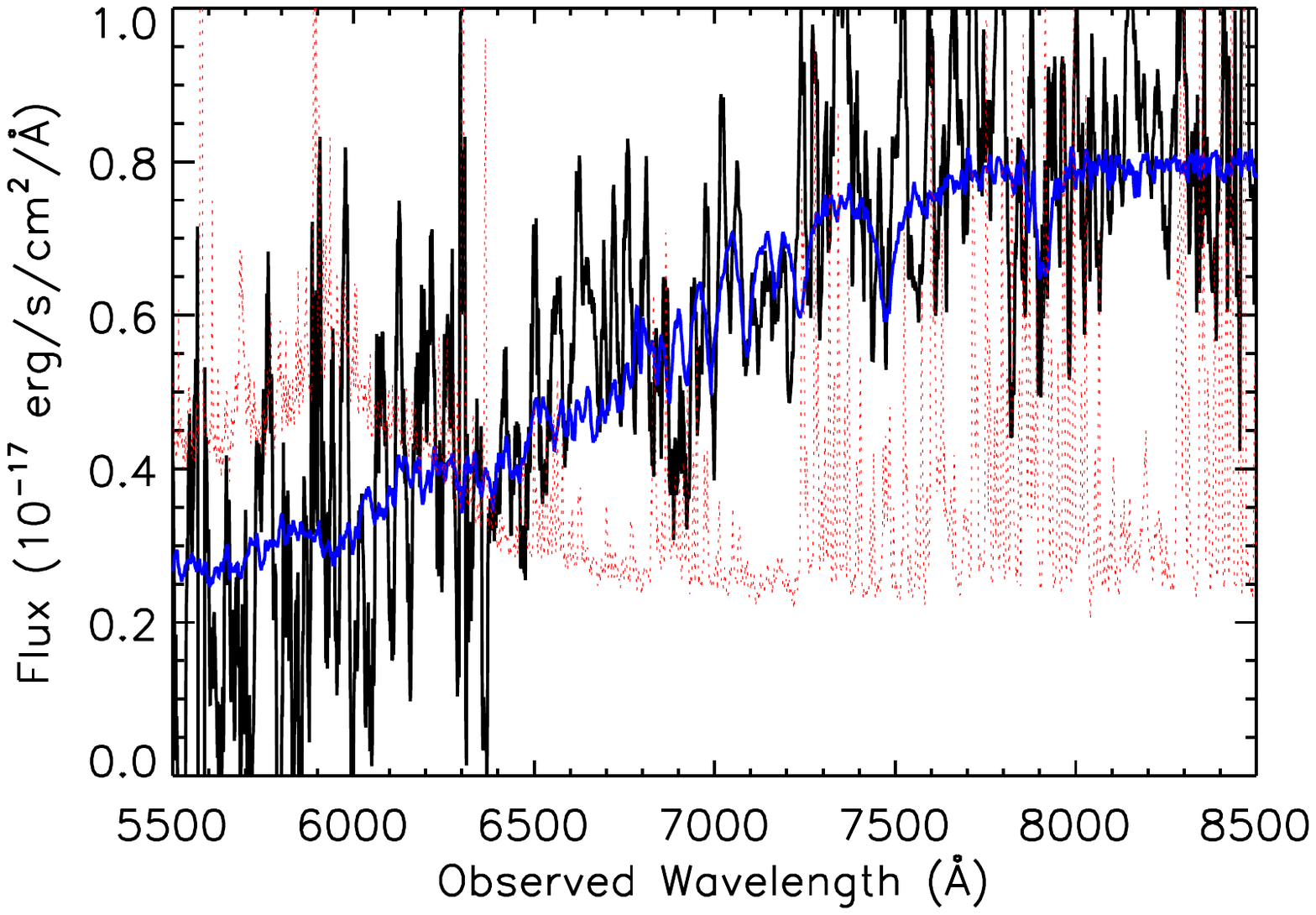}
\vspace{-6.5cm}
\caption{
Example of confusion between broad G-band absorption and the 4000 \AA\ break. 
The data are represented in black, the flux errors on each pixel in red, and the template 
in blue.  A boxcar smoothing kernel of width 11 pixels has been applied to the data 
for illustrative effect. {\bf Left:  }The template corresponding to the visual inspection redshift
is fit to the data with a $\chi^2$ per degree of freedom of 1.258 at a redshift $z=0.64375$.
{\bf Right:  }An incorrect template is fit to the data with $\chi^2$ per degree of freedom of
1.264 at $z=0.82202$, a difference from the first template small enough to trigger the {\tt ZWARNING\_NOQSO} flag.
}
\label{fig:gband}
\end{figure}

A second failure mode is shown in Figure~\ref{fig:gband}.
Because the sources in the LRG target sample are typically passive galaxies
at a high redshift, the significance of the 4000 \AA\ break can be diminished
due to low flux counts.
In these cases, it becomes difficult to discriminate between the
4000 \AA\ break and G-band absorption extending to 4304 \AA.
A spectral template of a higher redshift galaxy with a suppressed G-band feature
(right panel of Figure~\ref{fig:gband}) can sometimes fit the observed spectrum nearly as well as a
template with a stronger G-band feature and a correct fit to the 4000 \AA\ break.
While the template that produces the stronger G-band absorption profile
is more physically motivated than the higher redshift template,
the current reduction pipeline has no mechanism in place to impose a prior
in favor of the correct redshift.

\begin{figure}[h!]
\centering
\vspace{-2cm}
\includegraphics[width=0.45\textwidth, trim=4cm 0 2cm 0]{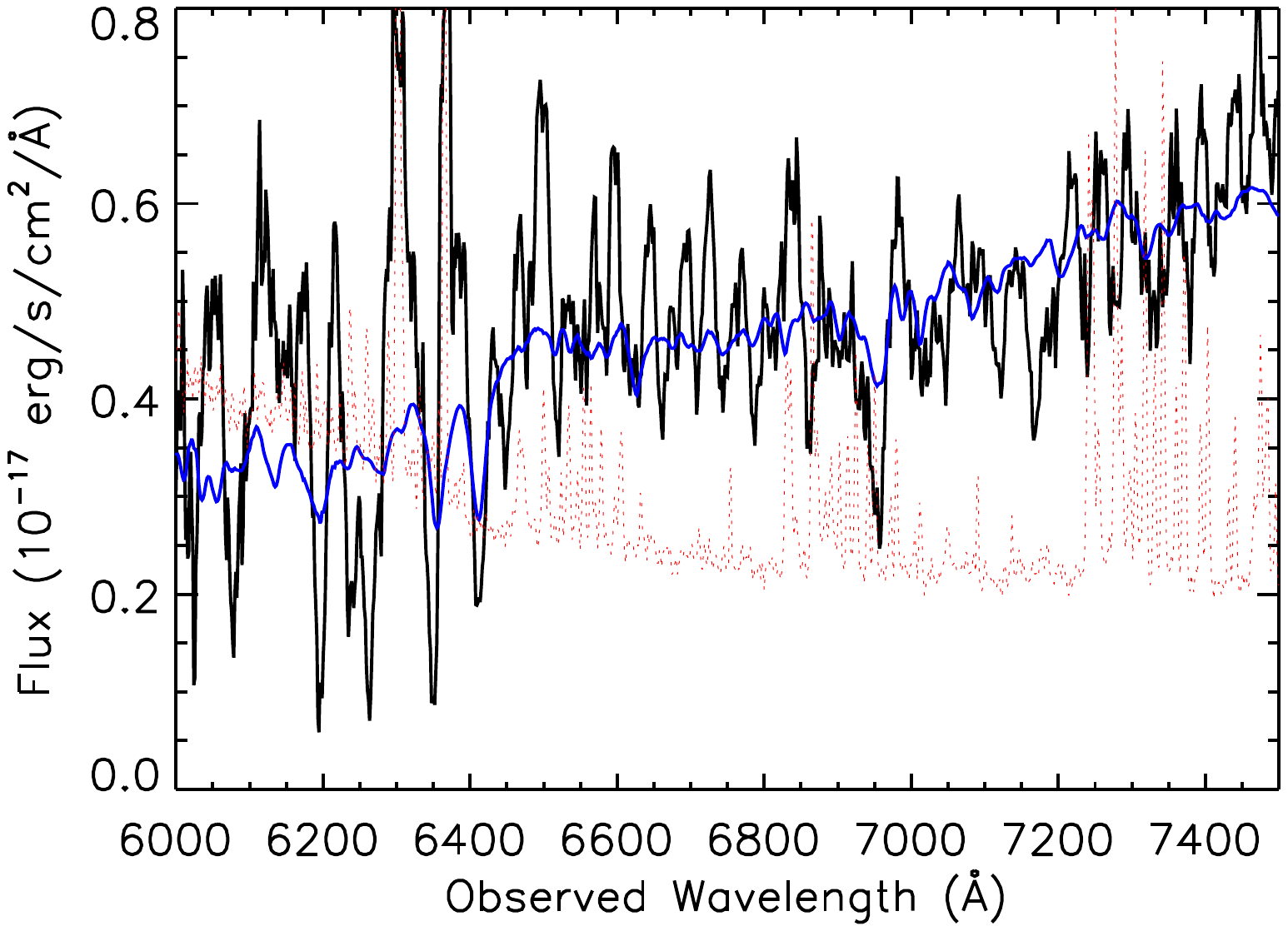}
\includegraphics[width=0.45\textwidth, trim=2cm 0 4cm 0]{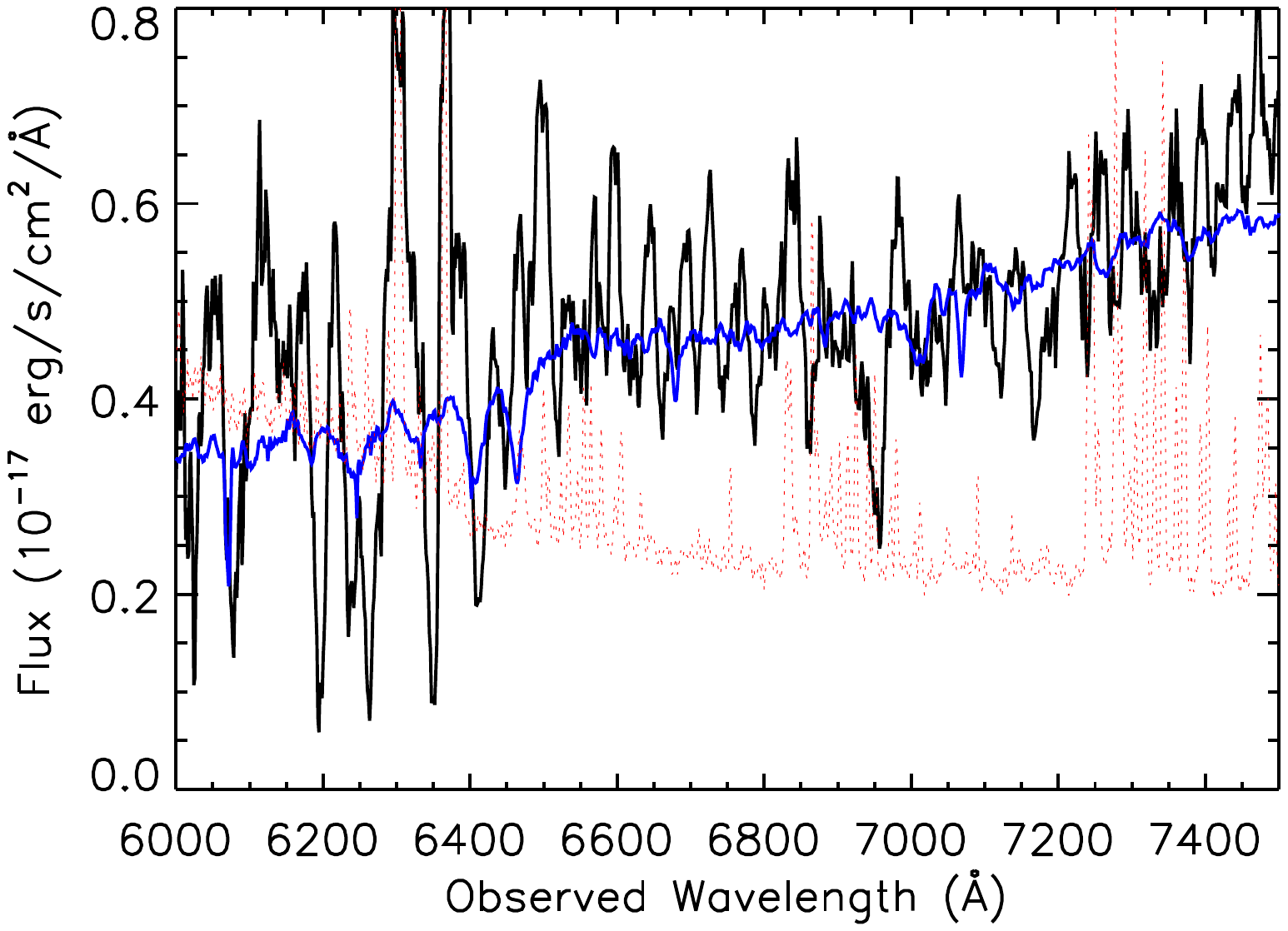}
\vspace{-6.5cm}
\caption{
Example of the spectra in which the templates are unable to differentiate the Ca H\&K features
at high significance.
The data are represented in black, the flux errors on each pixel in red, and the template
in blue.  A boxcar smoothing kernel of width 11 pixels has been applied to the data
for illustrative effect. {\bf Left:  }The template corresponding to the visual inspection redshift
is fit to the data with a $\chi^2$ per degree of freedom of 0.911 at a redshift $z=0.61516$.
{\bf Right:  }An incorrect template is fit to the data with $\chi^2$ per degree of freedom of 
0.915 at $z=0.62806$, a difference from the first template small enough to trigger the {\tt ZWARNING\_NOQSO} flag.
}
\label{fig:CaHK}
\end{figure}

Finally, the third failure mode occurs because the majority of the signal is found in the broad band
flux as opposed to the monochromatic Ca H\&K features.  In cases such as the spectrum shown in
Figure~\ref{fig:CaHK}, there is little power to statistically discriminate between a fit that
correctly places both Ca H\&K features and a fit that staggers the template with an incorrect fit to a single
feature.  In many cases, the G-band absorption line (4304 \AA), Magnesium line (5175 \AA), and Sodium
line (5894 \AA) provide additional constraints on the redshift, but these lines become ambiguous at
higher redshifts where sky subtraction residuals are more significant. 

As will be discussed in Section~\ref{subsec:pipeline}, 
we are now improving the algorithms for spectral extraction and classification
to mitigate these classes of failure modes.  We are confident that we can
increase the rate of reliable classifications beyond what is presented in DR12 and are optimistic
that we can achieve statistics comparable to the $z_{\rm conf} > 0$ redshift identification rate
from the visual inspections.

\subsection{Quasar Spectra from SEQUELS}\label{subsec:QSOspectra}

Quasar targets in BOSS were observed at a rate of roughly 200 per plate, totaling
nearly 500,000 over the full program.
Each of these spectra was visually inspected and given a classification (quasar, star, or galaxy)
and a redshift estimate with a documented degree of confidence \citep{paris12a,paris14a}.
We also relied on visual inspection to flag quasar sightlines with peculiarities
such as damped lyman-alpha and BAL systems that could not be classified with the automated redshift 
classification \citep{bolton12a}.
The photometric and spectroscopic parameters of each object were recorded for use
by all members of the collaboration and released in public form on a regular basis.
The process was undertaken primarily by two members of the collaboration
and proved to be very time-consuming.

The quasar redshift distributions presented in Section~\ref{subsec:targets}
were derived from visual inspections of all SEQUELS plates observed during BOSS.
Even the faintest quasars ($r_{\rm PSF} = 22$) can be confidently
classified in visual inspections; 95.7\% of the full quasar sample (including TDSS, SPIDERS,
and PTF-selected objects) can be identified with high confidence.
A comparison of the automated pipeline to the results of visual inspection reveal a high
level of agreement.
The redshift estimates from the spectroscopic pipeline are consistent with the visual
inspection estimates to within $1000\,$kms$^{-1}$ for 98.7\% of {\tt QSO\_CORE} quasars
($r_{\rm PSF} < 22$).
At this early stage, the spectroscopic pipeline nearly meets the requirement
presented in Section~\ref{subsec:uniformity} that $<1\%$ of objects are given catastrophically
incorrect redshifts.

The density in eBOSS and SEQUELS increases to roughly 600 quasar targets per plate 
which motivated the collaboration to revisit the automated classification schemes and
significantly decrease the level of visual inspection.
To fully transition to this new scheme, we evaluated the pipeline
performance against visual inspections to determine
the source of catastrophic failures and systematic uncertainties in redshift estimates.
We found patterns in catastrophic failures associated with the class of object reported by the 
pipeline, the redshift estimate, and the second, third, fourth and fifth best estimates.
By applying the following criteria, we are able to improve the consistency between
the pipeline classifications and the visual inspections.
\begin{itemize}
\item Objects identified as a star in the best estimate are assumed to be a star.
\item Objects identified as a galaxy with $z<1$ in the best estimate are assumed
to be a galaxy at $z<1$.
\item Objects identified as a galaxy with $z>1$ in the best estimate are assumed
to be a galaxy if one of the next four classifications is also a galaxy.
\item Objects identified as a quasar with ZWARNING$=0$ are assumed to be a quasar
unless two or more alternate identifications are given a class of star. In these
cases, the object is assumed to be a star.
\item Objects identified as a quasar with ZWARNING$>0$ are assumed to be a star
if two or more alternate identifications are given a class of star.
\item Objects that fail to meet these criteria are followed up with visual inspection.
\end{itemize}

The results of this classification scheme applied to SEQUELS data is presented in
Table~\ref{tab:visualinspections}.
This new classification provides identifications for 93\% of \textit{all} the quasar targets. 
Using the visual inspections as a baseline, among those automatically classified targets, 96.9\% are correctly characterized
by our new algorithm. It is specifically designed to pick up actual quasars and to limit the
fraction of contaminants in the quasar sample and lost quasars: 54.7\% are actual quasars that 
are confirmed after visual inspection. 
A total of 0.42\% quasar targets are wrongly classified as quasars by our automated
scheme. Those contaminants are largely dominated by spectra with very low signal-to-noise ratios,
and could also not be identified after visual inspection.
Our procedure misidentifies 0.40\% of quasar targets that are confirmed to be quasars after visual inspection.
Among the 7\% of all quasar targets that still require visual inspection, 70.3\% are true quasars. Assuming
that the identification error from the visual inspection is negligible, the total fraction of contamination of the
quasar sample and loss is below the 1\% of the overall sample as required in Section~\ref{subsec:uniformity}.\\

\begin{table}[htp]
\centering
\caption{
\label{tab:visualinspections}
Performance of the automated classification scheme applied to the SEQUELS quasar sample.
Percentages are computed with respect to the total number of automatically identified objects for the columns STAR, QSO and GALAXY.
For objects without an automated classification (UNKNOWN), percentages are calculated with respect to the total number of UNKNOWN
objects. Quasar targets classified as UNKNOWN by the automated classification are visually inspected.
}
\begin{tabular}{l c c c | c }
\hline\hline
                                   & \multicolumn{4}{c}{Automatic classification} \\
                                   & STAR             &  QSO                & GALAXY            & UNKNOWN        \\
\hline
Visual inspection result           &                  &                     &                   &                \\
?/BAD                              & 78 (0.2\%)       & 102 (0.3\%)         & 427 (1.3\%)       & 486 (19.9\%)    \\
STAR                               & 9,845 (29.2\%)   & 19 (0.05\%)         & 159 (0.5\%)       & 189 (7.7\%)     \\
QSO                                & 37 (0.1\%)       & 18,475 (54.7\%)     & 103 (0.3\%)       & 1,718 (70.3\%)   \\
GALAXY                             & 13 (0.04\%)      & 24 (0.07\%)         & 4,390 (13.0\%)    & 50 (2.1\%)      \\
\hline
\end{tabular}
\end{table}

So we can expect that eBOSS quasar redshifts will be robust, but an additional requirement noted in Section~\ref{subsec:uniformity}
is that they should also have a precision of order 300--600 kms$^{-1}$.
Based on repeat spectroscopy, the RMS scatter of BOSS pipeline redshift estimates for quasars is a factor of 2 higher than the
reported statistical errors \citep{bolton12a}.
Taking into account these underestimated errors, the typical statistical precision is between $\sigma_v \sim 100$ and $\sigma_v \sim 300$ 
km s$^{-1}$
in the redshift range $1 < z < 2$ , depending on the luminosity of the source.
If truly representative of the redshift errors, this level of precision would be sufficient to avoid degradation in the clustering signal.

The reported statistical precision does not necessarily reflect the true accuracy on a quasar redshift due to possible systematic uncertainties in the 
redshifting templates \citep[e.g.][]{hewett10a}.
Highly ionized gas in the broad-line region of quasars is subject to radiation-driven winds \citep[e.g.][]{proga03a}.
It is therefore likely that the measured redshifts largely determined by these emission lines are offset from the systemic redshift.  
Such an effect has been reported in the redshift estimates of $z>2.1$ quasars used in \lya\ forest -- quasar cross correlations
\citep{font-ribera13a}.
Because MgII is ionized at lower energies and is largely present at larger separations from the central black hole, MgII emission
lines are likely the 
least affected broad emission lines by systematic offsets.
Various studies have estimated that the MgII emission line is offset by an amount varying from -100 to +50 kms$^{-1}$ with respect to the 
quasar systemic redshift \citep[e.g.][]{mcintosh99a,hewett10a,font-ribera13a}.
The upper bound on the statistical errors of quasar redshift estimates can be therefore estimated directly from fits to the MgII emission 
lines. 

\begin{figure}[h!]
\centering
% \vspace{-2cm}
\includegraphics[width=0.49\textwidth,trim=1in 3.5in 1in 1in, clip]{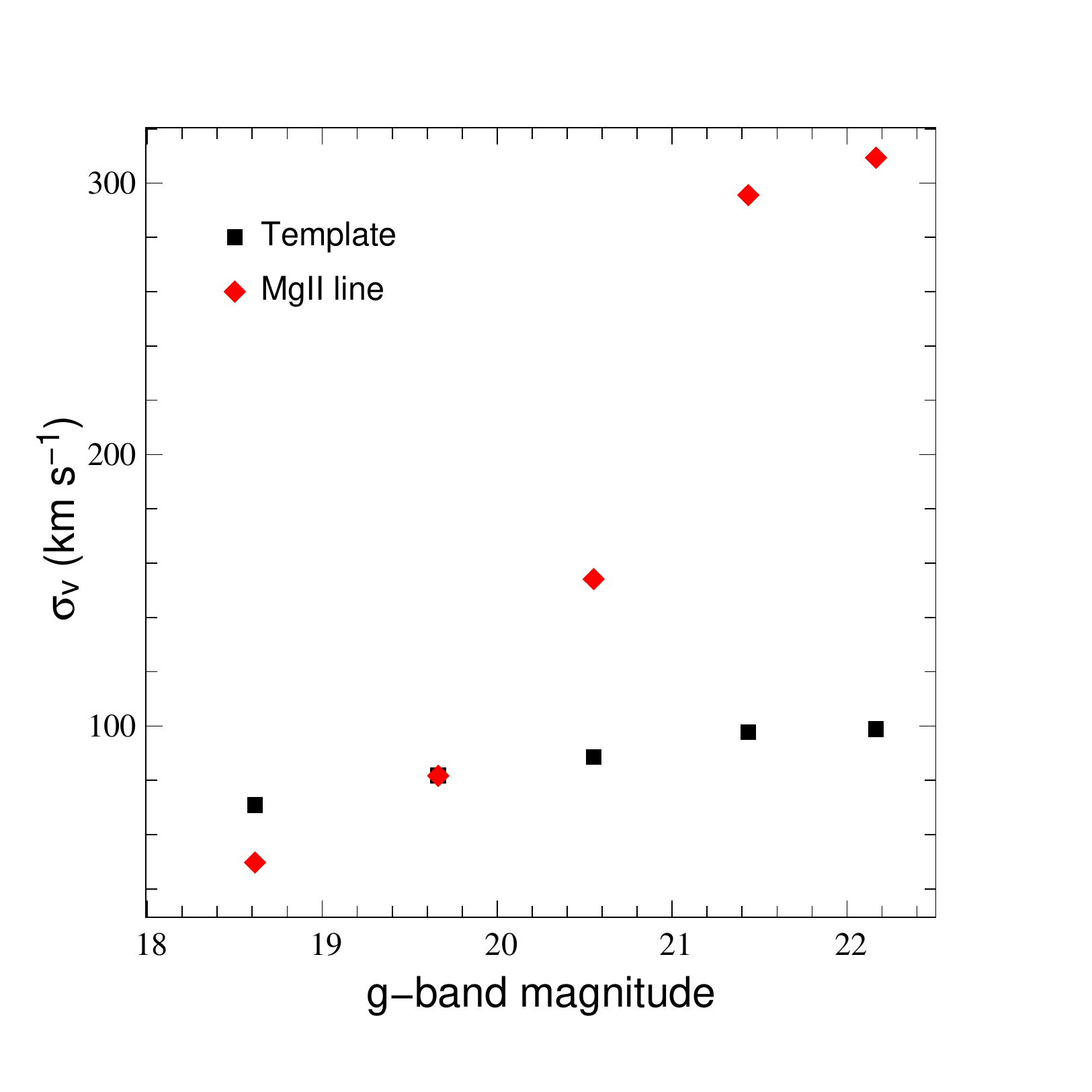}
\includegraphics[width=0.49\textwidth,trim=1in 3.5in 1in 1in, clip]{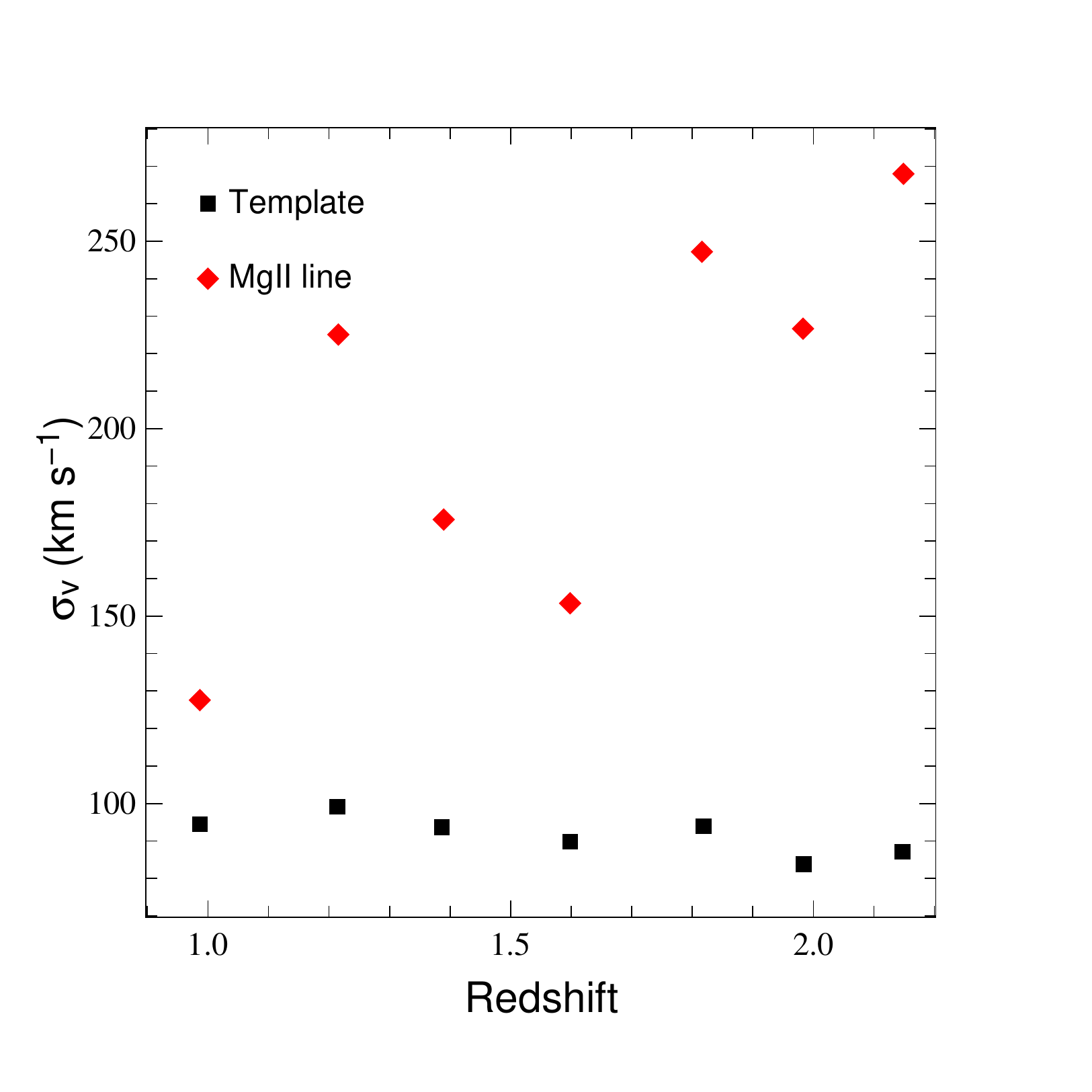}
% \vspace{-6.5cm}
\caption{
The statistical error on redshift estimates for the quasar sample for template-based (black squares) and MgII-based (red diamonds)
redshift estimates.
{\bf Left:  }The redshift error as a function of $g$-band magnitude.
{\bf Right:  }The redshift error as a function of redshift.
}
\label{fig:qso_errs}
\end{figure}

In the redshift range of interest for eBOSS quasar clustering, the MgII emission line is always covered in the spectra.
Hence, it can be directly used to estimate quasar redshifts. 
In order to estimate the statistical error on MgII-based redshifts, we use the spectra of 472 quasars in the redshift
range $0.9 < z < 2.2$ taken as part
of the SDSS Reverberation Mapping Project \citep{shen15a}. 30 spectroscopic epochs were taken in SDSS-III for each of those quasars with
a homogeneous S/N distribution corresponding to twice the normal BOSS and eBOSS exposure depth.
The redshift and $g$-band magnitude distributions are comparable with the eBOSS sample.
By fitting the MgII line with a linear combination of five principal components, its redshift is measured at the position of the
maximum flux of the emission line as described in \cite{paris12a}.
We estimate the RMS scatter of the MgII-based redshifts as a function of $g$-band magnitude and quasar redshift in
Figure~\ref{fig:qso_errs} (red diamonds).
The redshift errors increase with $g$-band magnitude and vary from $\sim$80 to $\sim$300 kms$^{-1}$.
This behavior is expected since the S/N decreases for fainter objects.
Redshift errors also increase from $\sim$130 kms$^{-1}$ at z~=~1 to $\sim$270 kms$^{-1}$ at z~=~2.2.
Larger redshift errors for quasars at $z \geq 2$ are
measured because the MgII line lies in the red part of eBOSS spectra where the sky subtraction is noisy and/or imperfect and
makes the line fitting more difficult.
Redshift errors are also larger at $z \sim 1.2$ when the MgII emission line moves
from the blue arm to the red arm of the eBOSS spectrograph.\\

\begin{figure}[h!]
\centering
\includegraphics[width=0.75\textwidth,trim=1in 3.5in 1in 1in, clip]{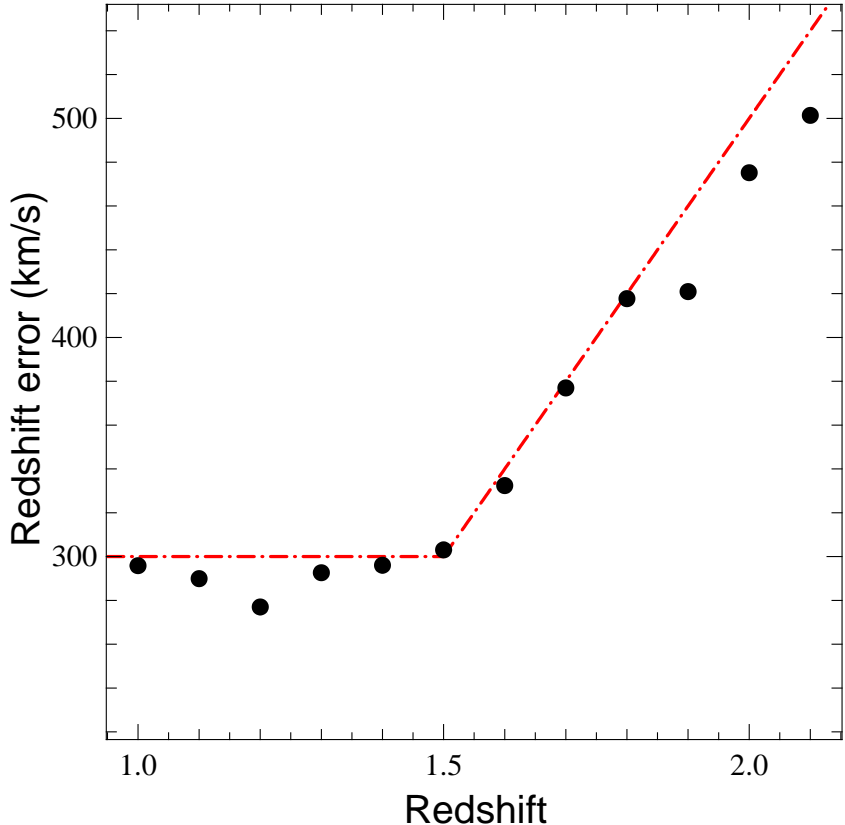}
\caption{
Redshift evolution of the statistical error on systemic redshift estimates for the
quasar sample based on the comparison of template-based and MgII-based redshit estimates. Black points are
the statistical errors derived from the data. The red line shows the redshift evolution used for the
cosmological forecasts described in Section~\ref{subsec:uniformity}.
}
\label{fig:qso_err_template}
\end{figure}

The statistical errors on redshift measurements based on MgII emission line fits are well within the requirements
described in Section~\ref{subsec:uniformity}.
However, this emission line is not always detected, especially in the spectrum of faint quasars at $z \geq 2$
when the MgII line is at its red end.
We thus consider redshifts estimated with quasar templates in order to overcome this issue. These templates have been
calibrated with the MgII emission line \citep[see details in ][]{paris12a}.
We estimate the statistical error on those redshift measurements in the same manner as for the MgII-based estimates.
As expected, template-based redshifts are more stable (black squares in Figure~\ref{fig:qso_errs}).
The RMS scatter increases from $\sim$60 to $\sim$100 kms$^{-1}$ for $18 < g < 22$ (left panel).
There is no obvious redshift dependence of the statistical error with an average of 90 kms$^{-1}$ (right panel).
Despite this apparent better behavior of statistical errors for template-based redshifts,
one significant drawback comes from astrophysical effects.
Intrinsic emission line shifts depend on quasar luminosity \citep[e.g. ][]{hewett10a}.
Offsets can be accurately reproduced with templates if the whole luminosity range of interest
is covered by the training sample, which is not the case of our current templates.
The most affected emission line is CIV$\lambda$1550 that is blueshifted by
several hundreds of kms$^{-1}$ with respect to the systemic redshift \citep[e.g.][]{richards11a}.
When this line enters in the spectrum, it may affect the accuracy of template-based redshift measurements and thus,
it may cause an additional source of systematics that
is redshift dependent. In order to quantify this effect, we measure MgII- and template-based redshifts
of quasars in the SEQUELS sample. We then measure the redshift evolution
of the scatter between MgII-based and template-based redshifts (Figure~\ref{fig:qso_err_template}).
Between z~=~0.9 and z~=~1.5, redshift errors are below 300 kms$^{-1}$.
At larger redshift, the CIV line is covered by eBOSS spectra and redshift errors increase up to 600 kms$^{-1}$.
This demonstrates that template-based redshifts are dominated by astrophysical effects.
Templates need to be improved to reduce these systematics.
Nevertheless, redshift requirements (and high level BAO projections)
noted in Section~\ref{subsec:uniformity} are set to our current redshift accuracy (red line in Figure~\ref{fig:qso_err_template}).

\subsection{\lya\ Forest Spectra from SEQUELS}\label{subsec:lyaspectra}

The \lya\ forest sample differs from the LRG, ELG, and clustering quasar samples in that it will be used to
increase the density of a BOSS sample that is already being used to measure the BAO distance scale.
The analysis tools and spectral classification are well advanced for this sample of targets and are described
in recent results.  For example, three techniques for determining the underlying quasar continuum in the \lya\ forest region
were compared in the latest BAO measurement \citep{delubac15a}.  \citet{blomqvist15a} explore a
model in Fourier space to correct broadband distortion introduced by continuum fitting.
Implementing the technique on simulated spectra, they find that 
the linear bias parameter $b_{F}$ and the redshift-space distortion parameter $\beta_{F}$
can be recovered with systematic errors less than 0.5\%.
Astrophysical effects (quasar continuum diversity and high-density absorbers) and 
instrumental effects (noise, spectral resolution, and data reduction artifacts)
that affect estimates of the \lya\ forest auto-correlation function were quantified through catalogs of
mock spectra \citep{bautista15a}.
Analysis in SDSS-IV will be performed jointly on the BOSS and eBOSS \lya\ forest samples;
modeling of the astrophysical and instrumental contributions to the measured clustering signal
will continue to evolve.

The sample of $z>2.1$ quasars also provided the first opportunity in SDSS to measure BAO through
cross-correlation of different tracers.  The target sample produces both direct tracers of
the underlying density field (quasars themselves) and illumination of neutral hydrogen through
the \lya\ forest.  The large-scale cross-correlation of quasars with the \lya\ forest absorption
was first measured to comoving separations of 80 $h^{-1}$Mpc \citep{font-ribera13a}.
The technique was later scaled to larger separations with a larger sample of quasars
to measure BAO and provide constraints on $H(z)$ and $d_A(z)$ as discussed in Section~\ref{subsec:bao}.

\begin{table}[b]
\caption{
Densities of quasars and lines of sight to \lya\ forest in BOSS and the various eBOSS samples.
The density of quasars ($n_Q$) includes BAL quasars at redshifts $z>2.0$.
The density of lines of sight ($n_F$) include only quasars with $z>2.1$
that are not BAL quasars.
BOSS numbers include the true non-BAL population while
the eBOSS numbers include a 13\% reduction in the observed
number density to account for the predicted rate of BAL contamination.}
\label{tab:lya}
\begin{center}
\begin{tabular}{l|c|c}
\hline
\hline
Description & Symbol & Value \\
\hline
Total area of BOSS survey & $A_{BOSS}$ & 10 000 deg$^2$ \\
BOSS area covered after 2 years of eBOSS & $A_{eBOSS1}$ & 3 000 deg$^2$ \\
BOSS area covered after 4 years of eBOSS & $A_{eBOSS2}$ & 4 500 deg$^2$ \\
BOSS area covered after 6 years of eBOSS & $A_{eBOSS3}$ & 7 500 deg$^2$ \\
\hline
Density of quasars in BOSS & $n^Q_{BOSS}$ & 19.7 deg$^{-2}$ \\
Density of new quasars in eBOSS from CORE selection & $n^Q_{CORE}$ & 10.1 deg$^{-2}$ \\
Density of new quasars in eBOSS from PTF selection & $n^Q_{PTF}$ & 2.8 deg$^{-2}$ \\
\hline
Density of \lya\ lines of sight in BOSS & $n^F_{BOSS}$ & 16.7 deg$^{-2}$ \\
Density of \lya\ lines of sight in BOSS not re-observed in eBOSS & $n^F_{bright}$ & 8.7 deg$^{-2}$ \\
Density of \lya\ lines of sight in BOSS re-observed in eBOSS & $n^F_{faint}$ & 8.3 deg$^{-2}$ \\
Density of new \lya\ lines of sight in eBOSS from CORE selection & $n^F_{CORE}$ & 5.5 deg$^{-2}$ \\
Density of new \lya\ lines of sight in eBOSS from PTF selection & $n^F_{PTF}$ & 2.3 deg$^{-2}$ \\
\hline
Mean value of a \lya\ quasar in BOSS & $\langle \nu_{BOSS} \rangle $ & 0.60 \\
Mean value of a \lya\ quasar in BOSS not re-observed in eBOSS & $\langle \nu_{bright} \rangle $ & 0.85 \\
Mean value of a \lya\ quasar in BOSS before re-observed in eBOSS & $\langle \nu_{before} \rangle $ & 0.35 \\
Mean value of a \lya\ quasar after re-observed in eBOSS & $\langle \nu_{after} \rangle $ & 0.55 \\
Mean value of a new \lya\ quasar in eBOSS from CORE selection & $\langle \nu_{new} \rangle $ & 0.39 \\
Mean value of a new \lya\ quasar in eBOSS from PTF selection & $\langle \nu_{new} \rangle $ & 0.50 \\
\hline
Effective density of lines of sight in BOSS & $n_{\rm eff}^{BOSS}$ & 10.0 \\
Effective density of lines of sight in eBOSS (+BOSS) & $n_{\rm eff}^{eBOSS}$ & 15.3 \\
\hline
\end{tabular}
\end{center}
\end{table}

The measurement of BAO with \lya\ forest in auto-correlation and in cross-correlation with quasars
are both shot-noise limited and produce nearly independent constraints on the distance scale.
The final BOSS DR12 sample is expected to produce combined constraints on $H(z)$ and $d_A(z)$
with a precision of 2.0\% and 2.5\%, respectively.
One can project the relative improvement offered by the new eBOSS $z>2.1$ quasar
sample by evaluating the number density and depth of spectra compared to those in BOSS.
The observed number density of quasars can be computed from the redshift distributions 
found in Table~\ref{tab:LRGQSOdensities} and the tiling efficiencies 
in Table~\ref{tab:completeness}.
The density of quasars from each sample after accounting for tiling efficiency is shown in
Table~\ref{tab:lya}.
SEQUELS observations are used to assess the depth of spectra.

We follow the formalism presented in \citet{mcquinn11a} to estimate the fractional improvement
on the BOSS BAO distance precision from the auto-correlation when adding eBOSS \lya\ forest spectra.
The BAO uncertainties are forecast based on the number of modes available given the survey volume
and the signal-to-noise per mode (S/N).
In the case of \lya\ forest spectra, the S/N scales as
\begin{equation}
(S/N)^2 = \frac{P^2_F (k)}{2 P^2_{tot} (k)}
\end{equation}
where $P_F (k)$ is the flux power spectrum. $P_{tot} (k)$ is the variance of a given mode approximated by the combination
\begin{equation}
P_{tot}(k) = P_F(k) + \frac{P_{1D}(k_\parallel)}{n_{\rm eff}}~,
\end{equation}
where $P_{1D}(k_\parallel) = P_{1D}$ is the 1D power along the line of sight.
The effective angular density of lines of sight, $n_{\rm eff}$, is defined
\begin{equation}\label{eq:definenu}
n_{\rm eff} = n^F < \nu > = n < P_{1D} / ( P_{1D} + P_N^F ) >.
\end{equation}
The average surface density of quasars used to measure absorption in the \lya\ forest is represented by
$n^F$ and $P_N$ is a weighted average of the noise power ($P_N = dx / (SNR)^2$).
The mean pixel width within the forest is recorded as $dx$ (units of Mpc/h) and SNR is the mean signal-to-noise
ratio per pixel within the forest. 
In a survey that is dominated by shot noise, the S/N per Fourier mode should be roughly proportional to $n_{\rm eff}$
at a given redshift.  Although the calculations are redshift dependent, we assume that the redshift distributions in eBOSS are similar
to BOSS so the uncertainties of eBOSS relative to BOSS should scale as the inverse ratio of $n_{\rm eff}$ in the areas of overlap. 

In studying the precision of BAO measurements in the \lya\ auto-correlation with the addition of eBOSS,
we simply determine $n_{\rm eff}^{eBOSS}$ in the area ($A_{eBOSS}$) covered by BOSS and eBOSS
and $n_{\rm eff}^{BOSS}$ in the area ($A_{BOSS}$) covered uniquely by BOSS.
Since the two regions will be independent, we can compute the uncertainty on the BAO in eBOSS relative to BOSS according to
\begin{equation}\label{eq:ebosserr}
 \frac{\sigma^2_{eBOSS}}{\sigma^2_{BOSS}} = \frac{A_{BOSS}~(n_{\rm eff}^{BOSS})^2}
     {A_{eBOSS}~(n_{\rm eff}^{eBOSS})^2+(A_{BOSS}-A_{eBOSS})~(n_{\rm eff}^{BOSS})^2} 
  = \frac{1}{f_A~f_n^2 + (1-f_A)} ~,
\end{equation}
where we have defined $f_A = A_{eBOSS} / A_{BOSS}$ and
$f_n = n_{\rm eff}^{eBOSS} / n_{\rm eff}^{BOSS}$.

The depth of these SEQUELS measurements for each target sample was used to compute the average effective value of each \lya\ forest
line of sight according to Equation~\ref{eq:definenu}.
Using the observed densities $n^F$ found in Table~\ref{tab:lya}, we find $n_{\rm eff}^{BOSS} = 10.0$,
$n_{\rm eff}^{eBOSS} = 15.3$, and $f_n = 1.53$.  
The area covered in eBOSS will progress according to the survey strategy presented in Section~\ref{subsec:strategy}.
$f_A$ will be equal to 0.3, 0.45, and 0.75 after the first, second, and third public data releases, respectively.

In the case of the cross-correlation, the S/N per Fourier mode should be roughly proportional to the sampling of the density field.
Since both the number density of quasars and the measurements of the \lya\ forest contribute,
the term $n_{\rm eff}$ from the auto-correlation gets replaced by what is effectively the geometric mean
between the effective number density of \lya\ forest lines of sight and the quasar density.
The last form of Equation~\ref{eq:ebosserr} can be simply rewritten as
\begin{equation}
 \frac{\sigma^2_{eBOSS}}{\sigma^2_{BOSS}} = 
  = \frac{1}{f_A~f_n~f_q + (1-f_A)}
\end{equation}
where we have defined the additional parameter as the density of quasars in eBOSS
relative to BOSS, $f_q = n^{Q}_{eBOSS} / n^{Q}_{BOSS}$.
The values of $f_A$ for each incremental data release are of course the same as in the auto-correlation.
Using the number density of quasars in the second block of Table~\ref{tab:lya},
we find $n^{Q}_{BOSS}=19.7$ deg$^{-2}$, $n^{Q}_{eBOSS}=32.6$ deg$^{-2}$, and  $f_q = 1.65$.
The additional quasars from eBOSS will therefore have a slightly higher impact on the cross-correlation
measurement than the auto-correlation measurement.

For the auto-correlation, we expect the error on the BAO distance scale from BOSS to be reduced by factors
1.18, 1.27, and 1.42 for the two year, four year, and six year increments in the program.
For the cross-correlation, we expect the error to be reduced by factors
1.21, 1.30, and 1.46.
Because the improvements in the auto- and cross-correlation are so similar,
we average the two and assume improvements
of 1.19, 1.28, and 1.44 on the distance scale determined from the joint analysis.
These values are used to project the BAO distance precision from \lya\ forest data in Section~\ref{sec:projections}.

\section{Survey Metrics and Future Development Efforts}
\label{sec:performance}

\subsection{Data Quality Compared to Requirements}\label{subsec:quality}

The high-level goals of eBOSS are to obtain BAO distance measurements with 1\% precision
using an LRG sample, with 2\% precision using an ELG sample, and with
2\% precision using a quasar sample.
A number of low-level metrics must be satisfied to meet these goals.
The redshift range and number of spectroscopically confirmed objects required
for each class of tracer was presented in Section~\ref{subsec:program}.
The requirements on spectroscopic classification and uniformity in targeting
were presented in Section~\ref{subsec:uniformity}.

Significant testing of the LRG and quasar samples has allowed us to determine
whether we meet all of the goals.  A summary of the target population
was outlined in Section~\ref{subsec:targets} while the expected rate of
fiber efficiencies was presented in Section~\ref{subsec:tiling}.
Combining the number density of the parent target population with the tiling
statistics, we can predict the number of spectroscopically confirmed targets that can be used
as tracers.
Pilot studies described in Section~\ref{sec:sequels} produced a sample of more than
10,000 LRG and more than 10,000 quasar spectra from which we can estimate the
performance of spectroscopic classification.
Assuming a joint LRG and quasar survey covering 7500 $\sqdeg$, we summarize the
expected performance of these two target classes with respect to requirements
in Table~\ref{tab:performance}.
As can be seen, we satisfy the pipeline requirements on both the LRG and quasar samples
and we exceed the required number of clustering quasars by 15\%.
We expect to miss the requirement for the number of new spectroscopically confirmed
LRGs by 12\%, leading to an expected degradation in BAO signal of roughly 6\% relative
to the requirement.
The overlap with the high redshift tail of the BOSS CMASS sample increases the total
number of $0.6<z<1.0$ passive galaxies by 66\%, thus allowing high precision
measurements of BAO over this redshift range. 

\begin{table}[htp]
\centering
\caption{
\label{tab:performance}
Comparison of expected eBOSS performance compared to requirements for high-level BAO measurements.
}
\begin{tabular}{l c c c c c }
\hline\hline
  & LRG  & CMASS & Clustering Quasars & UgrizW1 ELG & DECam ELG   \\ \hline
Redshift Range  & $0.6<z<1.0$  &  $0.6<z<1.0$  &  $0.9<z<2.2$  &  $0.6<z<1.0$   &  $0.7<z<1.1$   \\
\hline
\multicolumn{6}{l}{{\bf Target Density and Fiber Density}} \\
\hline
\hline
Density of Targets (deg$^{-2}$) & 60  & 23        & 115 & 180 & 240   \\
Density of Previously Observed Objects & 0  & 23 & 25 & 0 & 0   \\
Density of Objects Assigned Fibers & 52  & 0 & 85 & 171 & 228   \\
\hline
\multicolumn{6}{l}{{\bf Efficiency of Redshift Classification}} \\
\hline
\hline
Density Expected Confirmed Tracers (deg$^{-2}$) & 36  & 23 & 67 & 121 &  179  \\
Number of Required Confirmed Tracers & 300,000  &  --       & 435,000 & 190,000 & 190,000   \\
Number of Expected Confirmed Tracers & 266,000  & 174,000 & 500,000 & 182,000 & 197,000   \\
\hline
\multicolumn{6}{l}{{\bf Statistical Accuracy of Redshift Estimates}} \\
\hline
\hline
Required Redshift Accuracy (kms$^{-1}$) & $< 300$  & $< 300$ & $< 300+400[z-1.5]$ & $< 300$ & $< 300$   \\
Expected Redshift Accuracy  (kms$^{-1}$) & 58   & 35 & $< 300+400[z-1.5]$  & $< 300$ & $< 300$   \\
\hline
\multicolumn{6}{l}{{\bf Unrecognized Redshift Classification Failures}} \\
\hline
\hline
Catastrophic Failures Requirement & $<1$\%  & $<1$\% & $<1$\%  & $<1$\% & $<1$\%   \\
Catastrophic Failures Expected & 0.6\%  & $<1$\% &  0.8\%  & TBD & TBD   \\
\hline
\multicolumn{6}{l}{{\bf Uniformity Over Targeting Area}} \\
\hline
\hline
Uniform Area in Imaging Systematics  & 92\%  & --  & 90\%  & TBD  & TBD   \\
$\Delta n/\Delta(0.01 mag)$ Zeropt error & 6.2\%  & --  & 0.86\%  & TBD  &  TBD  \\
Uniform Area in zeropoint errors  & 86.7\%  & --  & 100\%  & TBD  & TBD   \\
\hline
\end{tabular}
\end{table}

The final ELG selection remains undecided, so it is not yet possible to present the
expected performance for this target class.  We do have estimates for the
number density using several techniques presented in Section~\ref{subsec:targets}.
We present the statistics we can estimate for two of those samples in Table~\ref{tab:performance}.
We report the statistics for the Fisher UgrizW1 covering 1500 $\sqdeg$
and the high density DECam selection covering 1100 $\sqdeg$.
Because the emission lines are narrow, we expect to easily meet the requirement
of $< 300\,$kms$^{-1}$ redshift precision, although we do not have a specific estimate at this time.
More challenging will be robust identification of sources; catastrophic failures due to line
confusion and sky subtraction residuals pose a risk to this class of target.
We postpone discussion of the catastrophic failure rate and the
expected uniformity until a future publication.

\subsection{Future Improvements in Automated Data Reduction}\label{subsec:pipeline}

The eBOSS spectroscopic pipeline will be based on the BOSS spectroscopic
pipeline that was adapted from the original SDSS ``idlspec2d''
pipeline.  It consists of two main parts: the ``spectro2d'' portion that
extracts two-dimensional raw data into one-dimensional flux calibrated
spectra; and the ``spectro1d'' portion that measures scalar quantities
such as classifications and redshifts from those one-dimensional
spectra \citep{bolton12a}.
For the most part, eBOSS requires the same spectro2d performance as achieved in BOSS,
but extended to fainter targets.  As discussed in Section~\ref{subsec:LRGspectra},
a number of problems in extraction and/or flux calibration have arisen that
degrade the redshift classification.

New development on the data reduction pipeline has begun to address these problems.
This renewed effort includes removing a S/N dependent bias in the extractions,
coadditions, and sky subtraction; improving the sky model in the presence of variations
in brightness and fiber resolution over the focal plane;
improving the error propagation in the coadditions; and correcting occasional
failures of flux calibration at the extreme red or blue ends of the spectrum.  
New flux calibration that accounts for ADR across the focal plane
\citep{margala15a} will also be implemented in the data reduction pipeline.
These changes will provide more accurate spectra with
better error models.
Without making any changes to the spectral templates, we hope to resolve a significant
fraction of uncertain spectral classifications for the new faint targets in eBOSS
by making these changes to the spectro2d pipeline.

The analysis of quasar Lyman-alpha forest beyond the
baseline BAO measurements may require further developments to the spectro2d
pipeline, though the exact requirements are not specified at this point in time.
Characterization of the spectra and development of new algorithms falls under
the purview of the Lyman-alpha working group and not the team that develops the
data reduction pipeline.

Any new developments in the spectro2d component of the data reduction pipeline
will be backward compatible with BOSS data, enabling consistent reprocessing
for joint analyses of BOSS and eBOSS data.
We do not require SDSS-I and SDSS-II
data to be re-processed because the smaller wavelength range renders the early
$z>2.15$ quasar spectra obsolete with respect to the BOSS and eBOSS spectra.
Because of a lack of refined selection techniques and reduced system throughput,
no high redshift ELG or LRG spectra are found in the earlier programs.

The spectroscopic classifications and redshifts from spectro1d
provide the final map for the
clustering measurements and must meet the requirements laid out in Section~\ref{subsec:uniformity}.
As described throughout the text and in Table~\ref{tab:performance},
the BOSS spectral templates are nearly sufficient to meet these goals.
BOSS quasar identifications were
visually checked by two individuals to unambiguously classify the BOSS quasars
to a level of detail not currently possible by automated software.
There is a small sample of {\tt QSO\_CORE} objects that have spectra from previous
SDSS surveys that were never visually inspected.  To guarantee consistency with prior
observations, we will visually inspect each of those.
The vast number of quasar spectra in eBOSS require
robust spectral classification in an automated fashion. 
However, even with the new redshifting algorithm presented in Section~\ref{subsec:QSOspectra},
7\% of eBOSS quasar targets will need visual inspection.  Given that these
will be the most difficult spectra to characterize, the time investment will
not be terribly different from that in BOSS.
Improvements to the spectral templates are therefore required to improve the quasar classification in an
automated fashion.  Improvements to the LRG and ELG templates should also
mitigate catastrophic failures and provide better discrimination between
the true redshift and other local minima in the distribution of fits.
In addition to redshifts for clustering measurements, TDSS
and SPIDERS will produce a more heterogeneous spectroscopic sample than is
expected for the eBOSS LRG, ELG, and quasar clustering samples.
Thus new templates will likely be needed to meet the
requirements for automated redshift determination for the LRG, ELG, and quasar
samples, and to encompass the heterogeneity of the TDSS and SPIDERS data.

The BOSS spectro1d pipeline was based upon PCA templates
for quasars and galaxies and physically motivated templates for stars.
The BOSS quasar templates do not cover the full variation of actual quasars for BOSS,
much less the TDSS and SPIDERS samples.
Unphysical combinations of the BOSS PCA templates can also lead to erroneous
redshift estimates as shown in Section~\ref{subsec:LRGspectra}.  We are
exploring a purely archetype-based alternative where, by construction, every template is 
physically meaningful, thus eliminating the false redshift estimates
from unphysical combinations of PCA-based templates.
We also expect to introduce more linear combinations of templates to allow classification of
such as binary stars or AGN with significant galaxy light.

Any new developments in the spectro1d component of the data reduction
pipeline must be backward compatible with SDSS-I, SDSS-II, and BOSS. This
requirement is necessary to ensure that redshifts from all samples including objects observed
in SDSS and BOSS can be determined in a consistent fashion for clustering measurements.

The initial updates of spectral templates will be modeled on the SEQUELS data
to more accurately represent the true target spectral variations.
Redshifts for LRGs, ELGs, and quasars will be determined by visual inspection
and used as input in the generation of templates. Subsequent updates of these
templates will be possible as more spectra are taken (especially of rarer
objects) and any problems in the templates are uncovered.
We expect to present new templates in a future publication similar in philosophy
to \citet{bolton12a}.

\section{Cosmological Projections}
\label{sec:projections}

Following the methodology developed in \citet{font-ribera14b},
we forecast cosmological constraints for the eBOSS program.
We begin by using the expected density, redshift distribution, volume, and bias for
each tracer to independently predict the constraints on the
matter power spectrum at the relevant redshift.
From this, we predict the sensitivity of eBOSS to measure the cosmic distance scale through
BAO and to measure the rate of structure growth through RSD.
We conclude by presenting the power of the combined sample
to constrain the properties of dark energy, gravity, neutrino
masses, and inflation.

\subsection{Number Density of Spectroscopic Sample}

\begin{table}[t]
\centering
\caption{
\label{tab:nz}
Expected volume density of each target class, presented in units $10^{-4}\vunits$.
Entries highlighted in bold font denote the fraction of the sample that satisfies the high-level
requirement for the redshift distribution of the sample.
Because \lya\ quasars will be used as random sightlines to projected neutral hydrogen,
we report the surface density of those objects parenthetically.
}
\begin{tabular}{l c c c c c c c}
\hline\hline
Redshift  & CMASS & eBOSS & Clustering & \lya\ & Fisher & Low Density & High Density \\
& LRG & LRG  & Quasars\tablenotemark{a} & Quasars\tablenotemark{b} & ELG & DECam ELG & DECam ELG   \\
\hline
$0.6 < z < 0.7$ & {\bf 1.137} & {\bf 0.810} & 0.119 &  & {\bf 1.412} & 0.183 & 0.205   \\
$0.7 < z < 0.8$ & {\bf 0.170} & {\bf 0.678} & 0.130 &  & {\bf 2.165} & {\bf 1.908} & {\bf 2.068}   \\
$0.8 < z < 0.9$ & {\bf 0.010} & {\bf 0.350} & 0.154 &  & {\bf 1.654} & {\bf 2.673} & {\bf 3.034}   \\
$0.9 < z < 1.0$ & {\bf 0.001} & {\bf 0.097} & {\bf 0.171} &  & {\bf 0.624} & {\bf 1.135} & {\bf 1.605}   \\
$1.0 < z < 1.1$ &  &  & {\bf 0.163} &            & 0.218 & {\bf 0.373} & {\bf 0.568}   \\
$1.1 < z < 1.2$ &  &  & {\bf 0.170} &            & 0.081 & 0.159 & 0.241   \\
$1.2 < z < 1.4$ &  &  & {\bf 0.175} & &  &  &    \\
$1.4 < z < 1.6$ &  &  & {\bf 0.166} & &  &  &    \\
$1.6 < z < 1.8$ &  &  & {\bf 0.151} & &  &  &    \\
$1.8 < z < 2.0$ &  &  & {\bf 0.137} & &  &  &    \\
$2.0 < z < 2.1$ &  &  & {\bf 0.122} & &  &  &    \\
$2.1 < z < 2.2$ &  &  & {\bf 0.093} & {\bf 0.069 (2.75 deg$^{-2}$)} &  & &    \\
$2.2 < z < 2.4$ &  &  &  & {\bf 0.063 (5.02 deg$^{-2}$)} &  &  &   \\
$2.4 < z < 2.6$ &  &  &  & {\bf 0.042 (3.35 deg$^{-2}$)} &  &  &   \\
$2.6 < z < 2.8$ &  &  &  & {\bf 0.027 (2.16 deg$^{-2}$)} &  &  &   \\
$2.8 < z < 3.0$ &  &  &  & {\bf 0.018 (1.39 deg$^{-2}$)} &  &  &   \\
$3.0 < z < 3.5$ &  &  &  & {\bf 0.009 (1.69 deg$^{-2}$)} &  &  &  \\ \hline
Observed Surface Area & 7000 $\sqdeg$ & 7000 $\sqdeg$ & 7500 $\sqdeg$ & 7500 $\sqdeg$ & 1500 $\sqdeg$ & 1400 $\sqdeg$ & 1100 $\sqdeg$ \\
Uniform Surface Area & -- &  6120 $\sqdeg$ &  6750$\sqdeg$ & -- & -- & -- & -- \\
\hline
\end{tabular}
\tablenotetext{a}{The clustering quasars are uniformly selected and include BAL and non-BAL quasars
previously observed in SDSS-I, -II, and -III.}
\tablenotetext{b}{The \lya\ forest quasars are selected by various means and only include non-BAL quasars
that will be observed in eBOSS, hence the reduction in surface density relative to clustering quasars over $2.1 < z < 2.2$.}
\end{table}

The density and redshift distributions used in projections are taken from the parent target samples
described in Table~\ref{tab:LRGQSOdensities} and Table~\ref{tab:ELGtargets}.
Estimates for the number densities of the final spectroscopic sample are corrected for incompleteness introduced
in the tiling process as explained in Section~\ref{subsec:tiling}.
The assumed volume density for each spectroscopic sample for the appropriate
redshift ranges and the expected survey area is shown in Table~\ref{tab:nz}.
In addition to reporting the observed area, we report the total area that meets the requirement
for 15\% uniformity in target density.  As discussed in Section~\ref{subsec:uniformity},
areas with larger fluctuations could be discarded in the cosmology analysis.
Because we currently have no way of tracking the regions with excess fluctuations,
we do not include the loss of area due to calibration uncertainty.  Instead,
zeropoint errors will need to be modeled in the analysis.
Likewise, the members of the eBOSS collaboration are developing a new bright star mask
using the WISE photometry.  While the final mask has not been established,
it is likely that 3--4\% of the sky will be removed due to proximity to bright stars and suspect photometry.
Throughout, a flat $\Lambda$CDM model is assumed with $H_0$=70 and $\Omega_M = 0.31$.
These predicted number densities allow us to predict the sensitivity to $D_A(z)$, $H(z)$,
and $f\sigma_8$ from each class of tracer.

We assume 100\% completeness on the clustering quasars observed prior to eBOSS,
95\% completeness on the new targeted clustering quasars, and 100\% completeness on the reobservations
of known \lya\ quasars.
For the PTF-selected quasars, we assume 92\% completeness in fiber assignments and then 
reduce the density by an additional 10\% to account for incomplete PTF coverage of the eBOSS footprint.
We simply average the effect of incomplete coverage over the full area that will be observed.

The surface density of the LRG target sample is taken to be the more conservative
estimate presented in Table~\ref{tab:LRGQSOdensities}.
Because the effective area of the LRG program is reduced by 7\% due to masking by
targets tiled in the first round, we take the average density of targets to be the
total number of galaxies assigned a fiber over the reduced footprint.  We assume a completeness
of 87\% of the LRG target sample over a final area of $\sim 7000 \, \sqdeg$ for the LRG sample instead of 7500 $\sqdeg$.
To account for isolated regions with decollided completeness
below 85\% (see Section~\ref{subsec:tiling}), we include a 5\% loss of area in addition to the 8\%
area that could be lost to targeting non-uniformities when reporting the ``Uniform Surface Area'' for the LRG sample.
Finally, because the high redshift tail overlaps with the new LRG sample, we include the
$n(z)$ distribution for CMASS in the redshift range of interest.  In the cosmological projections
that follow, we combine the two samples for optimal constraints over the
redshift range $0.6<z<1.0$.

The ELG program is not yet finalized so we include the volume density for three
potential selection schemes.
We do not report the results for the $gri + Uri$ selection as it produces tracers at an efficiency
of only 52.5\%.
We assume a completeness of 95\% over 1500 $\sqdeg$ for the
Fisher, over 1400$\sqdeg$ for the low density DECam selections, and over
1100 $\sqdeg$ for the high density DECam selection.
As it produces a sample with the highest median redshift, we assume the high density DECam-based
ELG selection in the projected BAO and RSD measurements and for
the cosmological projections presented in Section~\ref{subsec:cosmology}.

\subsection{BAO and RSD Forecasts}

In order to delineate clustering
measurements between samples and ensure the predictions are
independent, we use only the $0.9<z<2.2$ component of the quasar
sample for direct clustering predictions and only the $z>2.1$
component of the quasar sample for \lya\ forest predictions.  In
reality, the {\tt QSO\_CORE} sample will extend over both redshift regimes
and will allow for direct clustering measurements at redshifts beyond
those used in these projections.

The \lya\ quasar sample in eBOSS will complement the BOSS survey, improving the
BAO measurement by providing deeper spectra of known $z>2.1$ quasars
and spectra of new $z>2.1$ quasars.
As explained in \citet{font-ribera14a}, the auto-correlation
\lya\ forest clustering measurement is shot-noise limited and therefore
nearly uncorrelated with the \lya\ forest -- quasar cross-correlation measurement.
We therefore use the combined precision from these two measurements to forecast
the final BAO distance measurement.
Rather than applying direct Fisher projections, we instead scale the expected DR12 results
according to the scheme presented in Section~\ref{subsec:lyaspectra}.
We do this for $H(z)$ and $d_A(z)$ but not $R$ because the weighting of the
radial modes and transverse modes for determining the optimal distance indicator
is so different between the cross-correlation and auto-correlation analysis.
The measurement of RSD from the \lya\ sample is considerably more challenging for
both theoretical and instrumental considerations, and we do not project its precision in this document.

We present the projections for each sample in two year increments,
according to the data release schedule presented in Section~\ref{subsec:strategy}.
For BAO forecasts we assume measurements using modes with $k<0.5 h $Mpc$^{-1}$;
for RSD forecasts we use modes with $k<0.2 h $Mpc$^{-1}$.
This sequence of projections roughly corresponds to the expected schedule
for public data releases and the likely publications of clustering measurements.
For the LRG, quasar, and \lya\ forest samples, we assume that
eBOSS will complete 3000 deg$^2$ in the first release, an additional 1500 deg$^2$
for the second release, and an additional 3000 deg$^2$ in the final release.
The usable LRG area is somewhat less than the observed area due to loss
of targets in regions that are masked by higher priority TDSS, SPIDERS, and quasar targets.
We report the area that is expected to be observed in each case.
For the ELG sample, we only present the results expected after the second
release, in which 300 plates are expected to be observed.
The assumed bias for each sample is the same as that presented in the beginning
of Section~\ref{sec:eboss}.

The time series of predictions on BAO distance precision and RSD growth precision
are shown in Table~\ref{tab:BAORSD}.
After six years, the LRG sample is expected to produce the highest redshift, sub-percent precision 
distance measurement obtained by any means.
The quasar sample is expected to produce a 1.8\% BAO distance measurement,
the first percent-level distance measurement achieved in the interval $1<z<2$.
The ELG sample will produce the highest precision BAO measurement to date using this
tracer of the matter density field.
If one were to assume that the area that fails to meet 15\% uniformity in target selection
cannot be used for clustering measurements, the precision of the BAO and RSD measurements
will be degraded.  For the LRG sample, reducing the area to account for
low completeness sectors and excessive systematic fluctuations results in a 7\% increase
in the projected errors.
For the clustering quasar sample, the potential loss of area would result in a 5.5\% increase
in the projected errors.

\begin{table}[b]
\centering
\caption{
\label{tab:BAORSD}
Basic parameters expected for each eBOSS sample, together with predictions for
the effective volumes and fractional constraints on BAO distance measurements
and growth of structure.}
\begin{tabular}{l c c c c c c}
\hline\hline
Sample & Epoch  & Area (deg$^{-2}$) & $\sigma_H/H$ &  $\sigma_{D_A}/D_A$ &   $\sigma_R/R$ & $\sigma_{f\sigma_8}/f\sigma_8$ \\ \hline
LRG  & year 2  & 2790   & 0.032 & 0.017 & 0.012 & 0.040\\
     & year 4  & 4185   & 0.026 & 0.015 & 0.010 & 0.034\\
     & year 6  & 6975   & 0.021 & 0.012 & 0.008 & 0.026\\
\hline
ELG (High Density DECam)       & year 4 & 1100 & 0.047 & 0.031 & 0.020 & 0.038 \\         
\hline
Quasar   & year 2  & 3000     & 0.066 & 0.043  & 0.028 & 0.050\\
         & year 4  & 4500     & 0.054 & 0.036  & 0.023 & 0.041\\
         & year 6  & 7500     & 0.042 & 0.028  & 0.018 & 0.032\\
\hline
BOSS \lya\ Quasars           &  & 10,400       & 0.02  & 0.025 & -- & --\\
\hline
BOSS $+$ eBOSS               &  year 2 & 3000 & 0.017   & 0.021 & -- & --\\
\lya\ Quasars                &  year 4 & 4500 & 0.016   & 0.020 & -- & --\\ 
                             &  year 6 & 7500 & 0.014   & 0.017 & -- & --\\ \hline
\end{tabular}
\end{table}

The six-year predictions for the eBOSS distance measurements
are shown in Figure~\ref{fig:baodistance}.
It can be seen in the bottom panel that SNe~Ia offer high precision measurements
of distance out to redshifts $z<0.6$ or so, while the BAO probes offer
constraints comparable to SNe~Ia around $z\sim 0.6$ but offer much stronger constraints
for redshifts beyond.
As emphasized in \citet{kim15a},
BAO and SNe provide distinct cosmological constraints
even at the same redshift because of the difference
between absolute and Hubble flow calibration, making the combination
much more powerful than either probe in isolation.
The combination of SNe from DES with the final data sets from
SNLS and SDSS-II will provide a far larger sample for reducing
both statistical and systematic errors.
However, because the leading supernova analyses are currently limited by systematic
uncertainties \citep[e.g.][]{sullivan11a,betoule14a},
it is difficult to forecast the performance of future surveys.

\subsection{Cosmological Constraints}\label{subsec:cosmology}

In what follows, we report the projections for
cosmological constraints from the eBOSS BAO distance, RSD, and power spectrum measurements
from the galaxies and quasars using modes with $k<0.1 h $Mpc$^{-1}$.
We report predicted cosmological constraints for a six year program using the combined
information of all four tracers.
We assume that clustering measurements can be recovered over the areas which display
deviations from the uniformity requirement.
For these projections, we
take the values of the parameters of the fiducial cosmology to be the
flat $\Lambda$CDM model with parameter values as shown in Table~1 of
\citet{font-ribera14b}.
We allow free parameters to describe the growth rate of structure ($\Delta\gamma$ and a normalization constant $G_9$),
the summed neutrino mass ($\sum m_\nu$), non-Gaussianity in the perturbations of the initial
density field ($f_{\rm NL}$), and a time-evolving equation of state for dark
energy ($w_0$, $w_a$).
We assume a baseline of Planck CMB measurements,
5\% $H_0$ constraint, and the BAO measurements from
the complete BOSS galaxy sample.
The projected constraints are found in Table~\ref{tab:cosmoconstraints}.
A brief interpretation of those predicted constraints in terms of dark energy,
modified gravity, neutrinos and inflation is presented below.
\begin{table}[htp]
\centering
\caption{
\label{tab:cosmoconstraints}
Predicted precision from the combination of CMB and large-scale structure measurements.
All values correspond to the estimated 1$-\sigma$ uncertainties.
}
\begin{tabular}{l c c c }
\hline\hline
Parameter & Constraint from & Constraint from & Constraint from \\
& CMB & BOSS and CMB & BOSS, eBOSS, and CMB \\ 
 \hline
$\Omega_Mh^2$ & 0.008  & 0.0028 & 0.0017 \\
$w_0$ & 0.52  & 0.17 & 0.15 \\
$w_a$ & 1.4  & 0.67 & 0.48 \\
$\gamma$ & 30.  & 0.13 & 0.10 \\
$\sum m_\nu$ & 0.81 eV & 0.29 eV & 0.16 eV \\
$n_s$  & 0.0045  & 0.0026 & 0.0022 \\
\hline
\end{tabular}
\end{table}

\subsubsection{Dark Energy}

\begin{figure}[b!]
\centering
%\vspace{-1.5cm}
\includegraphics[width=0.8\textwidth,trim=0 0 0 0, clip]{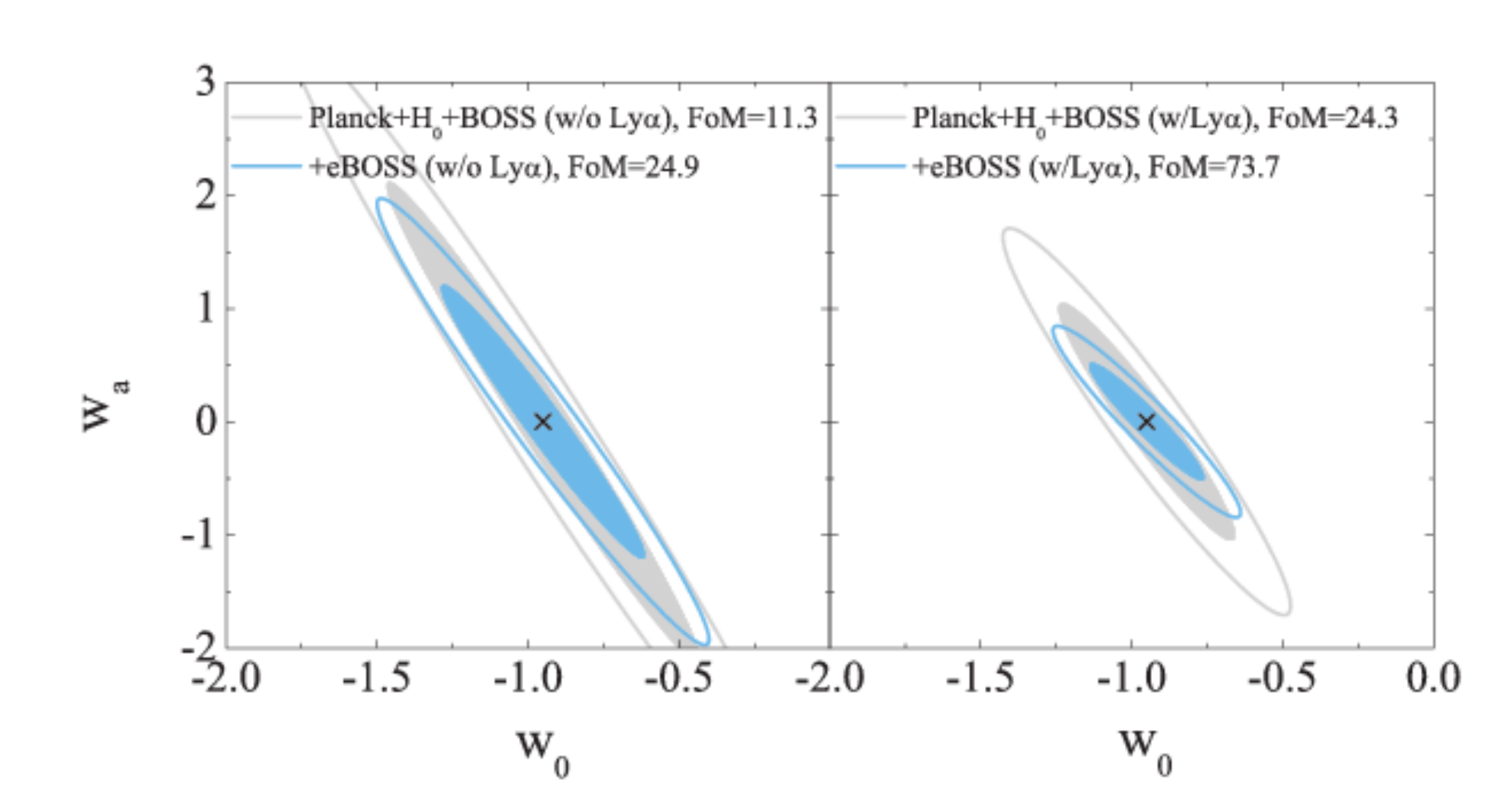}
\caption{\small Current constraints on the DETF model for time-varying
dark energy compared to projected constraints from eBOSS.
We report constraints from the BAO probes, Planck, and $H_0$ from HST observations of
SNe~Ia and Cepheid variables \citep{riess11a}.
For all measurements, the filled ellipse represents the 68\% confidence interval and
the open ellipse represents the 95\% confidence interval.
}
\label{fig:w0wa}
\end{figure}

To demonstrate the power of BAO as an independent probe of the cosmic
distance ladder, we report the constraints on the equation of state for dark energy
using BOSS and eBOSS BAO measurements
as the sole source of information from cosmic times between the CMB and today.
We fix neutrino mass and inflation parameters to highlight the relative constraints
on dark energy.
We predict constraints on a standard parameterization of dark energy in which the
equation of state evolves with time as $w(a) = w_0 + w_a(1-a)$.
The prediction for the combined constraints on $w_0$ and $w_a$
is shown in Figure~\ref{fig:w0wa}.

\subsubsection{Modified Gravity}

At a given redshift, RSD measurements constrain
the product $f\sigma_8$, but the extended redshift range of the combined eBOSS and BOSS
measurements will significantly reduce the degeneracy between $f$ and $\sigma_8$.
In predicting the power of eBOSS RSD measurements to constrain models of
modified gravity, we assume a parameterization $f(z) = \Omega_M^\gamma (z)$, where $\gamma$ is the
growth index \citep{linder05a}.
The value of $\gamma$ is approximately 0.55 when predicting the growth rate from General Relativity.
Measurements with RSD that differ from this value would indicate a model for gravity that deviates from
GR at cosmological scales.
The amplitude of $f(z)$ is normalized according to a free parameter referenced as $G_9$ in the
assumed model.

An example of the power of RSD is shown in Figure~\ref{fig:rsd},
where the growth rates are compared for a set of cosmological models that
predict the same background expansion (i.e. the same distance-redshift relation and comoving BAO position).
In the $\Lambda$CDM model, which assumes that GR correctly describes gravity on all scales,
the evolution of $f\sigma_8$ is determined entirely by the expansion history $H(z)$.
On the other hand, if GR requires modification on cosmological scales,
as demonstrated in the other two models plotted,
then measurements of $f\sigma_8$ over a large redshift interval will reveal
that inconsistency, thus indicating that cosmic acceleration arises from modified
gravity instead of dark energy.
The eBOSS measurements will significantly improve the constraints on $f\sigma_8$
between $0.6<z<2.0$, which will help constrain the amplitude of the growth rate 
at fairly early times where various models nearly converge.
More generally, the combination of BAO and
RSD measurements will enable eBOSS to simultaneously constrain models
of dark energy and modified gravity (e.g., \citealt{song09a}).
A review of the power of eBOSS and other spectroscopic surveys to measure the growth of structure in
tandem with imaging surveys is presented in \citet{huterer15a}.

\subsubsection{Neutrinos}

A global fit to solar and atmospheric neutrino flavor oscillations
implies a difference in the squares of masses
$m_2^2-m_1^2=7.54^{+0.26}_{-0.22} \times 10^{-5}$ eV$^2$ and
$|(m_3^2-m_1^2)/2 + (m_3^2-m_2^2)/2|= 2.42^{+0.07}_{-0.11} \times 10^{-3}$ eV$^2$
\citep{fogli12a}.
Assuming a normal hierarchy in which the lightest neutrino is massless,
one can infer the minimum sum of the masses to be greater than 0.05 eV.
In the case of a so-called inverted hierarchy, where $m_3=0$,
the minimum sum of the masses must be greater than 0.1 eV.
Both of these are well below the reach of the
present terrestrial experiments. For example, beta-decay experiments
currently constrain the effective electron-neutrino mass to be $\lesssim
2$eV \citep{otten08a} and even the next generation direct tritium decay experiments such as
the Karlsruhe Tritium Neutrino experiment
\citep[KATRIN;][]{katrin01a,wolf10a} will only have a sensitivity to
constrain $m(\nu_e) < 0.2$ eV (90\% upper limit).
When considering the small mass differences derived from the oscillation
experiments, the projected results from KATRIN can at best constrain
each individual mass eigenstate to $m < 0.2$ eV, or $\sum m_\nu < 0.6$ eV (90\% confidence).

\begin{figure}[t!]
\centering
\vspace{-3.0cm}
\includegraphics[width=0.75\textwidth]{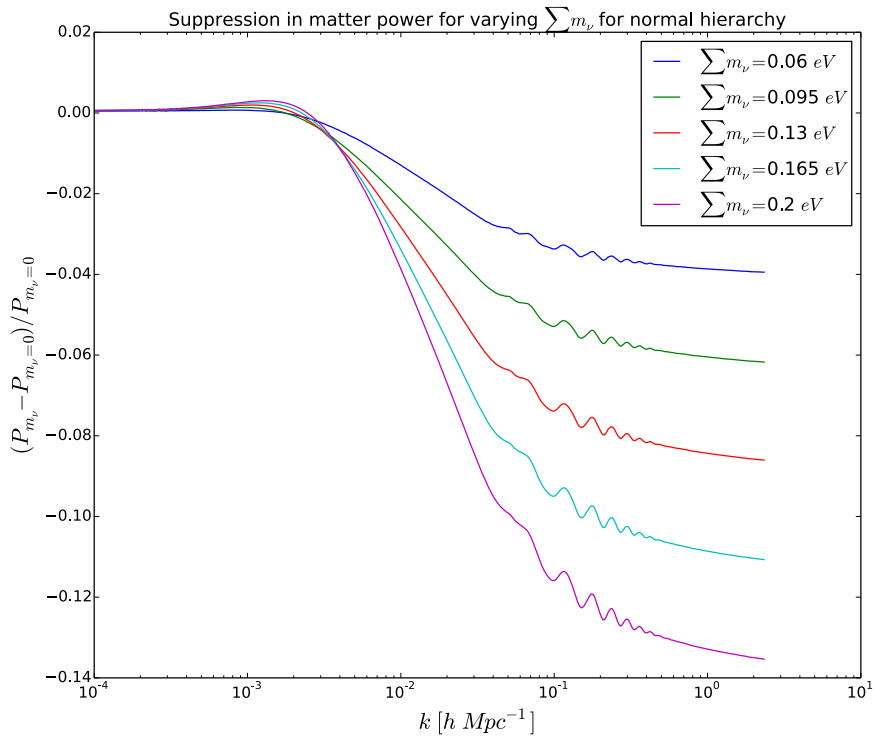}
\vspace{-3.5cm}
\caption{\small
The suppression of power at small angular scales due to free-streaming
of massive neutrinos.  A normal hierarchy for neutrino masses is assumed.
}
\label{fig:neutrino}
\end{figure}

The large effective volume covered by eBOSS makes it sensitive to neutrino masses.
The signature of neutrinos appear as modulations of
clustering in the same large-scale structure maps used for BAO and RSD constraints.
An example of the predicted suppression of power by massive neutrinos is shown in
Figure~\ref{fig:neutrino}.
eBOSS will therefore place tight new constraints on this fundamental quantity
without any changes to the BAO survey design.
The relative impact of eBOSS, CMB and other cosmology programs
to improve neutrino constraints is reviewed in \citet{abazajian15a}.

The best current cosmological constraints on $\sum m_\nu$ from large-scale clustering result from the
combination of Planck CMB, CMB polarization, lensing and BAO.
As reported in \citet{ade15a}, these combined probes produce an upper limit
$\sum m_\nu < 0.23$ eV (95\% confidence) when assuming zero curvature.
With eBOSS, we can provide comparable constraints
if one combines information derived from the shape of the broad-band power
spectrum (for neutrino masses) with the distance measurements derived from BAO.
Estimates from the clustering of BOSS galaxies
have already shown the potential in spectroscopic surveys to constrain
the neutrino mass \citep[e.g.][]{zhao13a,beutler14b}.

The projections in Table~\ref{tab:cosmoconstraints} include an estimated precision on the
neutrino mass that is highly degenerate with the modified gravity parameters.
If one were to assume a flat, $\Lambda$CDM cosmology described by GR on cosmological scales, we 
predict a 95\% upper limit $\sum m_{\nu} <0.104$ eV when combining the
results from the BOSS, eBOSS, and CMB.  The reported constraint
assumes that that measurements of large-scale modes can be made with the eBOSS
galaxies and quasars at all scales larger than $k_{max}=0.1 h $Mpc$^{-1}$.
If one were to assume measurements of large-scale modes with wavelengths
up to $k_{max}=0.2 h $Mpc$^{-1}$, the projections for the 95\% upper limit
improve to $\sum m_{\nu} <0.072$ eV.
The statistical power of eBOSS can potentially provide an upper limit on neutrino
masses comparable to the minimum allowed mass in an inverted hierarchy.

An alternative way to constrain neutrino masses with optical spectroscopy
is to use the information from the 1D flux power spectrum of \lya\ forest.
Measurements of the 1D flux power spectrum using BOSS quasars \citep{palanque-delabrouille13b}
yield the tightest constraints to date using any
cosmological tracer.
When combining the BOSS measurement with CMB and BAO measurements,
the sum of the neutrino masses is constrained with a 95\% upper
limit $\sum m_{\nu} <0.15$ eV \citep{palanque-delabrouille15a},
improved to $\sum m_{\nu} <0.12$ eV \citep{palanque-delabrouille15b}.
We do not include projections for new constraints from 1D power on the neutrino
mass from eBOSS.  The deeper spectra obtained by reobserving 1.2
deg$^{-2}$ known quasars and the discovery of 1.2 deg$^{-2}$ new
quasars at $z>3$ should allow tighter constraints on the redshift
evolution of the 1D flux power spectrum, and thus tighter
constraints on the neutrino mass.

\subsubsection{Inflation}

Inflation is the best candidate that we have for a
theory that simultaneously established the initial conditions for structure
formation while producing a homogeneous, nearly flat Universe.
Inflation can explain small super-horizon
fluctuations which are nearly Gaussian and scale-independent.
A review of inflation and the potential of upcoming cosmology programs to
improve inflationary models is found in \citet{abazajian15b}.

The eBOSS survey volume provides sensitivity to the standard
inflationary parameters such as the spectral index of primordial matter
fluctuations ($n_s$), defined according to $P(k) = k^{n_s}$.
Most inflationary models predict a value of $n_s$ slightly less than one.
One possible departure from standard inflationary scenarios is expected to appear as
small deviations from Gaussian fluctuations in the initial density field.
This signature of non-Gaussianity in the primordial fluctuations
could also be visible in the large-scale structure maps from eBOSS.

A form of potential non-Gaussianity that arises naturally in many
inflationary models \citep{wands10a} is the ``local'' form, with a gravitational
potential described by
\begin{equation}
  \phi = \phi_g + \fnl \left( \phi_g^2-\left< \phi_g^2 \right> \right),
\end{equation}
where $\phi_g$ is an auxiliary Gaussian field. In the limit of
$\fnl=0$, one recovers the Gaussian case.
It has been shown \citep{dalal08a} that such
non-Gaussianity will generate a deviation from the standard prediction
in which galaxies are a linear tracer of the underlying dark matter
field on large scales.  This correction scales inversely as the square
of the wavenumber in the matter power spectrum, and it thus
becomes important on the larges scales - precisely in the limit where
non-linearities in the dark-matter fluctuations are negligible.

We find that primordial non-Gaussianities of the local form can be constrained
to a precision $\sigma_{f_{\rm NL}}=12$ (68\% confidence) by the combination of
eBOSS LRG, ELG, and quasars.
This measurement
will be independent of the current Planck bispectrum limits of $\fnl=0.8\pm5.$ (68\% confidence)
\citep[Planck 2015 results. XXVII;][]{planck15b}.
While we do not have forecasts for constraints arising from measurements of
the galaxy bispectrum, we expect these to significantly tighten the constraints from eBOSS,
provided systematic effects can be brought under control \citep[e.g.][]{ross13a}.
In fact, the first analysis of the BOSS galaxy bispectrum \citep{gilmarin14a}
and cosmological interpretation \citep{gilmarin14b} were completed in 2014.

\newpage

\section{Conclusion}
\label{sec:conclusion}

As described throughout this paper, the eBOSS survey design has been thoroughly
evaluated and should meet a high-level goal of precise clustering measurements
using four distinct tracers over four different redshift intervals.
A sample of 52 deg$^{-2}$ targets will be observed and will produce
more than 35 deg$^{-2}$ LRGs with confident spectroscopic redshifts between $0.6<z<1.0$.
When combined with the $z>0.6$ tail of the BOSS CMASS sample, this LRG
sample has the statistical power to constrain the matter power spectrum
at BAO scales for better than a 1\% distance measurement.
Similarly, a uniform sample of quasar candidates selected at a density of 115 deg$^{-2}$
can be efficiently identified.
Roughly 30 targets deg$^{-2}$ have confident spectroscopic classification
from SDSS-I, -II, or -III, revealing quasars at a density 13 deg$^{-2}$ between $0.9<z<2.2$.
New observations will produce an average of 56 deg$^{-2}$ new quasars at $0.9<z<2.2$.
The final sample should be sufficient for a 1.8\% BAO distance measurement.
The range of redshifts covered by the LRG and quasar program is only sparsely sampled by supernova
surveys, but it covers the crucial epoch of transition from decelerating cosmic expansion to
accelerating expansion.
As with BOSS, the area covered by the LRG and quasar clustering samples approaches the limits
accessible by ground-based telescopes:  we will observe one-third of the extra-Galactic sky.

eBOSS will use a deeper
sample of \lya\ forest observations to improve the BOSS BAO distance measurements
by a factor of 1.44 and will introduce a new sample of highly-biased ELG targets.
Several potential ELG programs have been identified that can provide BAO measurements
at 2\% precision.
The final ELG program will be decided when the imaging data sets are better understood
and when uniformity of the target selection is fully assessed.
Beyond BAO, the clustering in LRG, ELG, quasars, and the \lya\ forest provide broad-band
power spectra to further explore the cosmological model, particularly in the realm
of modified gravity and neutrino mass estimates.
The primary analysis effort within the eBOSS collaboration will be to 
mitigate systematic errors due to the non-uniformities in the target selection and
improve modeling of the observed power spectrum at small scales to capitalize on this statistical power.

The cosmological projections presented in Section~\ref{subsec:cosmology} assume a standard
cosmological model.  Because of the overlap in redshift range between the ELG sample
and the LRG and quasar samples, the eBOSS data also enable techniques such as those
introduced in \citet{mcdonald09a} and \citet{seljak09a}
to combine multiple tracers and reduce the effects of sample variance.
Projections for $f_{\rm NL}$ and RSD from eBOSS following the multi-tracer technique are 
found in the work by \citet{zhao15a}.
\citet{zhao15a} also present Fisher forecasts for the dark energy equation-of-state
and modified gravity based on a principal component analysis parameterization.

eBOSS will provide unique spectroscopic information
that will complement imaging cosmological surveys such as DES.
Most imaging probes of dark energy
rely on photometric redshift estimates which must be calibrated to
extremely high accuracy to avoid degradation in dark energy inference.
This calibration is typically done using spectroscopically calibrated
redshifts of objects spanning the photometric properties of the sample of interest.
However, existing spectroscopic surveys are inherently limited
by the difficulty of securing redshifts for faint, high redshift galaxies.
Spectroscopy from eBOSS offers an alternative approach to photometric redshift
calibration through cross-correlation techniques \citep{newman08a,menard13a}.
The observed degree of correlation between a
well-measured spectroscopic sample and a photometrically-selected sample
provides information on the fraction of the imaging sample at that redshift.
The possibility of cross-correlation as a tool to calibrate photometric redshifts
for DES was the primary reason that the 500 deg$^2$ eboss2 region 
was tiled in the first year (see Figure~\ref{fig:ebosschunks}).

Improvements in the automated data reduction pipeline are underway to ensure that 
we meet the redshift efficiencies required to meet the assumed cosmological precision.
The introduction of new spectral templates that cover a more limited
parameter space should significantly improve the ability of the pipeline to 
differentiate the best fit redshift from contaminating interlopers.
When complete, the classifications will produce the largest sample of
$z>0.6$ LRGs to date, a high-redshift ELG sample exceeding that of
the WiggleZ Dark Energy Survey \citep{drinkwater10a,blake11a},
a sample of quasars exceeding that of all previous observations combined,
and an enhanced sample of \lya\ forest spectra.
These new samples will allow studies of galaxy evolution and quasar astrophysics
well beyond the cosmological studies that drove the survey design.

Several studies of galaxy evolution and quasar astrophysics have already begun with the
early release of the SEQUELS sample in DR12 and the first eBOSS data taken in Fall 2014.
At the redshifts of the LRG and quasar samples, many absorption lines are accessible in the eBOSS spectra.
Measurements of quasar absorption features in the vicinity of BOSS spectroscopic galaxies 
revealed the Mg II distribution surrounding LRGs at redshifts as low as $z=0.5$ \citep{zhu14a}.
Composite spectra constructed from \lya\ forest absorbers shed light on
circumgalactic regions and the intergalactic medium \citep{pieri14a}.
These studies will be extended to higher redshift with the eBOSS data.
Studies of the correlations in the \lya\ forest with other systems are underway following
early studies in BOSS \citep[e.g.][]{font-ribera12b},
while others have the shown the potential of eBOSS to offer new probes of large-scale 
structure and BAO \citep[e.g.][]{pieri14b}.
The composite spectrum of ELGs from limited pilot observations already shows rich spectral features
such as resonant absorption and non-resonant emission.
The first science result from the eBOSS program resulted from studies of this composite ELG
spectrum \citep{zhu15a}.
These features are rare in other spectroscopic samples; eBOSS will allow systematic investigation
of these features to explore the effect of gas processes on galaxy evolution.
The overlap in spectroscopic area covered by eBOSS and wide-field far infrared imaging from the
Herschel Space Telescope \citep{pilbratt10a} will enable an exploration of the far-IR luminosity
function of the quasar host galaxies.
Finally, the quasar sample will greatly exceed all prior work both in total numbers and in the coverage
of luminosity--redshift phase space.  The sample can be used to enhance BOSS constraints on
the luminosity function \citep[e.g.][]{ross13b,mcgreer13a}, provide halo occupation
statistics that can be used to constrain the duty cycle of quasars as a function
of halo mass and quasar luminosity, and to explore the redshift and luminosity evolution of quasars.
When complete, the eBOSS spectroscopic sample will results in a diverse range of findings
both for cosmology and for galaxy and quasar science.

\acknowledgements

KD acknowledges support from the U.S. Department of Energy under Grant
DE-SC000995.
JPK and TD acknowledge support from the ERC advanced grant LIDA.
WJP acknowledges support from the UK STFC through the consolidated grant ST/K0090X/1,
and from the European Research Council through grant Darksurvey.

This paper includes targets derived from the images of
the Wide-Field Infrared Survey Explorer, which is a
joint project of the University of California, Los Angeles,
and the Jet Propulsion Laboratory/California Institute
of Technology, funded by the National Aeronautics and
Space Administration.

This paper represents an effort by both the SDSS-III and SDSS-IV collaborations.
Funding for SDSS-III was provided by the Alfred
P. Sloan Foundation, the Participating Institutions, the
National Science Foundation, and the U.S. Department
of Energy Office of Science.   
Funding for the Sloan Digital Sky Survey IV has been provided by
the Alfred P. Sloan Foundation, the U.S. Department of Energy Office of
Science, and the Participating Institutions. SDSS-IV acknowledges
support and resources from the Center for High-Performance Computing at
the University of Utah. The SDSS web site is www.sdss.org.

SDSS-IV is managed by the Astrophysical Research Consortium for the 
Participating Institutions of the SDSS Collaboration including the 
Brazilian Participation Group, the Carnegie Institution for Science, 
Carnegie Mellon University, the Chilean Participation Group,
the French Participation Group, Harvard-Smithsonian Center for Astrophysics, 
Instituto de Astrof\'isica de Canarias, The Johns Hopkins University, 
Kavli Institute for the Physics and Mathematics of the Universe (IPMU) / 
University of Tokyo, Lawrence Berkeley National Laboratory, 
Leibniz Institut f\"ur Astrophysik Potsdam (AIP),  
Max-Planck-Institut f\"ur Astronomie (MPIA Heidelberg), 
Max-Planck-Institut f\"ur Astrophysik (MPA Garching), 
Max-Planck-Institut f\"ur Extraterrestrische Physik (MPE), 
National Astronomical Observatory of China, New Mexico State University, 
New York University, University of Notre Dame, 
Observat\'ario Nacional / MCTI, The Ohio State University, 
Pennsylvania State University, Shanghai Astronomical Observatory, 
United Kingdom Participation Group,
Universidad Nacional Aut\'onoma de M\'exico, University of Arizona, 
University of Colorado Boulder, University of Portsmouth, 
University of Utah, University of Virginia, University of Washington, University of Wisconsin, 
Vanderbilt University, and Yale University.

\bibliographystyle{mnras}
\bibliography{archive}

\end{document}